\documentclass[aps,prx,twocolumn, superscriptaddress, longbibliography,  nofootinbib]{revtex4-2}
\usepackage{amsmath,amssymb}
\usepackage{dsfont,yfonts}
\usepackage{tabularx}
\usepackage{multirow}
\usepackage{diagbox}
\newcolumntype{C}{>{\centering\arraybackslash}X}%
\usepackage{graphicx}
\usepackage{subfigure}
\usepackage[usenames,dvipsnames]{xcolor}
\usepackage{tikz}
\usepackage{pgffor}
\usepackage{verbatim}
\usepackage{bbm}
\usepackage{float}
\usepackage{mathrsfs}
\usepackage{xcolor,colortbl}
\usepackage[colorlinks, bookmarks=true, breaklinks=true,linkcolor=red, citecolor=blue, linktocpage=true, urlcolor=blue]{hyperref}
\usepackage{float}
\usepackage{graphicx}
\usepackage{amssymb}
\usepackage{amsthm}
\usepackage{mathtools}

\newcommand{\g}[1]{{\mathbf{g}}}

\def\Z{\mathbb{Z}}

\newcommand{\modulo}[1]{{\ (\mathrm{mod}\ #1)}}
\newcommand{\dd}{\mathrm{d}}

\newcommand{\bfg}{\mathbf{g}}
\newcommand{\bfh}{\mathbf{h}}
\newcommand{\bfk}{\mathbf{k}}
\newcommand{\bbfg}{\bar{\mathbf{g}}}

\newcommand{\gbf}{\mathbf{g}}
\newcommand{\hbf}{\mathbf{h}}
\newcommand{\kbf}{\mathbf{k}}
\newcommand{\lbf}{\mathbf{l}}
\newcommand{\calH}{\mathcal{H}}
\newcommand{\Aut}{\mathrm{Aut}}
\newcommand{\calO}{\mathcal{O}}
\newcommand{\bfI}{\mathbf{1}}
\newcommand{\calL}{\mathcal{L}}

\newtheorem{theorem}{Criterion}

\newtheorem{col}{Corollary}

\begin{document}

\title{{\fontfamily{ptm}\fontseries{b}\selectfont Enforced  symmetry breaking by  invertible topological order}}

\author{Shang-Qiang Ning}
\affiliation{Department of Physics and HKU-UCAS Joint Institute for Theoretical and Computational Physics, The University of Hong Kong, Pokfulam Road, Hong Kong, China}

\author{Yang Qi}
\affiliation{Center for Field Theory and Particle Physics, Department of Physics, Fudan University, Shanghai 200433, China}\affiliation{State Key Laboratory of Surface Physics, Fudan University, Shanghai 200433, China} 
\affiliation{Collaborative Innovation Center of Advanced Microstructures, Nanjing 210093, China}

\author{Zheng-Cheng Gu}
\email{zcgu@phy.cuhk.edu.hk}
\affiliation{Department of Physics, The Chinese University of Hong Kong, Shatin, New Territories, Hong Kong}

\author{Chenjie Wang}
\email{cjwang@hku.hk}
\affiliation{Department of Physics and HKU-UCAS Joint Institute for Theoretical and Computational Physics, The University of Hong Kong, Pokfulam Road, Hong Kong, China}

\begin{abstract}  
It is well known that two-dimensional fermionic systems with a nonzero Chern number must break the time reversal symmetry, manifested by the appearance of chiral edge modes on an open boundary. Such an incompatibility between topology and symmetry can occur more generally. We will refer to this phenomenon as \emph{enforced symmetry breaking} (ESB) by topological orders. In this work,  we systematically study ESB of a finite symmetry group $G_f$ by fermionic invertible topological orders. Mathematically, the group $G_f$ is a central extension over a bosonic symmetry group $G$ by the fermion parity group $\Z_2^f$, characterized by a 2-cocycle $\lambda\in \mathcal{H}^2(G,\Z_2)$.   
For given $G$ and $\lambda$, we are able to obtain a series of criteria on the existence or non-existence of ESB by the corresponding fermionic invertible topological orders. Using these criteria, we discover many ESB examples that are not known previously. For 2D systems, we define a set of physical quantities to describe symmetry-enriched invertible topological orders and derive  obstruction functions using both fermionic and bosonic languages. In the latter case which is done via gauging the fermion parity, we find that some obstruction functions are consequences of \emph{conditional anomalies} of the bosonic symmetry-enriched topological states, with the conditions inherited from the original fermionic system.  We also study ESB of the continuous group $SU_f(N)$ by 2D invertible topological orders through a different argument.
\end{abstract}
\date{{\small\today}}
\maketitle

\clearpage 

\section{Introduction}
Symmetry and topology play very important roles in modern physics. One of the most profound concepts in physics is spontaneous symmetry breaking\cite{Landau1937obd}. It is behind many physical phenomena, ranging from superconductivity and Bose-Einstein condensation in condensed matter systems to the unification of electromagnetic and weak forces in particle physics. For a long time, people believed that Landau symmetry-breaking theory could describe all possible phases and continuous phase transitions. The discovery of fractional quantum Hall effects (FQHE)\cite{Tsui1982} opened the door to the realm of topological phases of matter where interesting physics such as fractional charge and fractional statistics were uncovered\cite{Laughlin1983}. 

Very recently, great effort has been made in the study of the interplay between symmetry and topology in quantum many-body systems. 
The concept of topological insulators\cite{KaneMele2005PRL1,KaneMele2005PRL2,bernevig2006quantum,konig2007quantum,FuPRL2007,hasan10,qi11} has been extended to a large class of short-range entangled states of matter, namely symmetry-protected topological (SPT) phases\cite{gu09}. Complete classifications\cite{chenScience2012,chen13,kapustin14a,wen15, GuWen2014,kapustin14,chong14,kapustin14,freed14,Cheng2018PRB,gaiotto16,freed16,morgan16,Kapustin2017,WangGu2017, WangGu2020PRX} as well as various characterizations\cite{levin12,cheng2014,threeloop,ran14,vishwanath13,wangc13,chen14,WangPRX2016, bonderson13,wangc13b,fidkowski13,chen14a,chong14,metlitski15,wangj15,Juven2015,wanggu16,Ning2021fSPT} and model realizations\cite{levin12,chen13,Fidkowski1604,Chen17,Son2019,Kobayashi2020,YuAn2021} have been obtained for interacting bosonic and fermionic SPT phases.  Moreover, for systems with anyon excitations, the interplay between symmetry and topology gives rise to various symmetry-enriched topological (SET) phases\cite{BarkeshliPRB2019_SET,Essin2013PRB,Mesaros2013PRB,Tarantino2016,Teo2015}. The FQHE states actually can be regarded as the simplest SET state with $U(1)$ charge conservation symmetry. For bosonic systems, a complete classification of SET phases has been achieved by using the so-called $G$-crossed braided fusion category theory\cite{BarkeshliPRB2019_SET}. However,  fermionic SET phases are much more complicated, and their classification and physical characterization have not been fully understood so far. 

While the mutual appreciation between topology and symmetry has resulted in many interesting physics, it is also known that they are not always compatible. For example, 2D Chern insulators with a nonzero Chern number must break the time reversal symmetry\cite{Haldane1988}. It is manifested by the existence of chiral edge modes that reverse the direction under time reversal action and thereby cannot appear in time-reversal symmetric systems. In fact, time-reversal is broken in any chiral topological order. Such incompatibility can occur more generally in other systems and for other symmetries. We will refer to this phenomenon as \emph{enforced symmetry breaking} (ESB) by topological orders. Of course, depending on one's viewpoint, it can also be called symmetry constraints on topological order. 

To be possibly enforced to break by certain topological orders, there is a precondition on the symmetry: it must have a nontrivial action on the Hilbert space by its very definition. A few examples of these symmetries are: (1) the time-reversal symmetry $\mathcal{T}$, under which the wave function  must be complex-conjugated, (2) the mirror reflection $\mathcal{R}$, under which the spatial orientation  associated with the many-body wave function must be reversed, and  (3) the fermion parity $P_f$, under which the states containing an odd number of fermions must obtain a minus sign. More generally, a symmetry group $\mathcal{G}$ that contains $P_f$, $\mathcal{T}$, and/or $\mathcal{R}$, also satisfies this precondition. For example, it was known that $p_x+ip_y$ superconductor is incompatible with $U_f(1)$ particle number conservation, which is a nontrivial extension of the fermion parity group $\Z_2^f = \{1, P_f\}$. If this precondition is not imposed, then a symmetry can always be implemented as the identity operator, a rather trivial way, so that it is compatible with any topological order. So far, systematic exploration of ESB physics has not been done yet and a general framework is very much desired.


In this work, we study the ESB physics for a special class of topological orders, namely \textit{invertible topological orders} (iTOs)\cite{freed2021reflection}. They are topological phases somewhat in between SPT phases and the topological orders that host anyons: they cannot be smoothly connected to the trivial product state even in the absence of symmetry, but they do not host anyon excitations. While it sounds exotic, all iTOs in low dimensions are known, including the 1D Majorana chain\cite{Kitaev2001}, 2D $p_x+ip_y$ superconductors\cite{NRead2000,Ivanov2001} (stacking even copies of which are topologically equivalent to integer quantum Hall states). See Table \ref{tab1} for a summary of fermionic iTOs in low dimensions, which can all be realized in non-interacting fermionic systems. For bosonic systems, the only nontrivial invertible topological order is the 2D $E_8$ state.\cite{e8} They are called ``invertible'' because, for every state, there exists an inverse state such that stacking the two gives rise to the trivial state. The ESB physics for the bosonic $E_8$ state is simple, which is not compatible with all anti-unitary or orientation-reversing symmetries but compatible with all internal unitary symmetries. On the other hand, we will see that ESB physics for fermionic iTOs are extremely rich.  

\begin{table}
\caption{Invertible topological orders of fermionic systems in low dimensions.}\label{tab1}
\begin{tabular}{ccc}
\hline\hline
 Dimension $\ $ & $\ $ Classification $\ $ &  Root state  \\
\hline
0D & $\Z_2$ & complex fermion \\
1D & $\Z_2$ & Kitaev's majorana chain \\
2D & $\Z$ & $p_x + ip_y$ superconductor  \\
3D & $\Z_1 $ &  trivial \\
\hline 
\end{tabular}
\end{table}

\begin{table*}
\caption{Locations of ESB criteria for 0D, 1D and 2D fermionic invertible topological orders in this paper, and a few examples of ESB with unitary symmetries. For the dihedral group $D_8$ and quaternion group $Q_8$, the fermion parity group $\Z_2^f$ is identified as their $\Z_2$ centers. For $SU_f(N)$, the fermion parity $\Z_2^f$ is identified as the $\Z_2$ subgroup of its center which is $\Z_N$. We note that Criterion \ref{criterion2} was known previously in Ref.~\cite{FidkowskiPRB2011}. }\label{tab2}
\begin{tabular}{lll}
\hline\hline
 \ iTO &  ESB criteria  &   Examples \\
\hline
\ 0D complex fermion $\quad$ &  Criterion \ref{criterion1} in Sec.~\ref{sec:0d-criterion} &  $D_8$, $Q_8$  \quad \\
\ 1D Majorana chain  & Criterion \ref{criterion2} in Sec.~\ref{sec:esb-1D} &  $\Z_4^f$, any $G_f$ with nontrivial $\lambda$ \\
\ 2D iTO with odd $\nu$ &  Criterion \ref{criterion3} in Sec.~\ref{sec:criterion_odd_nu} &  $\Z_4^f$, any $G_f$ with nontrivial $\lambda$  \\
\ 2D iTO with $\nu=4k+2$&  Criterion \ref{criterion4} in Sec.~\ref{sec:criterion_4k+2}  &  $Q_8$,  $SU_f(2)$\\
\ 2D iTO with $\nu=8k+4$ \quad \quad \quad\quad  &  Criterion \ref{criterion5} in Sec.~\ref{sec:criterion_8k+4}  \quad \quad \quad \quad&  $SU_f(4)$, an order-32 group in Sec.~\ref{sec_example_nu4} \ \\
\ 2D iTO with $\nu=16k+8$ &  Criterion \ref{criterion6} in Sec.~\ref{sec:criterion_16k+8}   &  $SU_f(8)$\\
\hline 
\end{tabular}
\end{table*}

We will study 0D, 1D and 2D fermionic iTOs with a general finite group $G_f$ --- in other words, symmetry-enriched iTOs. Mathematically, the group $G_f$ is a central extension of a bosonic symmetry group $G$ by the fermion parity symmetry $\Z_2^f$. Different central extensions are characterized by non-trivial 2-cocycles $\lambda\in \mathcal{H}^2(G,\Z_2)$. However, we will also study an interesting example of the continuous symmetry group $SU_f(N)$  in Sec.~\ref{sec:SU(N)}. The main result of this work is a set of criteria on whether a given $G_f$ is enforced to break by 0D, 1D, or 2D iTOs (summarized in Table \ref{tab2}) and various examples that exhibit ESB physics. When deriving the 2D criteria, we develop the description of 2D symmetry-enriched fermionic iTOs using both fermionic and bosonic languages, with the latter achieved via gauging the fermion parity and turning the fermionic iTOs to bosonic SETs. In particular, we derive the formulas for the so-called obstruction functions, which are important components of the ESB criteria. It is unfortunate that in one of the cases, our explicit formula is not the most general. In the bosonic SET language, we also find that the obstruction functions are consequences of \emph{conditional anomalies}, with the conditions inherited from the preconditions in the definition of the fermionic group $G_f$.

Characterization of symmetry-enriched fermionic iTOs involves two categories of quantities. The first category is a triplet $(G, \lambda, \nu)$, where $G$ is the bosonic symmetry group, $\lambda$ is a cocycle in $\calH^2(G, \Z_2)$, and $\nu$ is an element of the iTO classification group (see Table \ref{tab1}). The first two quantities $G$ and $\lambda$ determine the fermionic group $G_f$, and the third quantity $\nu$ labels the iTO which can be viewed as a characterization of its intrinsic topological property. Even between these quantities, incompatibility may occur and lead to ESB phenomena. The ESB Criteria \ref{criterion1}, \ref{criterion2} and \ref{criterion3} are of this type. The second category involves the quantities that describe symmetry enrichment. For examples, we define two quantities $n_1$ and $n_2$ (which are actually functions) in Sec.~\ref{sec:def-n1n2} for 2D fermionic iTOs: $n_1$ characterizes if certain symmetry defects carry Majorana zero modes and $n_2$ characterizes how fermion parity conservation is achieved when defects fuse. Unlike the quantities in the first category which are given, they may vary. The symmetry enrichment quantities are  well-defined only when they are compatible with the given quantities in the first category. Otherwise, if no compatible symmetry enrichment quantities exist, ESB occurs. The ESB Criteria \ref{criterion4}, \ref{criterion5} and \ref{criterion6} are of this type. 

The rest of this paper is organized as follows. In Sec.~\ref{sec:0D1D}, we study ESB by 0D and 1D fermionic iTOs. We set up our convention, derive the ESB criteria, and explore a few examples. Various 2D ESB criteria and examples are given in Sec.~\ref{sec:2dESB}. In particular, we develop a description of symmetry-enriched fermionic iTOs in Sec.~\ref{sec:def-n1n2} using defects in the fermionic language. Also, an ESB example of $SU_f(N)$ group is discussed in Sec.~\ref{sec:SU(N)}. We derive the 2D criteria in Sec.~\ref{sec:derivation} by transforming fermionic iTOs into bosonic SET states via gauging the fermion parity. The obstruction functions are obtained. We make a summary and discuss some potential generalizations in Sec.~\ref{sec: discussion}. Appendices \ref{app:cohomology}, \ref{app:localization_anomaly} and \ref{sec_calculation_of_H3_nu2_abelian} contain some technical analyses.

\section{ESB by 0D and 1D iTOs}
\label{sec:0D1D}

To warm up, we discuss ESB by 0D and 1D fermionic iTOs in this section.  We begin with the definition of symmetry group in fermionic systems. Then, we discuss ESB criteria, examples, and derivations of the criteria for 0D and 1D fermionic iTOs.
 
\subsection{Symmetry} 
\label{sec:sym}

Let $G_f$ be the symmetry group of a general fermionic system. Throughout the paper, we assume all symmetries are internal and $G_f$ is finite, except in Sec.~\ref{sec:SU(N)} where a case of continuous symmetries will be discussed. In all fermionic systems, $G_f$ must contain a special element, the fermion parity $P_f$, with $P_f^2=1$. It must be preserved due to the requirement of locality. Moreover, $P_f$ should commute with all other symmetries. Let $\Z_2^f=\{1, P_f\}$ be the subgroup formed by the fermion parity, which sits inside the center of $G_f$. Mathematically, the relation between $G_f$ and $\Z_2^f$ is given by a short exact sequence 
\begin{equation}
1\rightarrow \Z_2^f \rightarrow G_f \rightarrow G \rightarrow 1,
\label{eq:ses}
\end{equation}
where $G$ is the quotient group $G_f/\Z_2^f$. For a given $G$, different $G_f$'s are said to be different central extensions of $G$ by $\Z_2^f$.

A central extension of $G$ by $\Z_2^f$ is determined by a 2-cocycle $\lambda\in \mathcal{H}^2(G,\Z_2)$.\footnote{The mathematically correct notation should be the cohomology class $[\lambda]\in \mathcal{H}^2(G,\Z_2)$. However, with abuse of notation, we simply use $\lambda\in \mathcal{H}^2(G,\Z_2)$ or say ``$\lambda$ is a 2-cocycle in $\mathcal{H}^2(G,\Z_2)$'' to denote that $\lambda$ is a representative 2-cocycle in the class $[\lambda]$. Similar notation is used for other cocycles.} A brief review on group cohomology is given in Appendix \ref{app:cohomology}. A 2-cocycle is a function $\lambda: G\times G \rightarrow \Z_2=\{0,1\}$ that satisfies the condition:
\begin{align}
0  & = \dd \lambda(\gbf,\hbf,\kbf) \nonumber\\
  & = \lambda(\hbf,\kbf) -\lambda(\gbf\hbf,\kbf) +\lambda(\gbf,\hbf\kbf)-\lambda(\gbf,\hbf),
\label{eq:cocycle_con}
\end{align}
where $\gbf,\hbf\in G$, $\dd$ is the coboundary operator, and ``modulo $2$'' is implicitly assumed for addition.\footnote{An implicit ``modulo'' is assumed in most additive expressions and in the expressions of the coboundary operator $\dd$, whenever we use additive convention for Abelian groups. If a ``modulo'' is not taken, we will use $\hat\dd$ for distinction.}  Two cocycles $\lambda$ and $\tilde{\lambda}$ are equivalent if $\tilde{\lambda}  = \lambda + \dd\epsilon$ and 
\begin{align}
\dd\epsilon(\gbf,\hbf)  =  \epsilon(\gbf)+\epsilon(\hbf)-\epsilon(\gbf\hbf),
\label{eq:cobound}
\end{align}
where $\epsilon(\gbf)$ is an arbitrary function $\epsilon:G\rightarrow \Z_2$,  and $\dd \epsilon(\gbf,\hbf)$ is called a 2-coboundary (see Appendix \ref{app:cohomology}). A 2-coboundary is automatically a 2-cocycle. The equivalence classes $[\lambda]$ form the cohomology group $\mathcal{H}^2(G,\Z_2)$, with the equivalence class of 2-coboundaries being the identity. Due to the cocycle condition \eqref{eq:cocycle_con} and coboundary transformation \eqref{eq:cobound}, it is always possible to take the convention 
\begin{equation}
\lambda(\mathbf{1},\gbf)=\lambda(\gbf,\mathbf{1})=0.
\label{eq:lambda-convention}
\end{equation}
where $\mathbf{1}$ is the identity element of $G$. This convention will simplify many of our discussions.

Given $G$ and $\lambda$, the group $G_f$ can be constructed as follows:  $G_f=\{\gbf_\sigma|\gbf\in G, \sigma \in \Z_2\}$ and the group multiplication is 
\begin{equation}
\gbf_\sigma \hbf_\tau = (\gbf\hbf)_{\sigma+\tau+\lambda(\gbf,\hbf)},
\label{eq:gmulti}
\end{equation}
where $\bfI_0$ is the identity and $\bfI_1$ is the fermion parity $P_f$. One can check that associativity of multiplication is guaranteed by the cocycle condition \eqref{eq:cocycle_con}. Two fermionic groups $G_f$ and $\tilde{G}_f$, constructed from equivalent cocycles $\lambda$ and $\tilde{\lambda}$  respectively, are isomorphic. The isomorphism is given by $\gbf_\sigma \leftrightarrow \gbf_{\sigma+\epsilon(\gbf)}$, where $\gbf_\sigma\in G_f$ and $\gbf_{\sigma+\epsilon(\gbf)}\in \tilde{G}_f$. In this work, we will use $(G,\lambda)$ and $G_f$ interchangeably to describe symmetries in fermion systems.\footnote{Two inequivalent cocycles may also produce isomorphic $G_f$'s. It occurs when the cocycles can be related by an automorphism of $G$. However, this subtlety is not important to our discussions.} Nevertheless, we would like to emphasize that given any microscopic system, the symmetries $\gbf_0$ and $\gbf_1$ are physically distinct, corresponding to different operators (a.k.a. observables). So, one should keep in mind that a coboundary transformation $\epsilon(\gbf)$ does have physical consequences.

The group $G$ may contain anti-unitary symmetries, such as time-reversal. To specify if a symmetry is unitary or anti-unitary,  we need another piece of data, a group homomorphism
\begin{equation}
s: G \rightarrow \Z_2^T,
\end{equation}
where $\Z_2^T=\{0,1\}$ (with group multiplication being addition modulo 2). The element $\gbf$ is unitary if $s(\gbf)=0$ and anti-unitary if $s(\gbf)=1$. It satisfies $s(\gbf)+s(\hbf)= s(\gbf\hbf) \modulo{2}$. The special case that $s(\gbf) = 0$ for every $\gbf$ corresponds to that all symmetries are unitary. Since $P_f$ is unitary, we can make the extension
\begin{equation}
s(\gbf_\sigma) = s(\gbf).
\end{equation}
After the extension, $s$ becomes a homomorphism from $G_f$ to $\Z_2^T$.


\subsection{ESB criterion for 0D iTO}
\label{sec:0d-criterion}

In the usual sense, 0D systems neither host topological order nor support spontaneous symmetry breaking. So, it is not very interesting to study the phenomenon of ESB. However, there is indeed a sense of ESB as we explain below, exploring which helps to establish a few basic concepts.

One way to define 0D fermionic iTO is as follows. Consider 0D fermionic systems with a gapped unique ground state, so that they are invertible. Two systems are said to have the same topological order if the ground states can be smoothly deformed to each other by a fermionic unitary transformation(or bosonic unitary transformation with a $\Z_2^f$ symmetry). Under this definition, there are two inequivalent fermionic iTOs: those with the ground state being even under $P_f$, and those with the ground state being odd under $P_f$. Simple examples are the states $|0\rangle$ and $c^\dag|0\rangle$, respectively, where $|0\rangle$ is the vacuum in the Fock space and $c^\dag$ is a fermion creation operator. The latter is considered to be topologically non-trivial. We will refer to it as the ``complex-fermion iTO''.

Now we ask if the complex-fermion iTO enforces certain symmetry group $G_f$ to break. It should be an explicit symmetry breaking, as there is no spontaneous symmetry breaking in 0D. Let $|\Psi_0\rangle$ be the ground state of any system that hosts the complex-fermion iTO. It follows from the definition that
\begin{equation}
P_f |\Psi_0\rangle = -|\Psi_0\rangle.
\label{eq:pfrep}
\end{equation}
In the presence of a symmetry group $G_f$, the ground state $|\Psi_0\rangle$ should form a one-dimensional representation of $G_f$, with the condition \eqref{eq:pfrep} satisfied. Therefore, if $P_f={1}$ in all its one-dimensional representations, $G_f$ is incompatible with the complex-fermion iTO. That is, $G_f$ is enforced to break by the complex-fermion iTO.

This understanding can be used to derive a quantitative criterion on whether ESB occurs for a given symmetry group. Let $G_f$ be determined by  the pair $(G,\lambda)$, where $\lambda\in \mathcal{H}^2(G, \Z_2)$ is a 2-cocycle. We show that
\begin{theorem}
$G_f$ is enforced to break by the complex-fermion iTO, if and only if $(-1)^{\lambda(\gbf,\hbf)}$ is a non-trivial cocycle in $\mathcal{H}^2(G, U_T(1))$. \label{criterion1}
\end{theorem}

Derivation of the criterion will be deferred to Sec.~\ref{sec:deri-0D}. Here, we elaborate the criterion. A 2-cocycle in $\calH^2(G,U_T(1))$ is  any function $\alpha: G\times G \rightarrow U(1)=\{ e^{i\theta}| \theta\in[0, 2\pi)\}$, which satisfies the condition
\begin{equation}
\alpha(\gbf,\hbf)\alpha(\gbf\hbf,\kbf) = \alpha(\gbf,\hbf\kbf)K^{s(\gbf)}[\alpha(\hbf,\kbf)],
\label{eq:u1-2cocycle}
\end{equation}
where $s(\gbf)=0,1$ denotes if $\gbf$ is unitary or anti-unitary, and $K$ is the operation of complex conjugation with $K[\alpha] = \alpha^*$.  Two cocycles $\alpha$ and $\tilde{\alpha}$ are equivalent if they differ by a 2-coboundary:
\begin{align}
\tilde{\alpha}(\gbf,\hbf) = \alpha(\gbf,\hbf) \frac{\epsilon(\gbf)K^{s(\gbf)}[\epsilon(\hbf)]}{\epsilon(\gbf\hbf)}.
\label{eq:u1-cobound}
\end{align}
Different from Eq.~\eqref{eq:cobound}, $\epsilon(\gbf)$ can now be any $U(1)$-valued function. The equivalence classes $[\alpha]$ define the cohomology group $\mathcal{H}^2(G,U_T(1))$ with the subscript ``$T$'' denoting the nontrivial action on $U(1)$ values by the complex conjugation.

It is easy to see that $(-1)^{\lambda(\gbf,\hbf)}$ satisfies Eq.~\eqref{eq:u1-2cocycle}, following from the fact that $\lambda(\gbf,\hbf)$ satisfies Eq.~\eqref{eq:cocycle_con}. Accordingly, $(-1)^{\lambda(\gbf,\hbf)}$ is indeed a 2-cocycle in $\mathcal{H}^2(G,U_T(1))$. However, due to the enlarged choice of $\epsilon(\gbf)$ in the co-boundary transformation \eqref{eq:u1-cobound} compared to that in \eqref{eq:cobound}, certain $\lambda$, which belongs to a nontrivial cohomology class in $\mathcal{H}^2(G,\Z_2)$,  may produce a cocycle $(-1)^{\lambda(\gbf,\hbf)}$ that is trivial in $\mathcal{H}^2(G,U_T(1))$. Criterion \ref{criterion1} states that ESB occurs if and only if $(-1)^{\lambda(\gbf,\hbf)}$ is a non-trivial 2-cocycle in $\mathcal{H}^2(G,U_T(1))$.

\subsection{Examples}
\label{sec:exampl_0D}

\begin{table}
\caption{Character table of the dihedral group $D_8$. The first row shows the conjugacy classes and the next five rows show the characters of the five irreducible representations of $D_8$. The fermion parity $P_f$ is identified with the elment $r^2$.}
\label{tab:dihedral}
\begin{tabular}{c|ccccc}
\hline 
dim & $\quad 1 \quad $ &  $\quad r^2\quad $ & $\ \{r, r^3\}\ $ & $\ \{s, sr^2\}\ $ & $\ \{sr, sr^3\}\ $ 
\\
\hline 1D  & 1 & 1 & 1 & 1 & 1 \\
1D & 1 & 1 & $1$ & $-1$ & $-1$\\
$\ $ 1D $\ $ & 1 & 1 & $-1$ & $1$ & $-1$\\
1D & 1 & 1 & $-1$ & $-1$ & $1$\\
2D   & $2$ & $-2$ & 0 & 0 & 0\\
\hline
\end{tabular}
\end{table}

We illustrate 0D ESB by exploring a few examples. The simplest example is $G=\Z_2^T= \{1, T\}$, where $T$ is the time-reversal symmetry, and $G_f =\Z_4= \{1, T, P_f, TP_f| T^2 = P_f\}$. In this case, the cohomology group $\calH^2(\Z_2^T, \Z_2) = \Z_2$. One can check that the two inequivalent cocycles are specified by a single quantity $\lambda(T, T) = 0$ and $1$, respectively. The cocycle with $\lambda(T,T)=1$ is non-trivial and corresponds to $G_f = \Z_4$. One can show that $(-1)^{\lambda(\gbf,\hbf)}$ is non-trivial in $\calH^2(G, U_T(1))$. So, $G_f$ is enforced to break by the complex-fermion iTO, according to Criterion \ref{criterion1}. This is actually the well-known Kramers theorem: for fermions with $T^2=P_f$, the odd-parity states must form a doublet. That is, a ground state satisfying \eqref{eq:pfrep} cannot be non-degenerate.

The simplest example of unitary symmetries is $G_f = D_8$, the dihedral group of order 8, with $\Z_2^f$ being the center of $D_8$. More explicitly, let $D_8 = \{s^nr^m|n=0,1, m=0,1,2,3, r^4=s^2=1, srs=r^{3},\}$ and the fermion parity $P_f =r^2$. All irreducible representations of $D_8$ are listed in Table \ref{tab:dihedral}. We observe that in all the 1D representations, $P_f=r^2$ is represented by $1$, which is inconsistent with the condition \eqref{eq:pfrep} for the complex-fermion iTO. Therefore, this $D_8$ is enforced to break by the complex-fermion iTO. 

The $D_8$ example can also be seen from Criterion \ref{criterion1}. Nevertheless, let us use the criterion to study a family of examples, with $D_8$ being one of them. We take $G$ to be a general Abelian group $G=\prod_{i=1}^k\Z_{N_i}$, with all symmetries being unitary. Without losing too much generality, we assume all $N_i$ are even. Then, the cohomology group $\mathcal{H}^2(G,\Z_2) = (\Z_2)^{k(k+1)/2}$. A class of representative 2-cocycles in  $\mathcal{H}^2(G,\Z_2)$ is
\begin{align}
 \lambda(a,b)=\sum_i p_i w_{i}(a,b)+\sum_{i<j} p_{ij}w_{ij}(a,b),
\label{eqn_general_parametrize}
\end{align} 
where the indices $i,j$ run in $1,2,\dots, k$, and the parameters $p_i,p_{ij}=0$ or 1. To define $w_i(a,b)$ and $w_{ij}(a,b)$, let us denote the group elements as $a=(a_1,a_2,..., a_k)$ where $a_i=0,1,...,N_i-1$.  Then, $w_i(a,b)$ and $w_{ij}(a,b)$ are given by
\begin{align}
w_i(a,b)&=\frac{1}{N_i}(a_i+b_i-[a_i+b_i]_{N_i}),\label{eqn_type_1_2cocycle}\\
w_{ij}(a,b)&=a_ib_j \modulo{2},  \label{eqn_type_2_2cocycle}
\end{align}
where $[n_1+n_2]_N = n_1+n_2 \modulo{N}$. Both $w_i$ and $w_{ij}$ only take a value $0$ or $1$. One can check $\lambda(a,b)$ in the form \eqref{eqn_general_parametrize} is indeed a 2-cocycle in $\calH^2(\prod_i \Z_{N_i}, \Z_2)$ for any choice of $p_i$ and $p_{ij}$. It is useful to define the following quantities: 
\begin{align}
\Omega_i & =\sum_{i=0}^{N_i-1}\lambda(n e_i,e_i), \nonumber\\
\Omega_{ij} & =\lambda(e_i,e_j)-\lambda(e_j,e_i), \label{eqn_invariant_group_extension}
\end{align}
where ``modulo 2'' is assumed,  $e_i = (0,\dots, 0,1,0,\dots,0)$ with only the $i$th entry being $1$ and others being $0$. These quantities have a nice property that they are invariant under coboundary transformations, so we will call them \emph{topological invariants} for $\mathcal{H}^2(G,\Z_2)$ (see Appendix \ref{app:invariant} for more details). Inserting the explicit cocycle in \eqref{eqn_general_parametrize}, we obtain
\begin{align}
\Omega_i = {p_i}, \quad 
\Omega_{ij} = {p_{ij}},
\end{align}
where $i<j$. By varying $p_i$ and $p_{ij}$, the set $\{\Omega_i, \Omega_{ij}\}$ can have $2^{k(k+1)/2}$ distinct values, saturating the number $|\mathcal{H}^2(G,\Z_2)|$ of cohomology classes. Therefore, it is a complete set of topological invariants and the representative cocycle in \eqref{eqn_general_parametrize} exhausts all inequivalent cocycles in $\calH^2(G,\Z_2)$. 

Now consider the cocycle $(-1)^{\lambda(\gbf,\hbf)}$ in $\calH^2(G,U(1))$ (we drop the subscript $T$ as we only consider unitary symmetries here). Under coboundary transformations of $U(1)$-valued functions, one can check that $(-1)^{\Omega_i}$ is not invariant, but  $(-1)^{\Omega_{ij}}$ remains a well-defined topological invariant. Moreover, $(-1)^{\Omega_{ij}}$ is complete in the sense that it is enough to tell if  $(-1)^{\lambda(\gbf,\hbf)}$ is trivial or non-trivial in $\calH^2(G,U(1))$\cite{WangPRB2015}. Therefore, $(-1)^{\lambda(\gbf,\hbf)}$  is non-trivial in $\calH^2(G,U(1))$ if and only if any $\Omega_{ij}=1$.  For convenience, here and after, we call a 2-cocycle in $\mathcal{H}^2(G,\Z_2)$ as Type-I if all invariants $\Omega_{ij}=0$, and otherwise we call it Type-II. Accordingly, the corresponding $G_f$ is enforced to break by the complex-fermion iTO if and only if $\lambda$ is Type-II.

Take the simplest case $G=\Z_2\times \Z_2$ and let $\Omega_{12}=1$. Then, ESB occurs for the corresponding $G_f$.  Depending on the values of $\Omega_1$ and $\Omega_2$, we have two cases: (i) when $\Omega_1=\Omega_2=1$, $G_f$ is the quaternion group $Q_8$; (ii) otherwise, $G_f$ is the dihedral group $D_8$.

\subsection{Derivation of 0D criterion}
\label{sec:deri-0D}

We prove an alternative statement that is equivalent to Criterion \ref{criterion1}: $G_f$ is compatible with the complex-fermion iTO, if and only if $(-1)^{\lambda(\gbf,\hbf)}$ is trivial in $\mathcal{H}^2(G, U_T(1))$. The equivalence is obvious. Derivation of this statement mainly involves the representation theory of groups. 

Let us start with the ``only if'' direction. To show that, we assume that $G_f$ is compatible with the complex-fermion iTO. Then, the ground state forms a one-dimensional representation of $G_f$.  Let the representation be $U(\gbf_\sigma)K^{s(\gbf)}$ for $\gbf_\sigma \in G_f$, where $U(\gbf_\sigma)$ is a unitary operator and $K$ is the operator of complex conjugation. In the special case that $s(\gbf)=0$ for all $\gbf\in G$, it reduces to the usual unitary representation of a group. The  operators satisfy
\begin{equation}
U(\gbf_\sigma) K^{s(\gbf)} U(\hbf_\tau) K^{s(\hbf)} = U[(\gbf\hbf)_{\sigma+\tau+\lambda(\gbf,\hbf)}] K^{s(\gbf\hbf)},
\label{eq:irrep}
\end{equation}
which follows from the group multiplication law \eqref{eq:gmulti}. To fulfill Eq.~\eqref{eq:pfrep} and the requirement $U(\mathbf{1}_0)=1$, we have $U(\mathbf{1}_\sigma) = (-1)^\sigma$.  The convention \eqref{eq:lambda-convention} implies $\gbf_0 \mathbf{1}_\sigma = \gbf_{\sigma}$. So, we have
\begin{align}
U(\gbf_\sigma) K^{s(\gbf)} &= U(\gbf_{0}) K^{s(\gbf)} U(\mathbf{1}_\sigma) K^{s(\mathbf{1})} \nonumber \\
&= (-1)^\sigma U(\gbf_0) K^{s(\gbf)}.
\end{align}
Combining this equation with Eq.~\eqref{eq:irrep} and taking the shorthand notation $U(\gbf_0) = U(\gbf)$, we immediately have
\begin{equation}
U(\gbf) K^{s(\gbf)}  U(\hbf) K^{s(\hbf)} = (-1)^{\lambda(\gbf,\hbf)} U(\gbf\hbf) K^{s(\gbf\hbf)}.
\end{equation}
It is a projective representation of $G$, where $(-1)^{\lambda(\mathbf{g},\mathbf{h})}\equiv \alpha(\gbf,\hbf)$ is called a factor set. In general, the factor set of projective representations can be any 2-cocycle $\alpha(\gbf,\hbf)\in\mathcal{H}^2(G,U_T(1))$. A well known result is that if $\alpha$ is a non-trivial 2-cocycle, the representation cannot be 1D. The fact that we have a 1D representation implies that $\lambda$ must be a trivial 2-cocycle in $\mathcal{H}^2(G,U_T(1))$. Therefore, we have proven that if $G_f$ is compatible with the complex-fermion iTO, $(-1)^{\lambda(\gbf,\hbf)}$ must be trivial  in $\mathcal{H}^2(G,U_T(1))$. 

To show the ``if'' direction, we explicitly construct a 1D representation of $G_f$ with Eq.~\eqref{eq:pfrep} satisfied.  If $(-1)^{\lambda(\mathbf{g},\mathbf{h})}$ is trivial in $\mathcal{H}^2(G,U_T(1))$, it can be written as $\epsilon(\mathbf{g})K^{s(\gbf)}[\epsilon(\mathbf{h})]/\epsilon(\mathbf{g}\mathbf{h})$, where $\epsilon(\gbf)$ is some $U(1)$-valued function. Then, we take the representation to be
\begin{equation}
U(\mathbf{g}_\sigma)=(-1)^\sigma\epsilon(\mathbf{g}) K^{s(\gbf)}.
\end{equation} 
One can check that it is a 1D representation of $G_f$ that satisfies both \eqref{eq:irrep} and \eqref{eq:pfrep}. This completes our proof.

\subsection{ESB by 1D iTO}
\label{sec:esb-1D}

Invertible topological orders in 1D fermionic systems are classified by $\Z_2$. A representative of the nontrivial iTO is the famous Majorana chain, first discovered by Kitaev\cite{Kitaev2001}. So, we will refer to the nontrivial iTO as the ``Majorana-chain iTO''. The salient feature of the Majorana chain is that, when it is open, there exist robust zero modes at both ends, known as the Majorana zero modes (MZM). More specifically, it means the existence of Majorana operators $\gamma_l$ and $\gamma_r$ at the left and right ends, respectively, such that $[\gamma_l, H] =[\gamma_r, H]  =0$, where $H$ is the Hamiltonian of the chain. Majorana operators are fermionic, self-adjoint, squared to 1, and guarantee a two-fold ground-state degeneracy. Intuitively, the latter means the MZM at each end carries a ``fractional'' Hilbert space of dimension $\sqrt{2}$. The degeneracy is topologically protected and cannot be lifted by any local perturbations that respect the fermion parity.

Now consider a symmetry group $G_f$,  determined by the pair $(G,\lambda)$ with $\lambda\in \mathcal{H}^2(G,\Z_2)$. We ask if $G_f$ is compatible with the Majorana-chain iTO. This question has already been answered in the seminal paper Ref.~\onlinecite{FidkowskiPRB2011}. In our language, the result of Ref.~\onlinecite{FidkowskiPRB2011} can be stated as the following criterion:
\begin{theorem}
$G_f$ is enforced to break by the Majorana-chain iTO, if and only if $\lambda$ is a non-trivial 2-cocycle in $\mathcal{H}^2(G, \Z_2)$.
\label{criterion2}
\end{theorem}
\noindent In other words, the Majorana-chain iTO is only compatible with $G_f=\Z_2^f\times G$. This criterion holds regardless if $G$ contains antiunitary symmetries.

To be self-contained, we briefly revisit the proof given in Ref.~\onlinecite{FidkowskiPRB2011}. Without loss of generality, systems hosting the Majorana-chain iTO can be viewed as a stack of $2n+1$  Majorana chains with interaction between the chains allowed.  For open boundaries, there are $2n+1$ Majorana operators at each end. Let $\gamma_1,\gamma_2\dots, \gamma_{2n+1}$ be those at the left end. Then, any local operator at the left end (left-local operator) can be written as a sum of products of the Majorana operators. An important feature of the algebra of left-local operators is that there exist and only exist two operators that commute with all left-local operators and that square to 1. The two operators are $Z$ and $-Z$, with $Z=i^n \gamma_1\dots\gamma_{2n+1}$ and $Z^2=1$. The operator $Z$ contains an odd number of Majorana operators so that $P_fZ=-Z P_f$. Note that the fermion parity $P_f$ is not left-local.

Now consider symmetries in $G_f$. For $\gbf_\sigma\in G_f$, let $U(\gbf_\sigma)K^{s(\gbf)}$ be the corresponding operator that acts on the whole low-energy Hilbert space that includes both left and right ends. Under the action of $U(\gbf_\sigma)K^{s(\gbf)}$,  left-local operators remain left-local, with their algebraic structure preserved. In particular, $Z$ can only be transformed to either $Z$ or $-Z$. More specifically,
\begin{align}
U(\gbf_\sigma)K^{s(\gbf)} Z [U(\gbf_\sigma)K^{s(\gbf)}]^{-1} = (-1)^{\mu(\gbf_\sigma)}Z.
\end{align}
where $\mu(\gbf_\sigma)=0,1$ specifies how $Z$ transforms under the action of $\gbf_\sigma$. The specific $\mu(\gbf_\sigma)$ depends on details, except $\mu(\mathbf{1}_{1})=1$ due to $P_fZ=-Z P_f$. The operators $U(\gbf_\sigma)K^{s(\gbf)}$ shall form a representation of $G_f$, so that $\mu(\gbf_\sigma)$ is a group homomorphism from $G_f$ to $\Z_2$. Accordingly, 
\begin{align}
\mu(\gbf_\sigma) = \sigma+ \mu(\gbf_0),  \modulo{2}.
\end{align}
Then, between $\mu(\gbf_0)$ and $\mu(\gbf_1)$, one of them must be $0$. Let us pick out all the group elements with the $\mu$ value being 0. They are closed under multiplication and form a subgroup $G'\subset G_f$. Since $\Z_2^f$ is central in $G_f$, we must have $G_f = \Z_2^f\times G'$ and accordingly $G'$ is isomorphic to $G$. Hence, we have shown that the Majorana-chain iTO is compatible with $G_f$ only if it is a trivial extension of $G$ by $\Z_2^f$, i.e., $\lambda$ is a trivial 2-cocycle in $\mathcal{H}^2(G, \Z_2)$.  On the other hand, if $\lambda$ is trivial, $G_f$ is always compatible with the Majorana-chain iTO --- one simply represents all elements in $G$ by the identity operator. This concludes the proof.

\section{ESB by 2D iTOs}
\label{sec:2dESB}

Two-dimensional fermionic invertible topological phases are classified by $\Z$. The iTO indexed by $\nu\in \Z$ is exemplified by a stack of $\nu$ layers of $p_x+ip_y$ superconductors ($\nu\ge 0$) or $|\nu|$ layers of $p_x-ip_y$ superconductors ($\nu<0$). It is characterized by the chiral central charge $c_-=\nu/2$ of the gapless theory that lives on its edge. Physically, $c_-$ can be measured by the quantized thermal Hall conductance.  We note that $c_-$ is odd under anti-unitary symmetries. If $G_f$ contains an anti-unitary symmetry, it is always incompatible with any nontrivial 2D iTOs. That is, any $G_f$ containing anti-unitary symmetries is enforced to break by any non-trivial fermionic iTOs. Hence, from now on, we will only consider $G_f$ of unitary symmetries. In this section, we will define the data to describe 2D symmetry-enriched iTOs and obtain some obstruction functions, after which we state the ESB criteria, followed with a few examples.  Detailed derivations of ESB criteria will be given in Sec.~\ref{sec:derivation}. Section \ref{sec:SU(N)} is a special subsection that discusses an ESB example with continuous  symmetry group $SU_f(N)$.

\begin{table}
\caption{$\mathcal{O}_2$, $\mathcal{O}_3$ or $\mathcal{O}_4$ obstructions for 2D fermionic iTOs with index $\nu$. The mark ``$\times$'' means that the corresponding obstruction is always trivial, and ``$\bigcirc$'' means it may be nontrivial. Note that $\mathcal{O}_3$ is meaningful only if $\mathcal{O}_2$ is trivial, and $\mathcal{O}_4$ is meaningful only if both $\mathcal{O}_2$ and $\mathcal{O}_3$ are trivial.}
\label{tab:o2o3o4}
\begin{tabular}{c|ccc}
\hline 
$\quad \nu\quad $ & $\quad \mathcal{O}_2 \quad $ &  $\quad \mathcal{O}_3 \quad $ & $\quad \mathcal{O}_4 \quad $ 
\\
\hline even  & $\times$ & $\bigcirc $ &  $\bigcirc$ \\
odd & $\bigcirc$ &  $\times$  & $\bigcirc $ \\
\hline
\end{tabular}
\end{table}

\subsection{Symmetry-enriched fermionic iTOs}
\label{sec:def-n1n2}
Different from usual 2D topological orders, iTOs do not support anyon excitations.  Characterization of symmetry-enriched iTOs is more like that of SPT phases rather than SET phases. We will discuss two equivalent descriptions to characterize 2D fermionic symmetry-enriched iTOs: (i) by studying properties of symmetry defects (i.e., static version of gauge fluxes) in the fermionic theory and (ii) by gauging the fermion parity $\Z_2^f$ and studying the resulting \emph{bosonic} SETs of symmetry group $G$.  In this section, we will use the former to define a set of data to describe 2D symmetry-enriched iTOs, as it is physically more intuitive. This description allows us to obtain two obstructions, $\calO_2$ and $\calO_3$, which are important quantities in the ESB criteria. However, it is theoretically less mature than the latter, so we will use the second description to provide more detailed derivations of the ESB criteria in Sec.~\ref{sec:derivation}. 

We define a triplet $(\nu, n_1,n_2)$ to describe 2D symmetry-enriched fermionic iTOs. The first quantity $\nu\in\Z$ is defined above. It is a quantity associated with the intrinsic topology, which indexes the iTO and determines the chiral central charge $c_-=\nu/2$.  Depending on whether $\nu$ is even or odd, the fermion-parity defect behaves differently: for $\nu$ being odd, it carries an odd number of Majorana zero modes (MZMs)\cite{Ivanov2001}, which cannot be completely annihilated by local perturbations; for $\nu$ being even, it carries an even number of MZMs, whose stability relies on the protection of other symmetries. A refined characterization is given by the dynamical fermion-parity gauge flux, which is an anyon, denoted as $v$. According to Kitaev's 16-fold way\cite{Kitaev2006}, braiding statistics of the fermion-parity fluxes exhibit a 16-fold periodicity. In particular, the topological spin $\theta_v = e^{i\pi \nu/8}$. In fact, the $\nu=16$ fermionic iTO is topologically equivalent to the bosonic $E_8$ state stacked with a trivial fermionic insulator. The group $G_f$ acts the same as $G$ on the the bosonic $E_8$ state, so they are always compatible --- we simply set all elements in $G_f$ by the identity operator for the bosonic $E_8$ state. Since $G_f$ is compatible with a trivial fermionic insulator, we arrive at a conclusion: the $\nu=16$ iTO is compatible with any $G_f$, i.e., no ESB will occur. Therefore, the phenomenon of ESB exhibits a 16-fold periodicity in $\nu$ for 2D iTOs.

\subsubsection{$n_1$ and $\mathcal{O}_2$}
\label{sec:n1_fermion}

The quantity $n_1$,  a function $G\rightarrow \Z_2$, describes the Majorana properties of symmetry defects: if $n_1(\gbf)=1$, the $\gbf_0$ defect carries an odd number of MZMs ; if $n_1(\gbf)=0$, the $\gbf_0$ defect carries an even (including zero) number of MZMs. The party of MZMs on the $\bfg_\sigma$ defect is $n_1(\gbf)+\nu\sigma \modulo{2}$.  Parity of MZM numbers should respect the law of group multiplication $\mathbf{g}_\sigma \mathbf{h}_\tau=(\mathbf{gh})_{\sigma+\tau+\lambda(\mathbf{g},\mathbf{h})}$,  so we immediately obtain the following constraint
\begin{align}
\dd n_1(\mathbf{g,h})&=\mathcal{O}_2(\mathbf{g,h})
\label{eqn_H2_obstuction_2}
\end{align}
where $\dd n_1(\gbf,\hbf)=n_1(\gbf)+n_1(\hbf)-n_1(\gbf\hbf)$ and 
\begin{align}
\mathcal{O}_2(\mathbf{g,h})&=\nu \lambda(\mathbf{g,h})
\label{eqn_H2_obstuction_3}.
\end{align}
Here, modulo 2 is assumed in every equation. We observe that $\dd n_1(\gbf)$ is a coboundary in $\mathcal{B}^2(G,\Z_2)$ and $\mathcal{O}_2(\gbf,\hbf)$ is a 2-cocycle in $\mathcal{Z}^2(G,\Z_2)$. If $\mathcal{O}_2$ is a nontrivial 2-cocycle, Eq.~\eqref{eqn_H2_obstuction_2} can never hold, i.e., no valid $n_1(\bfg)$ exists. This gives the first obstruction to a valid symmetry-enriched iTOs, which we simply call it the ``$\mathcal{O}_2$ obstruction''. 

It is easy to see that $\mathcal{O}_2$ obstruction is trivial if and only if (i) $\nu$ is even, or (ii) $\nu$ is odd and $\lambda$ is a coboundary. In Sec.~\ref{sec:O2andSF}, we will derive the same result in the context of bosonic SETs.

\begin{figure*}
\begin{tikzpicture}[scale=0.5]
\begin{scope}
\node at (-1.5,3){(a)};
\draw [dashed](0,0)--(0,3);
\fill (0,0)circle(0.15);
\node at (0,-0.7){$\gbf_0^0$};
\draw [dashed](2,0)--(2,3);
\fill (2,0)circle(0.15);
\node at (2,-0.7){$\hbf_0^0$};
\draw [dashed](7,0)--(7,3);
\fill (7,0)circle(0.15);
\node at (6.8,-0.7){$(\gbf\hbf)_0^{0}$};
\draw [dashed](9,0)--(9,3);
\fill (9,0)circle(0.15);
\node at (9.3,-0.8){$\bfI^{n_2(\gbf,\hbf)}_{\lambda(\gbf,\hbf)}$};
\draw [line width=2pt, -stealth](4,1.5)--(5,1.5);
\end{scope}

\begin{scope}[xshift=18cm]
\node at (-1.5,3){(b)};
\draw [dashed](0,0)--(0,3);
\fill [red](0,0)circle(0.15);
\node at (0,-0.7){$\gbf_0$};
\draw [dashed](2,0)--(2,3);
\fill (2,0)circle(0.15);
\node at (2,-0.7){$\hbf_0^0$};
\draw [dashed](7,0)--(7,3);
\fill [red](7,0)circle(0.15);
\node at (7,-0.7){$(\gbf\hbf)_0$};
\draw [dashed](9,0)--(9,3);
\fill (9,0)circle(0.15);
\node at (9.3,-0.8){$\bfI^{n_2(\gbf,\hbf)}_{\lambda(\gbf,\hbf)}$};
\draw [line width=2pt, -stealth](4,1.5)--(5,1.5);
\fill [red](-0.5,3)circle(0.15);
\node [anchor=north] at (-0.5,3){$Z_0$};
\fill [red](6.5,3)circle(0.15);
\node [anchor=north] at (6.5,3){$Z_0$};
\end{scope}

\begin{scope}[yshift=-6cm]
\node at (-1.5,3){(c)};
\draw [dashed](0,0)--(0,3);
\fill (0,0)circle(0.15);
\node at (0,-0.7){$\gbf_0^0$};
\draw [dashed](2,0)--(2,3);
\fill [red](2,0)circle(0.15);
\node at (2,-0.7){$\hbf_0$};
\draw [dashed](7,0)--(7,3);
\fill [red](7,0)circle(0.15);
\node at (7,-0.7){$(\gbf\hbf)_0$};
\draw [dashed](9,0)--(9,3);
\fill (9,0)circle(0.15);
\node at (9.3,-0.8){$\bfI^{n_2(\gbf,\hbf)}_{\lambda(\gbf,\hbf)}$};
\draw [line width=2pt, -stealth](4,1.5)--(5,1.5);
\fill [red](-0.5,3)circle(0.15);
\node [anchor=north] at (-0.5,3){$Z_0$};
\fill [red](6.5,3)circle(0.15);
\node [anchor=north] at (6.5,3){$Z_0$};
\end{scope}

\begin{scope}[yshift=-6cm,xshift=18cm]
\node at (-1.5,3){(d)};
\draw [dashed](0,0)--(0,3);
\fill [red](0,0)circle(0.15);
\node at (0,-0.7){$\gbf_0$};
\draw [dashed](2,0)--(2,3);
\fill [red](2,0)circle(0.15);
\node at (2,-0.7){$\hbf_0$};
\draw [dashed](7,0)--(7,3);
\fill (7,0)circle(0.15);
\node at (6.5,-0.7){$(\gbf\hbf)_0^{0}$};
\draw [dashed](9,0)--(9,3);
\fill (9,0)circle(0.15);
\node at (9.3,-0.8){$\bfI^{n_2(\gbf,\hbf)}_{\lambda(\gbf,\hbf)}$};
\draw [line width=2pt, -stealth](4,1.5)--(5,1.5);
\end{scope}

\end{tikzpicture}
\caption{Four different situations to define $n_2(\gbf,\hbf)$ for even $\nu$. Black dots represent non-Majorana defects, and red dots represent  Majorana defects. The dashed lines are branch cuts needed to insert static gauge fluxes, the other end of which are not shown. In every case, we fuse and then re-split the defects using an operator $W$, and the operator $W$ is chosen in a way that produces the states depicted here (see the main text for discussions). In (b) and (c), $Z_0$ is a Majorana operator associated with some other defect. }
\label{fig:n2}
\end{figure*}
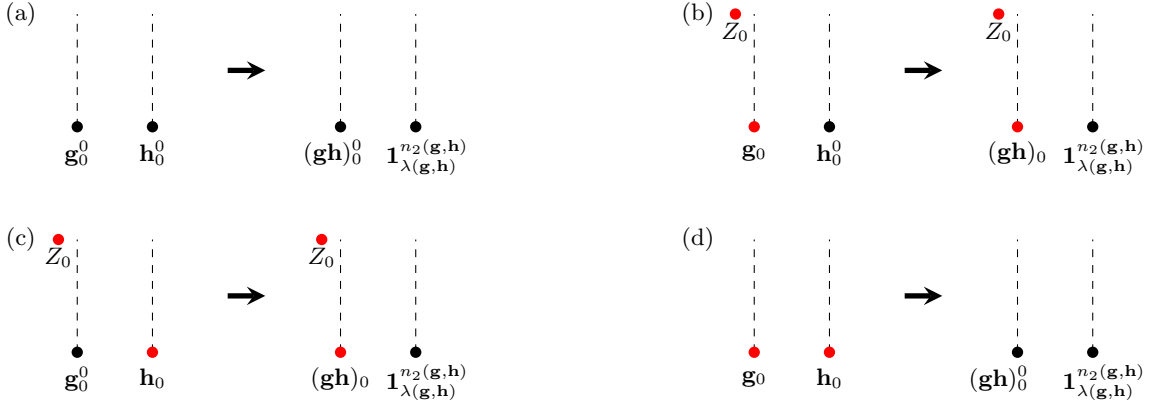

\subsubsection{Definition of $n_2$}
\label{sec:n2_fermion}

Now we focus on the case that $\nu$ is even; we will comment on the odd $\nu$ case at the end of Sec.~\ref{sec:fo3}. In this case, $\mathcal{O}_2(\mathbf{g,h})$ is always trivial and $n_1$ is well-defined. Then, we  move on to define the third quantity $n_2$. Mathematically, it is a function, $n_2: G\times G\rightarrow \Z_2$, subject to certain ambiguities and conditions that we will describe. The physical definition is slightly involved, depending on whether the relevant symmetry defects carry MZMs or not (see Fig.~\ref{fig:n2}). Below, we will refer to the defects that carry MZMs as Majorana defects, and otherwise as non-Majorana defects. 

Let us start with simplest situation that  $n_1(\gbf)=n_1(\hbf)=0$ for the group elements $\gbf$ and $\hbf$. Both $\gbf_\sigma$ and $\hbf_\sigma$ defects are Abelian and non-Majorana. Due to the existence of the local fermion $f$, $\gbf_\sigma$ defects come in two types: $\gbf_\sigma^0$ and $\gbf_\sigma^1$. (We use $\gbf_\sigma$ to denote both group elements and defects.) The two defects are related to each other by fusing the fermion $f$:
\begin{align}
    \gbf_\sigma^0\times f = \gbf_\sigma^1.
\end{align}
Nevertheless, the choice of which defect is $\gbf_\sigma^0$ is a convention. One can set up the convention as follows. Since the defects are non-Majorana, we choose a \emph{local} fermion parity operator, denoted as $P(\gbf)$ with $P(\gbf)^2=1$ for $\gbf_0$ defects.\footnote{One can calibrate different $\gbf_0$ defects, with the same $\gbf$ but sitting at different locations, such that they are associated with the same type of local fermion parity, e.g., by setting up a reference $\gbf_0$ defect for comparison. Then, $P(\gbf)$ depends only on $\gbf$ but not on the location.} If $P(\gbf)|\Psi\rangle = |\Psi\rangle$, where $|\Psi\rangle$ is the state that contains a $\gbf_0$ defect, we call this defect  $\gbf_0^0$; if $P(\gbf)|\Psi\rangle=-|\Psi\rangle$, we call it $\gbf_0^1$. The local fermion parity operator $P(\gbf)$ is ambiguous up to a sign. If one instead uses $\Tilde{P}(\gbf) = - P(\gbf)$ for setting up the convention, the two notations $\gbf_\sigma^0$ and $\gbf_\sigma^1$ will be swapped.

Since $\nu$ is even,  fermion-parity defects are also non-Majorana. We denote the two types as $\bfI_1^0$ and $\bfI_1^1$, conventionally determined by a local fermion parity operator $P(\bfI_1)$. We will denote the vacuum as $\bfI_0^0$ and the fermion $f$ as $\bfI_0^1$, and also refer to them as ``defects'' by abusing notation. Obviously, they are determined by measuring the actual fermion parity $P_f$. For convenience, let $P(\bfI)$ be either $P(\bfI_1)$ or $P_f$, measuring which allows us to determine a defect $\bfI_\sigma^\alpha$. They satisfy the fusion rule
\begin{align}
    \bfI_\sigma^\alpha\times \bfI_\rho^\beta = \bfI_{\sigma+\rho}^{\alpha+\beta + \sigma\rho \nu/2}.\label{eq:pf-fusion}
\end{align}
where $\alpha,\beta,\sigma,\rho=0,1$ and modulo 2 is implicitly taken for addition. The piece $\sigma\rho\nu/2$ implies different fusion rules of fermion-parity defects for $\nu=4k+2$ and $\nu=4k$, which is a well-known result (see e.g. Ref.~\cite{Kitaev2006}). That being said, we define a general $\gbf_\sigma^\alpha$ defect as
\begin{align}
    \gbf_\sigma^\alpha =  \gbf_0^0 \times \bfI_\sigma^\alpha.
    \label{eq:defect1}
\end{align}
The fusion outcome on the right is unique so it is a good definition. We remark that fusion rules of defects are \emph{non-commutative} and the order is important in \eqref{eq:defect1}. We always view a $\gbf_\sigma^\alpha$ defect as a composite of $\gbf_0^0$ and $\bfI_\sigma^\alpha$ defects, with $\bfI_\sigma^\alpha$ sitting on the right.

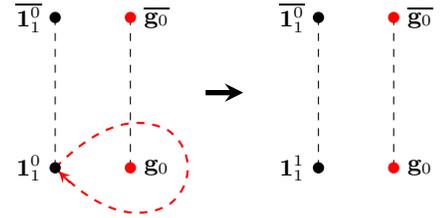
\begin{figure}
\begin{tikzpicture}[scale=0.5]
\begin{scope}
\begin{scope}[xscale=-1,xshift=-2cm]
\draw [red,dashed,thick] (2,0)..controls(1,1) and (-0.0,1.5)..(-1,1)..controls (-1.7,0.5) and (-1.7,-0.5)..(-1,-1)..controls(0,-1.5) and (1,-1)..(1.9,-0.1);
\draw [red, thick, -stealth] (1.6,-0.4)--(1.9,-0.1);
\end{scope}
\draw [dashed](0,0)--(0,4);
\fill (0,0)circle(0.15);
\fill (0,4)circle(0.15);
\node at (-0.7,0){$\bfI_1^0$};
\node at (-0.7,4){$\overline{\bfI_1^0}$};
\draw [dashed](2,0)--(2,4);
\fill [red](2,0)circle(0.15);
\fill [red](2,4)circle(0.15);
\node at (2.7,0){$\gbf_0$};
\node at (2.7,4){$\overline{\gbf_0}$};
\draw [dashed](7,0)--(7,4);
\fill (7,0)circle(0.15);
\fill (7,4)circle(0.15);
\node at (6.3,0){$\bfI_1^1$};
\node at (6.3,4){$\overline{\bfI_1^0}$};
\draw [dashed](9,0)--(9,4);
\fill [red](9,0)circle(0.15);
\fill [red](9,4)circle(0.15);
\node at (9.7,0){$\gbf_0$};
\node at (9.7,4){$\overline{\gbf_0}$};
\draw [line width=2pt, -stealth](4,2)--(5,2);
\end{scope}
\end{tikzpicture}
\caption{A configuration of four defects. When the fermion-parity defect $\bfI_1^0$ moves across the branch cut of the $\gbf_0$ Majorana defect (along the red dashed line), it turns into $\bfI_1^1$. }
\label{fig:Zparity}
\end{figure}

We now consider fusing two non-Majorana defects. Fusing $\gbf_0^0$ and $\hbf_0^0$ defects must give rise to one of the two defects, $(\gbf\hbf)_{\lambda(\gbf,\hbf)}^0$ or $(\gbf\hbf)_{\lambda(\gbf,\hbf)}^1$. Let us denote the possible superscript as $n_2(\gbf,\hbf)$. Then,
\begin{align}
    \gbf_0^0\times \hbf_0^0 = (\gbf\hbf)_{\lambda(\gbf,\hbf)}^{n_2(\gbf,\hbf)} = (\gbf\hbf)_{0}^0\times \bfI_{\lambda(\gbf,\hbf)}^{n_2(\gbf,\hbf)}.
    \label{eq:n2-1}
\end{align}
Figure \ref{fig:n2}(a) shows the corresponding image. The quantity $n_2(\gbf, \hbf)$ depends on the symmetry-enriched iTO. Let $|\Psi\rangle$ be the state containing the initial $\gbf_0^0$ and $\hbf_0^0$ defects, and let $W$ be the operator that does the fusion and splitting associated with \eqref{eq:n2-1}. Then, Eq.~\eqref{eq:n2-1} can be understood as follows: by properly choosing $W$, we first require
\begin{align}
W  P(\gbf)P(\hbf) |\Psi\rangle & =  P(\gbf\hbf)W|\Psi\rangle,
\label{eq:PPP}
\end{align}
and then take a measurement of $P(\bfI)$ on the state $W|\Psi\rangle$ such that
\begin{align}
P(\bfI) W |\Psi\rangle & = (-1)^{n_2(\gbf, \hbf)}W|\Psi\rangle.
   \label{eq:PP}
\end{align}
It means that we choose an operator $W$ such that it turns $\gbf_0^0$ and $\hbf_0^0$ into $(\gbf\hbf)_0^0$, and   $n_2(\gbf,\hbf)$ is defined by measuring $P(\bfI)$.   In general, the fusion and splitting can be done by other operators, say $W'$ that turns the original defects into  $(\gbf\hbf)_0^{n_2(\gbf,\hbf)}$ and $\bfI_{\lambda(\gbf,\hbf)}^0$. However, no matter which operator is used, the total fermion parity must be conserved. That is, the equation
\begin{align}
X P(\gbf)P(\hbf) |\Psi\rangle & = (-1)^{n_2(\gbf, \hbf)} P(\gbf\hbf)P(\bfI)X|\Psi\rangle,
\end{align}
holds for either $X=W$ or $X=W'$. Then, the quantity $n_2(\gbf,\hbf)$ must be the same, so it is well defined. In Fig.~\ref{fig:n2}(a), we have used $W$ instead of $W'$ for illustration. If we change the convention $P(\gbf) \rightarrow (-1)^{\epsilon(\gbf)}P(\gbf)$ (with the convention of $P(\bfI)$ fixed), the quantity $n_2(\gbf,\hbf)$ undergoes the following transformation
\begin{align}
    n_2(\gbf,\hbf) \rightarrow n_2(\gbf,\hbf) + \epsilon(\gbf\hbf)-\epsilon(\gbf) - \epsilon(\hbf)
    \label{n2:ambiguity}
\end{align}
Mathematically, it is a coboundary transformation. Accordingly, $n_2(\gbf,\hbf)$ is well-defined only up to the coboundary ambiguity \eqref{n2:ambiguity}. It is equivalent to use general defects $\bfg_0^\alpha$ and $\bfh_0^\beta$ to define $n_2(\bfg,\bfh)$.

Next, we consider the situation in Fig.~\ref{fig:n2}(b). In this case, $n_1(\gbf)=1$ and $n_1(\hbf)=0$, i.e., $\gbf_0$ defects are Majorana while the $\hbf_0$ defects are non-Majorana. The Majorana $\gbf_0$ defects come in only one type, which we simply denote as $\gbf_0$. In this case, we cannot define a local fermion parity operator. Instead, as discussed in Ref.~\cite{FidkowskiPRB2011} and also  Sec.~\ref{sec:esb-1D}, there exists a local involutory fermionic unitary operator $Z$ that commutes with all local operators. Similar to the local fermion parity operator, $Z$ is also ambiguous up to a sign. Since $Z$ is very like the usual Majorana operator, we will simply refer to it as the Majorana operator. For every $\gbf$ with $n(\gbf)=1$, we choose a Majorana operator $Z(\gbf)$, out of the two available, for the $\gbf_0$ defect. Once this convention is set, we can define a fermion parity operator out of these Majorana operators. Consider $2n$ Majorana defects, $(\gbf_j)_0$ with $j=1,\dots,2n$. There is a $2^n$ dimensional degenerate Hilbert space associated with these defects.  We define the fermion parity  as $\mathcal{P}=i^n\prod_j Z(\gbf_j)$. It is clear that change of the convention $Z(\gbf_j)\rightarrow -Z(\gbf_j)$ for any defect will result in a sign flip of $\mathcal{P}$. 

Fusing $\gbf_0$ and a fermion-parity defect $\bfI_1^\alpha$, we  obtain a $\gbf_1$ defect, which is also a Majorana defect. Regardless of $\alpha$, we always have $\gbf_0\times \bfI_1^\alpha=\gbf_1$. One may want to have a ``split view'' of $\gbf_1$ as in \eqref{eq:defect1}. However, such a splitting is not unique at the level of fusion rules. One needs to define it carefully. We will not define the splitting of an individual defect here, but instead will define a splitting  combined with a fusion of two defects below, for our our purpose of defining $n_2(\gbf,\hbf)$.

We are now ready to define $n_2(\gbf,\hbf)$ in Fig.~\ref{fig:n2}(b). Fusing $\gbf_0$ and $\hbf_0^0$ defects gives rise to a $(\gbf\hbf)_{\lambda(\gbf,\hbf)}$ defect, which is a Majorana defect as $n_1(\gbf\hbf)=n_1(\gbf)+n_1(\hbf)$. We further split it into $(\gbf\hbf)_0$ and $\bfI_{\lambda(\gbf,\hbf)}^{n_2(\gbf,\hbf)}$, with $n_2(\gbf,\hbf)$ determined using a similar procedure as in the first situation. Let us describe the procedure, which is slightly different, as Majorana operators are involved. Since a single Majorana operator does not form a Hilbert space, let us assume that there is an auxiliary Majorana operator $Z_0$, at somewhere not close to the two defects. Let $|\Psi\rangle$ be the initial state containing $\gbf_0$ and $\hbf_0^0$, and let $W$ be the operator associated with the whole fusing and splitting process. Then, similar to the first situation, we define $n_2(\gbf,\hbf)$ as follows. By properly choosing $W$, we first require
\begin{align}
    W [iZ_0 & Z(\gbf)] P(\hbf)|\Psi\rangle = [iZ_0 Z(\gbf\hbf)] W|\Psi\rangle,
    \label{eq:z0}
\end{align}
and then we take a measurement of $P(\bfI)$ on the state $W|\Psi\rangle$,
\begin{align}
    P(\bfI)W|\Psi\rangle = (-1)^{n_2(\gbf,\hbf)}W|\Psi\rangle.
\end{align}
The latter measurement defines $n_2(\bfg,\bfh)$. That is, we require the final state contains $(\gbf\hbf)_0$ and $\bfI_{\lambda(\gbf,\hbf)}^{n_2(\gbf,\hbf)}$ defects after fusion and splitting, as shown in Fig.~\ref{fig:n2}(b). $W$ is a fusion and splitting operator, so it should be bosonic to preserve fermion parity. Also it acts only near the defects. Because $Z_0$ is away from the defects, we must have $WZ_0=Z_0W$. Therefore, $Z_0$ in Eq.~\eqref{eq:z0} can be readily removed, so that the definition of $n_2(\gbf,\hbf)$ is independent of $Z_0$. Like in the first situation, we may choose another operator $W'$ that does a different fusion and splitting. However, the value of $n_2(\gbf,\hbf)$ cannot be changed for fixed $Z(\gbf)$, $P(\hbf)$, $Z(\gbf\hbf)$ and $P(\bfI)$ due to total fermion parity conservation. If we change our conventions of $Z(\gbf)$, $P(\hbf)$ and $Z(\gbf\hbf)$, the same ambiguity \eqref{n2:ambiguity} results. 

The situation in Fig.~\ref{fig:n2}(c) is similar to Fig.~\ref{fig:n2}(b). In the last situation of Fig.~\ref{fig:n2}(d), we have both $n_1(\gbf)=n_1(\hbf)=1$. The fusion outcomes are non-Majorana defects. In this case, the two Majorana operators $Z(\gbf)$ and $Z(\hbf)$ are enough to form a Hilbert space, so no auxiliary Majorana operator is needed. We define $n_2(\gbf,\hbf)$ by first requiring
\begin{align}
    W[i Z(\gbf) Z(\hbf)]|\Psi\rangle  =  P(\gbf\hbf) W|\Psi\rangle,
\end{align}
and then measuring $P(\bfI)$ on the state $W|\Psi\rangle$
\begin{align}
    P(\bfI)W|\Psi\rangle  = (-1)^{n_2(\gbf,\hbf)}W|\Psi\rangle,
\end{align}
where $|\Psi\rangle$ is the initial state and $W$ is a properly chosen fusion and splitting operator associated. One can similarly argue that $n_2(\gbf,\hbf)$ is well defined. The ambiguity on $n_2(\gbf,\hbf)$ due to change of convention is the same as above.

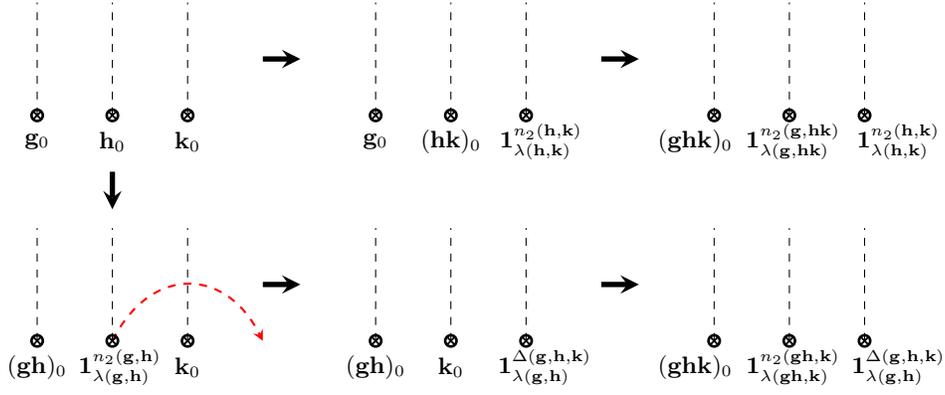
\begin{figure*}
\begin{tikzpicture}[scale=0.5]
\def\cross at (#1,#2){\draw [line width=1pt] (#1,#2)circle(0.15);
\draw [line width=0.7pt] (#1+0.1,#2+0.1)--(#1-0.1,#2-0.1);
\draw [line width=0.7pt] (#1-0.1,#20.1)--(#1+0.1,#2-0.1);}

\begin{scope}
\draw [dashed](0,0)--(0,3);
\cross at (0,0);
\node at (0,-0.7){$\gbf_0$};
\draw [dashed](2,0)--(2,3);
\cross at (2,0);
\node at (2,-0.7){$\hbf_0$};
\draw [dashed](4,0)--(4,3);
\cross at (4,0);
\node at (4,-0.7){$\kbf_0$};
\draw [line width=2pt, -stealth](6,1.5)--(7,1.5);
\draw [line width=2pt,-stealth](2,-1.5)--(2,-2.5);
\end{scope}

\begin{scope}[xshift=0cm, yshift=-6cm]
\draw [dashed](0,0)--(0,3);
\draw [red, dashed, thick, -stealth] (2,0)..controls (3,2) and (5,2)..(6,0);
\cross at (0,0);
\node at (0,-0.7){$(\gbf\hbf)_0$};
\draw [dashed](2,0)--(2,3);
\cross at (2,0);
\node at (2.2,-0.7){$\bfI_{\lambda(\gbf,\hbf)}^{n_2(\gbf,\hbf)}$};
\draw [dashed](4,0)--(4,3);
\cross at (4,0);
\node at (4,-0.7){$\kbf_0$};
\draw [line width=2pt, -stealth](6,1.5)--(7,1.5);
\end{scope}

\begin{scope}[xshift=9cm, yshift=-6cm]
\draw [dashed](0,0)--(0,3);
\cross at (0,0);
\node at (0,-0.7){$(\gbf\hbf)_0$};
\draw [dashed](2,0)--(2,3);
\cross at (2,0);
\node at (2,-0.7){$\bfk_0$};
\draw [dashed](4,0)--(4,3);
\cross at (4,0);
\node at (4.5,-0.7){$\bfI_{\lambda(\gbf,\hbf)}^{\Delta(\gbf,\hbf,\kbf)}$};
\draw [line width=2pt, -stealth](6,1.5)--(7,1.5);
\end{scope}

\begin{scope}[xshift=18cm, yshift=-6cm]
\draw [dashed](0,0)--(0,3);
\cross at (0,0);
\node at (-0.5,-0.7){$(\gbf\hbf\kbf)_0$};
\draw [dashed](2,0)--(2,3);
\cross at (2,0);
\node at (2.1,-0.7){$\bfI_{\lambda(\gbf\hbf,\kbf)}^{n_2(\gbf\hbf,\kbf)}$};
\draw [dashed](4,0)--(4,3);
\cross at (4,0);
\node at (4.9,-0.7){$\bfI_{\lambda(\gbf,\hbf)}^{\Delta(\gbf,\hbf,\kbf)}$};
\end{scope}

\begin{scope}[xshift=9cm, yshift=0cm]
\draw [dashed](0,0)--(0,3);
\cross at (0,0);
\node at (0,-0.7){$\gbf_0$};
\draw [dashed](2,0)--(2,3);
\cross at (2,0);
\node at (2,-0.7){$(\hbf\kbf)_0$};
\draw [dashed](4,0)--(4,3);
\cross at (4,0);
\node at (4.4,-0.7){$\bfI_{\lambda(\hbf,\kbf)}^{n_2(\hbf,\kbf)}$};
\draw [line width=2pt, -stealth](6,1.5)--(7,1.5);
\end{scope}

\begin{scope}[xshift=18cm, yshift=0cm]
\draw [dashed](0,0)--(0,3);
\cross at (0,0);
\node at (-0.5,-0.7){$(\gbf\hbf\kbf)_0$};
\draw [dashed](2,0)--(2,3);
\cross at (2,0);
\node at (2.1,-0.7){$\bfI_{\lambda(\gbf,\hbf\kbf)}^{n_2(\gbf,\hbf\bfk)}$};
\draw [dashed](4,0)--(4,3);
\cross at (4,0);
\node at (4.9,-0.7){$\bfI_{\lambda(\hbf,\kbf)}^{n_2(\hbf,\kbf)}$};
\end{scope}
\end{tikzpicture}
\caption{Associativity of defect fusion. For simplicity, we do not distinguish Majorana and non-Majorana defects in the diagrams. If some defects are non-Majorana, we have dropped the superscript and they should mean  $\gbf_0^0$, $\hbf_0^0$, or $\kbf_0^0$. The quantity $\Delta(\gbf,\hbf,\kbf) = n_2(\gbf, \hbf) + \lambda(\gbf, \hbf) n_1(\kbf)$, where the term $\lambda(\gbf, \hbf) n_1(\kbf)$ comes from the passing of  $\bfI_{\lambda(\gbf,\hbf)}^{n_2(\gbf,\hbf)}$ across the branch cut of $\kbf_0$.}
\label{fig:associativity}
\end{figure*}

This completes our definition of $n_2(\gbf,\hbf)$. Let us summarize that, for each $\gbf_0$ defect, we have chosen a local operator 
\begin{align}
    \calL(\gbf) = \left\{
    \begin{array}{ll}
    P(\gbf), & n_1(\gbf)=0  \\[5pt]
    Z(\gbf),& n_1(\gbf)=1 
    \end{array}
    \right.
\end{align}
Also, we have chosen a local fermion parity operator $P(\bfI)$ for determining defect $\bfI_\sigma^\alpha$. These local operators allow us to define the fermion parity in the low-energy Hilbert space associated with defects, which in turn allows us to define $n_2(\gbf,\hbf)$. As the local operator is subject to an ambiguity $\mathcal{L}(\gbf) \rightarrow -\mathcal{L}(\gbf) $, the quantity $n_2(\gbf,\hbf)$ is defined up to the transformation \eqref{n2:ambiguity}.

\subsubsection{Derivation of $\mathcal{O}_3$}
\label{sec:fo3}

Before showing the condition imposed on $n_2(\gbf,\hbf)$, we discuss a property regarding braiding between $Z(\gbf)$ and fermion parity defects. For concreteness, consider the defect configuration in Fig.~\ref{fig:Zparity}. There are four defects: a $\gbf_0$ Majorana defect, its anti-defect $\overline{\gbf_0}$ (which may be $\bar{\gbf}_0$ or $\bar\gbf_1$ depending on $\lambda$), a fermion-parity defect $\bfI_1^0$ and its anti-defect $\overline{\bfI_1^0}$ (which may be $\bfI_1^0$ or $\bfI_1^1$ depending on $\nu$).  Imagine adiabatically braiding the $\bfI_1^0$ defect across the branch cut between the $\gbf_0$ and $\overline{\gbf_0}$ defect. We claim that $\bfI_1^0$ turns into $\bfI_1^1$ after passing through the branch cut. One way to understand this is  through the bosonic SET language discussed in Sec.~\ref{sec:derivation}: when $\bfI_1^0$ passes through the branch cut, the state is acted by $\gbf$, whose action is permutation of two defects $\bfI_1^0$ and $\bfI_1^1$ ($v$ and $vf$ in the notation of Sec.~\ref{sec:derivation}). Here we consider a slightly different view. Let us assume that the braiding process makes a full loop. Then, we retract the loop back to the position of the fermion parity defect. The whole process is equivalent to an action of the fermion parity on the disk whose edge is the loop. Let this action be the operator $U(P_f)$. Since $Z(\gbf)$ is inside this disk and it is a fermionic operator, we must have $Z(\gbf)U(P_f)=-U(P_f) Z(\gbf)$. More specifically, let $|\Psi\rangle$ be the initial state before braiding and retracting. Then
\begin{align}
    U(P_f) Z(\gbf) |\Psi\rangle = - Z(\gbf) U(P_f) |\Psi\rangle.
\end{align}
Accordingly the fermion parity of the pair $\gbf_0$ and $\overline{\gbf_0}$ flips a sign after the braiding and retracting process. Since the total fermion parity must be conserved and the local parity of $\overline{\bfI_1^0}$ is untouched, the defect $\bfI_1^0$ must flip to $\bfI_1^1$. If the $\gbf_0$ defect is non-Majorana, the local fermion parity of $\bfI_1^0$ remains. Stating it compactly, we have: a $\bfI_\sigma^\alpha$ defect turns into $\bfI_\sigma^{\alpha + \sigma n_1(\gbf)}$, when passing through the branch cut of a $\gbf_0$ defect.

With the above preparation, we now prove the condition that $n_2(\gbf, \hbf)$ should satisfy. It follows from the associativity of defect fusion (Fig.~\ref{fig:associativity}). Consider fusing three defects, $\gbf_0$, $\hbf_0$ and $\kbf_0$. If $n_1(\gbf)=0$, we pick the $\gbf_0^0$ defect; if $n_1(\gbf)=1$, there is only a Majorana defect $\gbf_0$ to pick (similarly for $\hbf_0$ and $\kbf_0$ defects).  There are two ways that we can fuse the defects, as shown through the two paths in Fig.~\ref{fig:associativity}. Most steps only make use of the definition of $n_2(\gbf,\hbf)$ in Fig.~\ref{fig:n2} under a properly chosen fusion and splitting operator $W$, except the step associated with the red dashed line in the lower path that uses the property in Fig.~\ref{fig:Zparity}. At the end of both paths, we have two fermion-parity defects (or simply $1$, $f$). Further fusing the two defects according to Eq.~\eqref{eq:pf-fusion} and requiring the fermion parity to be equal, we obtain
\begin{align}
   &  n_2(\gbf,\hbf\kbf) + n_2(\hbf,\kbf) + \frac{\nu}{2}\lambda(\gbf,\hbf\kbf)\lambda(\hbf,\kbf) \nonumber\\
     = &  n_2(\gbf\hbf,\kbf)+n_2(\gbf,\hbf) + \lambda(\gbf,\hbf)n_1(\kbf) + \frac{\nu}{2}\lambda(\gbf\hbf,\kbf)\lambda(\gbf,\hbf)
     \label{eq:fo3}
\end{align}
where ``modulo 2'' is assumed for the addition. We define the function
\begin{align}
    \mathcal{O}_3(\gbf,\hbf,\kbf)  = &\lambda(\gbf,\hbf)n_1(\kbf)  +  \frac{\nu}{2}\lambda(\gbf,\hbf\kbf)\lambda(\hbf,\kbf) \nonumber\\
    &- \frac{\nu}{2}\lambda(\gbf\hbf,\kbf)\lambda(\gbf,\hbf)\nonumber\\
     = & \left(\lambda\cup n_1 + \frac{\nu}{2}\lambda\cup_1\lambda\right)(\gbf,\hbf,\kbf)
\end{align}
where the higher cup product $\cup_1$ is reviewed in Appendix \ref{app:cohomology}.  Then, Eq.~\eqref{eq:fo3} can be compactly written as
\begin{align}
    \mathcal{O}_3(\gbf,\hbf,\kbf) = \dd n_2(\gbf,\hbf,\kbf), 
    \label{eq:fo3-2}
\end{align}
where we have adjusted a few irrelevant minus signs to match the definitions. According to the mathematical properties of $\cup$ and $\cup_1$ products, $\mathcal{O}_3$ is a cocycle in $\calH^3(G,\Z_2)$. It may be a nontrivial cocycle. However, the right-hand side of Eq.~\eqref{eq:fo3-2} is always a coboundary. Accordingly, for given $\nu, n_1$ and $\lambda$, there might be no solution of $n_2$ to Eq.~\eqref{eq:fo3-2}, hence leading to an obstruction to have a valid $n_2$. Once $\mathcal{O}_3$ is trivial, Eq.~\eqref{eq:fo3-2} is a condition that $n_2$ should satisfy. Different solutions, subject to the coboundary transformation \eqref{n2:ambiguity}, correspond to different symmetry-enriched fermionic iTOs. Note that for $\nu=4k$, the $\lambda\cup_1\lambda$ term in $\calO_3$ vanishes (after taking modulo 2). This is an important difference between $\nu=4k+2$ and $\nu=4k$.

Finally, we comment that $\mathcal{O}_3$ obstruction is absent for odd $\nu$'s. First of all, the triviality of $\mathcal{O}_2$ requires $\lambda$ to be a coboundary. Then, let  $\lambda(\gbf,\hbf) = \epsilon(\gbf)+\epsilon(\hbf)-\epsilon(\gbf\hbf)$. One may use the same way as above to define $n_2$. However, let us use the defects $\gbf_{\epsilon(\gbf)}$, instead of $\gbf_0$, for picking the local fermion parity operator $P(\gbf)$ or Majorana operator $Z(\gbf)$. Then, in the definition of $n_2$,  we always have
\begin{align}
    \gbf_{\epsilon(\gbf)}\times \hbf_{\epsilon(\hbf)} = (\gbf\hbf)_{\epsilon(\gbf\hbf)} \times \bfI_0^{n_2(\gbf,\hbf)}.
\end{align}
Note that $\bfI_0^{n_2(\gbf,\hbf)} = f^{n_2(\gbf,\hbf)}$,  i.e., there is no fermion-parity defect involved. A simple check of the fusion process in Fig.~\ref{fig:associativity} shows that  $\mathcal{O}_3$ obstruction is absent. A summary of possible obstructions is shown in Table \ref{tab:o2o3o4}.
 
\subsection{ESB criteria}

We have defined the triplet $(\nu,n_1,n_2)$ for the description of 2D symmetry-enriched fermionic iTOs. In principle, there is another piece of data, corresponding to stacking bosonic SPT states. However, that data is irrelevant to our study of ESB physics, so we will not discuss it. We have also derived the $\calO_2$ and $\calO_3$ obstruction functions, by checking the consistency between the MZM composition rule or fermion parity conservation and the defect fusion rules. In fact, there is another level of obstruction, denoted as $\mathcal{O}_4$, which follows from the consistency between symmetry action on local Hilbert spaces of defects and defect fusion rules. We will not derive $\calO_4$ in the fermionic language in this work, but instead turn to the bosonic SET language and discuss it in Sec.~\ref{sec:derivation}. A symmetry group $G_f$ is enforced to break by a fermionic iTO of index $\nu$, if there are no valid $n_1$ and $n_2$  that make  $\mathcal{O}_2, \calO_3$ and $\calO_4$ all trivial, i.e., if there are no valid 2D symmetry-enriched fermionic iTOs. In this section, we will state and elaborate the ESB criteria, and illustrate the criteria via various examples.  Derivations of the criteria, including re-derivation of $\calO_2$ and $\calO_3$ and the derivation of $\calO_4$ in bosonic SET language, will be given in Sec.~\ref{sec:derivation}.

\subsubsection{odd $\nu$}
\label{sec:criterion_odd_nu}

We first consider 2D fermionic iTOs with $\nu$ being odd. Consider symmetry ground $G_f$, determined by the pair $(G,\lambda)$.  We show that
\begin{theorem}
$G_f$ is enforced to break by the 2D fermionic iTOs with odd $\nu$, if and only if $\lambda$ is a non-trivial 2-cocycle in $\mathcal{H}^2(G, \Z_2)$.
\label{criterion3}
\end{theorem}
\noindent In other words, the odd-$\nu$ iTOs are only compatible with $G_f=\Z_2^f\times G$. This criterion can be easily seen from the $\calO_2$ obstruction alone. For odd $\nu$, we have $\calO_2=\lambda$. If $\lambda$ is nontrivial in $\calH^2(G,Z_2)$, so is $\calO_2$. Then, $G_f$ is enforced to break. On the other hand, if $\lambda$ is trivial, $G_f = \Z_2^f\times G$ and there is always a valid symmetry-enriched iTO --- the one with all symmetries in $G$ represented by the identity operator.  

\begin{figure}
\begin{tikzpicture}[scale=0.5]
\fill [gray!30] (0, 1)--(8,1)--(8,-1)--(0,-1)--cycle;
\draw [fill=white](0,0) ellipse (0.5 and 1);
\draw [very thick] (-1.5,0)--(0,0);
\draw [-stealth, very thick] (8.5,0)--(9.7,0);
\begin{scope}
\clip (8,-1.1) rectangle+(0.6,2.2);
\draw [fill=gray!30](8,0) ellipse (0.5 and 1);
\end{scope}
\draw (0, 1)--(8,1);
\draw (0,-1)--(8,-1);
\node at (-1.5,0.5){$P_f$};
\node at (0,3){};
\node at (0,-3){};

\draw [gray, line width=10pt, -stealth](4,-1.7)--(4,-3.7);

\begin{scope}[yshift=-4.5cm]
\draw [line width =1pt](0,0)--(8,0);
\draw [fill](0,0) circle (0.15);
\draw [fill](8,0) circle (0.15);
\end{scope}
\end{tikzpicture}
\caption{Dimensional reduction of 2D cylindrical system with a fermion-parity flux inserted.}
\label{fig:dimreduct}
\end{figure}
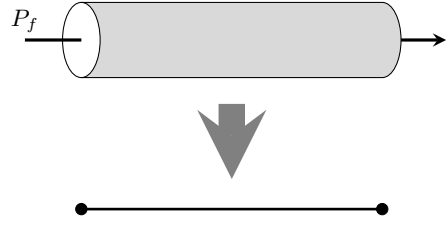

One may notice that this criterion is the same as Criterion \ref{criterion2} for the Majorana-chain iTO. It is not a coincidence.  One can understand this connection by dimensional reduction. Imagine we insert a pair of fermion parity defects in a cylindrical geometry (Fig.~\ref{fig:dimreduct}). There are two MZMs located at the two ends when $\nu$ is odd. If we reduce the cylinder to a 1D system, it becomes a 1D Majorana-chain iTO. It is important to note that $P_f$ is in the center of $G_f$ so that $G_f$ remains the symmetry group of the effective 1D system. Then, if we apply Criterion \ref{criterion2}, the ``only if'' direction of Criterion \ref{criterion3} results. One may extend this method to argue if a general $\gbf$ defect is allowed to carry MZMs. One subtlety is that inserting a $\gbf$ defect will break the symmetry group $G_f$ down to the centralizer $C_\gbf$ of $\gbf$ in $G_f$. If $C_\gbf$ is a nontrivial extension by $\Z_2^f$, then $\gbf$ cannot carry MZM as it contradicts with Criterion \ref{criterion2}. However, if $C_\gbf$ is a trivial extension of $\Z_2^f$, we cannot give a definite answer.

\subsubsection{$\nu=4k+2$}
\label{sec:criterion_4k+2}

For even $\nu$, the obstruction $\calO_2$ is always trivial and $n_1$ is a cocycle in $\calH^1(G,\Z_2)$. We need to consider $\calO_3$ and $\calO_4$.  Here, we discuss the ESB criterion for $\nu=4k+2$. Given a symmetry $G_f$ determined by $(G,\lambda)$,  we claim that
\begin{theorem}
$G_f$ is enforced to break by the 2D fermionic iTOs with $\nu=4k+2$, if one of the two situations occurs: (i) the quantity
\begin{equation}
\mathcal{O}_3=\lambda\cup n_1+\lambda\cup_1\lambda
\label{eqn_O3_obstruction_formula_nu2}
\end{equation}
is a nontrivial cocycle in $\mathcal{H}^3(G,\Z_2)$, for any $n_1\in \mathcal{H}^1(G,\Z_2)$; or (ii) another  quantity $\mathcal{O}_4=\calO_4[\lambda,\nu, n_1,n_2]$
is a nontrivial cocycle in $\mathcal{H}^4(G,U(1))$, for all $n_1$ that make $\mathcal{O}_3$ a coboundary and for all $n_2$ that satisfy $\calO_3 = \dd n_2$. We conjecture that there is no other situation such that $G_f$ is enforced to break for $\nu=4k+2$.
\label{criterion4}
\end{theorem}

A few comments are in order. First, the quantities $\lambda$, $n_1$, $n_2$ and $\calO_3$ are valued in $\{0,1\}$. We repeat the definitions of the cup products for convenience:
\begin{align}
    \lambda\cup n_1(\gbf,\hbf,\kbf) & = \lambda(\gbf,\hbf) n_1(\kbf), \nonumber\\
    \lambda\cup_1\lambda(\mathbf{g,h,k}) &=\lambda(\mathbf{h,k})\lambda(\mathbf{g,hk})-\lambda(\mathbf{g,h})\lambda(\mathbf{gh,k}), \nonumber
\end{align}
where again additions are defined modulo 2. More details on cohomology operations are reviewed in Appendix \ref{app:cohomology}. According to properties of cup products, $\calO_3$ is always a (trivial or nontrivial) 3-cocycle in $\calH^3(G, \Z_2)$ as long as $n_1\in \calH^1(G, \Z_2)$ and $\lambda\in \calH^2(G, \Z_2)$. Second, we do not have the most general expression of $\calO_4$ yet. However, we know that it is a functional of $\lambda,n_1$, $n_2$, and $\nu$. It can be defined only if $\calO_3$ is a coboundary. We also know that it should be a $U(1)$-valued  4-cocycle in $\calH^4(G,U(1))$ from our knowledge of bosonic SET physics. In the special case that $n_1=0$ and $\calO_3=\lambda\cup_1\lambda$ is trivial, $\calO_4$ has the following expression:
\begin{align}
\mathcal{O}_{4}& (\gbf,\hbf,\kbf,\lbf) \nonumber\\
 = & e^{i\nu\frac{\pi}{8} [\lambda\cup \lambda +\hat{\dd } \lambda \cup_1\lambda](\gbf,\hbf,\kbf,\lbf)}\times e^{i\pi (n_2+\lambda)\cup n_2 (\gbf,\hbf,\kbf,\lbf)}
\label{eq:h4nu2}
\end{align}
where ``$\hat{\dd}$'' is used to denote the usual coboundary operation but without taking modulo 2, and additions in the exponents do not assume modulo 2 either. We will argue in Sec.~\ref{sec:derivation} that a more general (but still incomplete) expression of $\calO_4$ can be obtained by stacking fermionic iTOs with fermionic SPT states. 

While we do not have the general expression of $\calO_4$, the $\calO_3$ obstruction alone is enough to give rise to many ESB examples. In particular, we give a complete answer to ESB phenomena that come from the $\calO_3$ obstruction for general finite Abelian groups $G=\prod_i \Z_{N_i}$. This is discussed in Appendix \ref{app:O3}. A few simple examples will be discussed in Sec.~\ref{sec_example_nu2}.

\subsubsection{$\nu=8k+4$}
\label{sec:criterion_8k+4}

Now we consider the case $\nu=8k+4$. As discussed in Sec.~\ref{sec:fo3}, when $\nu=4k$, the $\calO_3$ obstruction reduces to $\calO_3 = \lambda\cup n_1$. Then, $\calO_3$ obstruction alone cannot give rise to enforced symmetry breaking, because there always exists the case $n_1=0$ such that $\calO_3=0$. Accordingly, ESB may occur only if both $\calO_3$ and $\calO_4$ are considered. We claim that 
\begin{theorem}
$G_f$ is enforced to break by the 2D fermionic iTOs with $\nu=8k+4$, if a quantity $\calO_4=\calO_4[\lambda,\nu, n_1,n_2]$
is a nontrivial cocycle in $ \mathcal{H}^4(G,U(1))$, for all $n_1$ that make $\mathcal{O}_3=\lambda\cup n_1$ a coboundary and for all $n_2$ that satisfy $\calO_3 = \dd n_2$. We conjecture that there is no other situation such that $G_f$ is enforced to break for $\nu=8k+4$.
\label{criterion5}
\end{theorem}

Regarding the expression of $\calO_4$, we will argue in Sec.~\ref{sec:derive_4k} that a general and complete expression can be obtained by a stacking trick. Here, we give a simpler version of the formula for the case that $n_1=0$:
\begin{align} \mathcal{O}_4=e^{i\pi\mathcal{Q}_4/2}
\label{eqn_O4_obstruction_formula_4}
\end{align}
and
\begin{align}
    \mathcal{Q}_4 = \frac{\nu}{4}(\lambda\cup\lambda+\hat\dd\lambda\cup_1\lambda) +2n_2\cup (n_2+\lambda),
\label{eqn_O4_obstruction_formula_4-2}
\end{align}
where both $\lambda$ and $n_2$ are valued in $\{0,1\}$,  ``modulo 2'' is not taken for $\hat{\dd}$, and additions elsewhere are taken to be ``modulo 4''.  That $\calO_4$ is 4-cocycle in $\calH^4(G,U(1))$ follows from the fact that $\mathcal{Q}_4$ is actually a 4-cocycle in $\calH^4(G,\Z_4)$. To see that, first, it is straightforward to check that $2n_2\cup (n_2+\lambda)$ is $\Z_4$-valued 4-cocycle, following the general properties of cup products. Second, the piece $\mathcal{P}(\lambda)\equiv \lambda\cup\lambda+\hat{\dd}\lambda\cup_1\lambda$ is mathematically known as the Pontryagin square of a $\Z_2$-valued cocycle $\lambda$. It is known that $\mathcal{P}(\lambda)$ is a $\Z_4$-valued cocycle (see Appendix \ref{app:cohomology}). Accordingly, $\mathcal{Q}_4$ is a 4-cocycle in $\calH^4(G,\Z_4)$. This special expression of $\calO_4$ allows us to discover a few ESB examples for $\nu=8k+4$ fermionic iTOs with a finite Abelian group $G$, which we will discuss in Sec.~\ref{sec_example_nu4}. We note that \eqref{eqn_O4_obstruction_formula_4} and \eqref{eq:h4nu2} are  the same but $\nu$ takes different values.

\subsubsection{$\nu=16k+8$}
\label{sec:criterion_16k+8}

In the case $\nu=16k+8$, again both $\calO_3$ and $\calO_4$ need to be taken into account to support ESB. The statement of ESB criterion is the same as in Criterion \ref{criterion5}:
\begin{theorem}
$G_f$ is enforced to break by the 2D fermionic iTOs with $\nu=16k+8$, if a quantity $\calO_4=\calO_4[\lambda,\nu, n_1,n_2]$
is a nontrivial cocycle in $ \mathcal{H}^4(G,U(1))$, for all $n_1$ that make $\mathcal{O}_3=\lambda\cup n_1$ a coboundary and for all $n_2$ that satisfy $\calO_3 = \dd n_2$. We conjecture that there is no other situation such that $G_f$ is enforced to break for $\nu=16k+8$.
\label{criterion6}
\end{theorem}
\noindent Compared to the $\nu=8k+4$ case, the difference lies in the expression of $\calO_4$. We will argue in Sec.~\ref{sec:derive_4k} that a general and complete expression of $\calO_4$ can be obtained by stacking fermionic SPTs and using the SPT formulas. Here, we give a simpler version of the formula for the case that $n_1=0$:
\begin{align}
{\mathcal{O}}_4=e^{i\pi(\lambda \cup \lambda +n_2\cup n_2+ n_2 \cup \lambda)}.
\label{eqn_O4_obstruction_formula_8}
\end{align}
Note that if one sets $\nu=16k+8$ in \eqref{eqn_O4_obstruction_formula_4-2} and considers that $\hat{\dd}\lambda = 0\modulo{2}$, then Eq.~\eqref{eqn_O4_obstruction_formula_4} reduces to Eq.~\eqref{eqn_O4_obstruction_formula_8}. 

The above special expression of $\calO_4$ allows us to obtain a general result for finite Abelian $G$. We show that for finite Abelian group $G$, ESB will never occur at $\nu=8k$, regardless of $\lambda$. That is, if $G_f$ is determined by a finite Abelian group $G$ and an arbitrary $\lambda\in \calH^2(G,\Z_2)$, it will never be enforced to break by the $\nu=8k$ fermionic iTOs. (The case of $\nu=16k$ has been generally discussed at the beginning of Sec.~\ref{sec:def-n1n2}.)  To see that, we first notice that the piece $(-1)^{\lambda\cup \lambda}$ or $(-1)^{n_2\cup n_2}$ has appeared in the study of 2D fermionic SPT states\cite{Cheng2018PRB}. It was known that this piece is always a trivial 4-cocycle for finite Abelian $G$\cite{WangPRB2017}. One may use the topological invariants in \eqref{eq:H4invairants} to directly check this fact.  Therefore, $\mathcal{O}_4$ reduces to $(-1)^{n_2 \cup \lambda}$. Then, we can choose $n_2=0$ such that $\mathcal{O}_4$ is trivial. That is, if $n_1=n_2=0$, the obstruction functions $\calO_3$ and $\calO_4$ are both trivial, which gives rise to a valid symmetry-enriched iTO.

Therefore, to look for ESB by $\nu=16k+8$ fermionic iTOs, we must go to non-Abelian $G$. It is usually not easy to check if $\calO_4$ is a trivial or non-trivial cocycle for non-Abelian group $G$, even for the expression in \eqref{eqn_O4_obstruction_formula_8}. Fortunately, we will argue in Sec.~\ref{sec:SU(N)} that $SU_f(8n)$ ($n$ being any positive integer) will be enforced to break by 2D fermionic iTOs with index $\nu = 16k+8$, through a different argument. That argument actually applies to iTOs with any even $\nu$.

Finally, we remark that although ESB will never occur for $\nu=16k$,  there are still nontrivial obstruction functions. These obstruction functions were obtained previously in the study of SPT states.\cite{Kapustin2017,WangGu2017, WangGu2020PRX}

\subsection{Examples}
In this subsection, we give a few examples of ESB by 2D fermionic iTOs. For odd-$\nu$ iTOs, any $G_f$ with $\lambda$ being a non-trivial 2-cocycle is enforced to break. For $\nu=16k+8$, we do not have an example of finite groups. However, in Sec.~\ref{sec:SU(N)}, we give an example of Lie group. So, we discuss the cases of $\nu=4k+2$ and $\nu=8k+4$ below.

\subsubsection{$\nu=4k+2$}
\label{sec_example_nu2}

We will consider examples with $G$ being Abelian and focus on those ESB solely due to $\calO_3$ obstructions for $\nu=4k+2$. We will discuss some general setup regarding group cohomology of finite Abelian groups and topological invariants (quantities that are invariant under coboundary transformations). Then, we will specialize to  $G=\Z_2\times \Z_2$ and $\Z_2\times \Z_4$.  Discussions on general Abelian $G$ as well as some discussions on $\calO_4$ are given in Appendix \ref{sec_calculation_of_H3_nu2_abelian}.

Recall that $\lambda\in \calH^2(G,\Z_2)$ can be generally parametrized in Eq.~\eqref{eqn_general_parametrize} and the topological invariants $\{\Omega_i,\Omega_{ij}\}$  in \eqref{eqn_invariant_group_extension} are complete for finite Abelian group $G=\prod_{i=1}^k \Z_{N_i}$ (we assume $N_i$ being even without loss of generality). We note that $\Omega_i,\Omega_{ij}$ are valued in $\{0,1\}$, and $\Omega_{ij}=\Omega_{ji}$, $\Omega_{ii}=0$. Also, a general 1-cocycle $n_1\in \calH^1(G,\Z_2)$ can be parametrized as
\begin{align}
n_1(a)=\sum_i q_i v_i(a)
\label{eqn_1cocycle_parametrization}
\end{align}
where $q_i=0$ or $1$, $v_i(a)=a_i \modulo{2}$, and $a=(a_1, \dots, a_k)$ is an integer vector to denote group elements of $G$. The cohomology group $ \calH^1(G,\Z_2)=2^k$ and different choices of $\{q_i\}$ exhaust all cohomology classes of 1-cocycles.

We show in Appendix \ref{sec_calculation_of_H3_nu2_abelian} that one can define a complete set of topological invariants for 3-cocycles in $\calH^3(G,\Z_2)$. Given $u\in \calH^3(G,\Z_2)$, the topological invariants can be defined as follows.  Let $\chi_a(b,c) = u(a,b,c)-u(b,a,c)+u(b,c,a)$, then
\begin{align}
\Xi_{ij} & = \sum_{i=0}^{N_j-1} \chi_{e_i}(e_j, ne_j), \nonumber\\
\Xi_{ijk} & = \chi_{e_i}(e_j,e_k) - \chi_{e_i}(e_k,e_j),
\label{eq:Xi}
\end{align}
where $e_i$ is the $i$th generator of $G$ and ``modulo 2'' is again assumed. We remark that  $\Xi_{ij}$ and $\Xi_{ji}$ are independent, while $\Xi_{ijk}$ is a fully antisymmetric tensor.  For the special $3$-cocycle $\calO_3 = \lambda\cup n_1 + \lambda\cup_1\lambda$, we show in Appendix \ref{app:O3} that
\begin{align}
\Xi_{ij} & = q_i \Omega_j + \frac{N_j}{2}(q_j-1)\Omega_{ij},  \nonumber\\
\Xi_{ijk} & = q_i \Omega_{jk} + q_j\Omega_{ki} +q_k \Omega_{ij}.
\label{eq:inv-O3}
\end{align}
Accordingly, for given $\{\Omega_i,\Omega_{ij}\}$ (i.e., given $\lambda$), the corresponding $G_f$ is enforced to break if there is no solution of $\{q_i\}$ to the equations $\Xi_{ij}=\Xi_{ijk}=0$. This result applies to an arbitrary finite Abelian group. If there are solutions, then one needs to further check the $\calO_4$ obstruction. 

Now we specialize to $G=\Z_2\times\Z_2$. In this case, Eq.~\eqref{eq:inv-O3} becomes
\begin{align}
\Xi=\begin{pmatrix} 
q_1\Omega_1 \\
q_2\Omega_2 \\ 
q_1\Omega_2+(q_2-1)\Omega_{12}\\
q_2\Omega_1+(q_1-1)\Omega_{12} \end{pmatrix}
\end{align}
where $\Xi=(\Xi_{11}, \Xi_{22}, \Xi_{12}, \Xi_{21})^T$. We ask if there is any choice of $\Omega_1$, $\Omega_2$ and $\Omega_{12}$ such that there is no $q_1$ and $q_2$ making $\Xi=0$.  It turns out that when $\Omega_1=\Omega_2=\Omega_{12}=1$, there is indeed no solution. In this case, the fermionic symmetry group $G_f = Q_8$, the quaternion group with its center identified as $\Z_2^f$.  Therefore, $G_f=Q_8$ is enforced to break by $\nu=4k+2$ fermionic iTOs.

Similarly, for $G=\Z_2\times \Z_4$, Eq.~\eqref{eq:inv-O3} becomes
\begin{align}
\Xi=\begin{pmatrix} 
q_1\Omega_1 \\
q_2\Omega_2 \\ 
q_1\Omega_2\\
q_2\Omega_1+(q_1-1)\Omega_{12} \end{pmatrix}
\end{align}
It is not hard to check that there is ESB by $\nu=4k+2$ iTOs for two cases: (1) $\Omega_1=0$ and $\Omega_2=\Omega_{12}=1$, (2) $\Omega_1=\Omega_2=\Omega_{12}=1$.

\subsubsection{$\nu=8k+4$}
\label{sec_example_nu4}

For $\nu=8k+4$, we consider examples with $G=(\Z_2)^n$. Again, we use Eq.~\eqref{eqn_general_parametrize} to parameterize $\lambda\in\calH^2(G,\Z_2)$ and use the topological invariants $\{\Omega_i,\Omega_{ij}\}$  in \eqref{eqn_invariant_group_extension} to characterize it. We use \eqref{eqn_1cocycle_parametrization} to parameterize $n_1\in \calH^1(G,\Z_2)$. In the current case, evaluating the topological invariants in \eqref{eq:Xi}  for $\calO_3=\lambda\cup n_1$ gives
\begin{align}
\Xi_{ij} & = q_i \Omega_j + q_j\Omega_{ij},  \nonumber\\
\Xi_{ijk} & = q_i \Omega_{jk} + q_j\Omega_{ki} +q_k \Omega_{ij}.
\label{eq:inv-O3-8k}
\end{align}
To limit the possible solutions $\{q_i\}$, we will constrain ourselves to the case that $\Omega_{i}=1$ for all $i$. Then, we have $\Xi_{ii} = q_i$. The only solution to $\Xi_{ii}=0$ is $q_i=0$ for all $i$, i.e, $n_1=0$. This makes the special expression of $\calO_4$ in \eqref{eqn_O4_obstruction_formula_4} applicable. In this case $\dd n_2=\calO_3=0$ is a cocycle in $\calH^2(G, \Z_2)$.

Therefore, ESB will occur if $\calO_4$ in \eqref{eqn_O4_obstruction_formula_4} is nontrivial for all possible $n_2$. Like $\lambda$ in \eqref{eqn_general_parametrize}, we will parameterize $n_2$ as follows
\begin{align}
  n_2(a,b)=\sum_i x_i w_{i}(a,b)+\sum_{i>j} x_{ij}w_{ij}(a,b)
\end{align}
where $x_i$ and $x_{ij}$ can also take 0 or 1.  Using these parametrization, $\calO_4$ in \eqref{eqn_O4_obstruction_formula_4} can be parametrized by the integers $p_i,p_{ij}$ and $x_i$, $x_{ij}$. It can be proven that for Abelian $G$, $(-1)^{n_2\cup n_2}$ is a trivial U(1)-valued 4-cocycle. Therefore, we only need to consider the rest part of $\mathcal{O}_4$, which fortunately is linear in $x_i,x_{ij}$.

For Abelian group $G=\prod_i\Z_{N_i}$, one efficient way to check whether a 4-cocycle $\chi(a,b,c,d)\in \mathcal{H}^4(G,U(1))$ is nontrivial or not is to evaluate the following set of topological invariants\cite{WangPRB2015}:
\begin{align}
e^{i\Theta_{i,l}} &=\text{$\prod_{n=1}^{N_i}$} \chi_{e_l,e_i}(e_i,ne_i),\nonumber \\
e^{i\Theta_{ij,l}} &=\text{$\prod_{n=1}^{N^{ij}}$} \chi_{e_l,e_i}(e_j,ne_j) \chi_{e_l,e_j}(e_i,ne_i),\nonumber\\
e^{i\Theta_{ijk,l}}&=\frac{\chi_{e_l,e_i}(e_k,e_j)}{ \chi_{e_l,e_i}(e_j,e_k)}
\label{eq:H4invairants}
\end{align}
where $\chi_{a,b}(c,d)$ is two-steps  slant product of $\chi(a,b,c,d)$ over $a$ and $b$ recursively and we denote the group element of $G$ as $a=\sum_i n_i e_i$ with $e_i$ as the generator of subgroup $\Z_{N_i}$. The sufficient and necessary condition that $\chi(a,b,c,d)$ is a trivial 4-cocycle is that  all these topological invariants are  equal to one.
 
We consider two cases: $G=\Z_2^n$ with $n=3$ and $n=4$. We will show that the former case does not lead to ESB but the latter  might do. First, we consider $n=3$. For $G=(\Z_2)^3$, there are many different $\lambda$. Assisted by computer, we can show that for none of these $\lambda$'s,  $\mathcal{O}_4$ is nontrivial for all  $n_2$. Here, we specifically focus on the case $\lambda=\sum_i w_i + \sum_{i<j} w_{ij}$, the calculation of which helps for the $n=4$ case. We need to consider 
eight independent topological invariants --- six $e^{i\Theta_{i, l}}$ with $i\neq l$ and two $e^{i\Theta_{ij, l}}$ with $i\neq j\neq l$. To look for ESB, it is equivalent to ask whether there exists at least one set of $x_i$ and $x_{ij}$ suc that
 \begin{align}
 e^{i\Theta_{i,l}} & =1 \nonumber\\
 e^{i\Theta_{ij,l}} & =1,
\end{align}
or equivalently
 \begin{align}
\Theta_{i,l} &=0, \quad \modulo{2\pi} \nonumber\\
\Theta_{ij,l} &=0, \quad \modulo{2\pi}.
 \end{align}
Recall that we can ignore $n_2\cup n_2$ in $\mathcal{O}_4$. So, these equations are a set of (modular) linear equations respect to $x_i$ and $x_{ij}$ and can be solved straightforwardly. It turns out that there exsit only two solutions, that is 
\begin{align}
 &(1)\, x_1=x_2=x_3=x_{12}=x_{13}=x_{23}=0,\nonumber \\
& (2)\, x_1=x_2=x_3=x_{12}=x_{13}=x_{23}=1.\nonumber
\end{align}
Therefore, there exists  $n_2$ such that all the invariants are trivial. In other words, there is no ESB for $G=(\Z_2)^3$ with  $\lambda=\sum_i w_i + \sum_{i<j} w_{ij}$.
 
Now we consider the case $n=4$ and $\lambda=\sum_i w_i+\sum_{i<j} w_{ij}$. Now there are in total 21 independent topological invariants --- twelve $e^{i\Theta_{i, l}}$, eight $e^{i\Theta_{ij, l}}$ and one $e^{i\Theta_{ijk, l}}$. First, we find out the solution space of $x_i$ and $x_{ij}$ such that all the 20 topological invariants $e^{i\Theta_{i, l}}$ and $e^{i\Theta_{ij, l}}$ are equal to one.  Similar to the case $n=3$, we find that there are only two such solutions of $x_i, x_{ij}$:
  \begin{align}
 &(1) \ x_i=x_{ij}=0\nonumber \\
& (2) \ x_i=x_{ij}=1\nonumber
 \end{align} 
for all $i$ and $j$. However, neither of the two solutions make the last topological invariant  $e^{i\Theta_{ijk, l}}$ to be 1.  In other words, there is no such $n_2$ that all the 21 topological invariants are equal to 1. Therefore, we find an example of ESB by 2D fermionic iTOs with $\nu=8k+4$: $G=\Z_2^4$ with $\lambda=\sum_i w_i+\sum_{i<j} w_{ij}$. The fermionic group $G_f$ is of order 32.

\subsection{$G_f=SU_f(N)$}
\label{sec:SU(N)}

In this subsection, we consider 2D fermionic iTOs with a continuous symmetry group $SU_f(N)$ ($N$ being even). The main motivation is to look for examples of ESB by fermionic iTOs with $\nu=16k+8$. It turns out that this example is quite  neat and gives rise to ESB for all even $\nu$'s, when $N$ varies. Our argument in this example does not follow other parts of this work. It is a generalization of an argument given in Ref.~\onlinecite{Ning2021PRB} for $SU_f(2)$. 

Let us first explain some basic structures of $SU_f(N)$. It is the usual $SU(N)$ group, and the subscript ``$f$'' denotes that the fermion parity group $\Z_2^f$ is embedded. To be concrete, let us use a representation of $N\times N$ matrices. The group $SU_f(N)$ has a center $\Z_N^f$, represented by the diagonal matrices $e^{in 2\pi/N} \mathbb{I}$, with $n=0,1,\dots, (N-1)$ and $\mathbb{I}$ being the $N\times N$ identity matrix. We consider even $N$, and  $\Z_2^f$ is identified to the group $\{\mathbb{I}, -\mathbb{I}\}$. An important fact that we will use is that there exists a $U_f(1)$ subgroup and the center $\Z_N^f$ is its subgroup. Such a group is not unique. One of the choices is $U_f(1) = \{U_\theta| 0\le \theta <2\pi\}$ and
\begin{align}
  U_\theta=  \begin{pmatrix}
   e^{i\theta }&0&\cdots & 0\\
   0& e^{i\theta }&\cdots & 0\\
   \vdots &\vdots &\ddots & \vdots\\
    0&0&\cdots & e^{i(1-N)\theta}\\
   \end{pmatrix}
\end{align}
where the last diagonal term is $e^{i(1-N)\theta}$ and other diagonal terms are $e^{i\theta}$. It is easy to check that $U_{2\pi/N}$ generates the center and the fermion parity is $U_\pi$. In short, we have $\Z_2^f \subset \Z_N^f \subset U_f(1) \subset SU_f(N)$.

Now we consider an $SU_f(N)$-enriched 2D fermionic iTO with even $\nu$. First, with respect to $U_f(1)$, we can define a Hall conductance $\sigma_H$. It is known that, for fermionic iTOs with even $\nu$, we have
\begin{align}
    \sigma_H=\frac{\nu}{2} \modulo{8}.
    \label{eq:sigmaH0}
\end{align}
The ``modulo $8$'' is due to the $E_8$ state which does not contribute to $\sigma_H$ but changes the chiral central charge $c_-=\nu/2$ by 8. Next, we compute $\sigma_H$ in two ways. On one hand, by Laughlin's argument, $\sigma_H$ is equal to the charge $Q_{2\pi}$ accumulated on an adiabatically inserted $2\pi$ flux. On the other hand, we may also adiabatically insert $N$ copies of $2\pi/N$ fluxes. Let us denote the defect corresponding to a $2\pi/N$ flux as $v$, and $Q_v$ is the $U_f(1)$ charge carried by $v$. The total charge carried by the $v$ defects should be equal to $Q_{2\pi}$. Then, we have
\begin{align}
    \sigma_H=N Q_v.
    \label{eq:qv}
\end{align}
In general,  $Q_v$ is fractional, because $v$ is a defect instead of a local excitation.  

We argue that in the presence of $SU_f(N)$, the charge $Q_v$ must be an integer. Let us first give a simple argument and then justify it more rigorously. Recall that $U_{2\pi/N}$ lies in the center of $SU_f(N)$. Accordingly, inserting a $2\pi/N$ flux (the $v$ defect) does not break $SU_f(N)$. So, $v$ must carry a projective representation of $SU_f(N)$. However, $SU_f(N)$ is a connected and simply connected compact Lie group, which does not support nontrivial projective representations. Accordingly, $v$ carries a linear representation of $SU_f(N)$, which is also a linear representation of the subgroup $U_f(1)$. That means, $v$ must carry an integer charge of $U_f(1)$. The linear representation carried by $v$ is generally irreducible due to energy consideration.

Two clarifications are needed to make the above argument justified. First, from our understanding of SETs, symmetry fractionalization goes beyond projective representations. Technically speaking, we need to replace $\calH^2(G,U(1))$ which classifies projective representations with $\calH^2(G, \mathcal{A})$, where $\mathcal{A}$ is an Abelian group formed by Abelian anyons under fusion. However, for $G=SU_f(N)$, it can be shown mathematically that both $\calH^2(G,U(1))$ and $\calH^2(G, \mathcal{A})$ are trivial. Since $v$ is a defect, not an anyon, readers may still not be convinced. There is another way to see this: one can gauge the $\Z_N^f$ center and turn the fermionic iTO into a bosonic topological order enriched by the quotient group $PSU(N)=SU_f(N)/\Z_N^f$. In this way, $v$ becomes a true anyon. Symmetry fractionalization on $v$ is classified by $\calH^2(PSU(N), \mathcal{A})$, whre $\mathcal{A}$ is the fusion group associated with the bosonic topological order whose structure depends on $\sigma_H$. Using the universal coefficient theorem, one can show that $ \calH^2(PSU(N), \mathcal{A}) = \mathrm{Tor}\{\calH^2(PSU(N), U(1)),\mathcal{A}\}$, where $\mathrm{Tor}$ is a cohomological operation and $\calH^2(PSU(N), U(1))=\Z_N$. Regardless of what group $\mathrm{Tor}\{\calH^2(PSU(N), U(1)),\mathcal{A}\}$ is, what is important to us is the physical meaning:  $\mathrm{Tor}$ picks out those projective representations of $PSU(N)$ that are compatible with the fusion group $\mathcal{A}$. This interpretation has been widely used for $SO(3)$, which is equal to $SU_f(2)/\Z_2^f$ (see, e.g., Ref.~\cite{WangChong2014PRB}).  We believe it is applicable generally. Accordingly, symmetry fractionalization on $v$ can all be characterized by projective representations of $PSU(N)$, before considering the compatibility to fusion rules. Finally, it is well known that projective representations of $PSU(N)$ are simply linear representations of $SU_f(N)$. So, it goes back to our argument above. 

Second, the irreducible linear representation of $SU_f(N)$ carried by $v$ in the fermionic picture is usually multi-dimensional. Then, when we insert multiple $2\pi/N$ fluxes, they may stay in different states inside the irreducible linear representation, as these states are energetically degenerate. However, $U_f(1)$ charges carried by different states in an irreducible representation of $SU_f(N)$ can only differ by a multiple of $N$. This is again because $\Z_N^f\subset U_f(1)$ is the center of $SU_f(N)$, so that all states in an irreducible representation of $SU_f(N)$ must carry the same $\Z_N^f$ charge. This is equivalent to say that $U_f(1)$ charges can only differ by  a multiple of $N$. Accordingly, even if this subtlety is taken into account, Eq.~\eqref{eq:qv} is relaxed to
\begin{align}
    \sigma_H= 0 \modulo{N}
    \label{eq:qv2}
\end{align}
which is enough for our purpose.

With these clarifications, we now combine \eqref{eq:sigmaH0} and \eqref{eq:qv2}. We immediately have
\begin{align}
    \frac{\nu}{2} = n\gcd(N,8),
    \label{eq:SUN_condition}
\end{align}
where $n$ is any integer and ``$\gcd$'' stands for greatest common divisor. Equation \eqref{eq:SUN_condition} must be satisfied by all symmetry-enriched fermionic iTOs. If it cannot be satisfied, that means the symmetry group $SU_f(N)$ is enforced to break by the iTO. More explicitly, we have
\begin{enumerate}
\item $SU_f(2l)$ is enforced to break by 2D fermionic iTOs with $\nu=4k+2$;

\item $SU_f(4l)$ is enforced to break by 2D fermionic iTOs with $\nu=8k+4$;

\item $SU_f(8l)$ is enforced to break by 2D fermionic iTOs with $\nu=16k+8$;
\end{enumerate}
where $k,l$ are any positive integers. For $\nu=16k$, \eqref{eq:SUN_condition} is always satisfied, which is consistent to the 16-fold periodicity of ESB physics argued above.

\section{Derivation of 2D ESB criteria}
\label{sec:derivation}

In this section, we give alternative derivations of $\mathcal{O}_2$ and $\mathcal{O}_3$ obstructions, and derive the $\mathcal{O}_4$ obstruction and 2D ESB criteria. Compared to Sec.~\ref{sec:def-n1n2} where we define $n_1$ and $n_2$ using defects in the fermionic language,  the main strategy here is to gauge the fermion parity group $\Z_2^f$ and turn the fermionic iTO into a bosonic topological order. 

Let us denote the resulting bosonic topological order as $\mathcal{C}_{\nu}$, for the fermionic iTO of index $\nu$. Detailed properties of $\mathcal{C}_\nu$ can be found in Ref.~\onlinecite{Kitaev2006}. When $\nu$ is odd, $\mathcal{C}_\nu$ contains three anyons:
\begin{align}
\mathcal{C}_\nu= \{1,f,\sigma\},\nonumber
\end{align}
where $f$ corresponds to the original fermion and $\sigma$ is a non-Abelian fermion-parity flux of quantum dimension $d_\sigma=\sqrt{2}$. When $\nu$ is even, $\mathcal{C}_\nu$ contains four anyons:
\begin{align}
\mathcal{C}_\nu= \{1,f,v,vf\},\nonumber
\end{align}
where $v$ and $vf$ are two Abelian fermion-parity fluxes satisfying the fusion rule $v\times f = vf$. The even-$\nu$ cases can be further distinguished by the fusion rule of fermion-parity fluxes,
\begin{align}
v\times v = \left\{
\begin{array}{ll}
f, & \ \text{if } \nu=4k+2\\
1, & \ \text{if } \nu=4k
\end{array}
\right.
\label{eq:fp-fusion-even}
\end{align}
where $k$ is an integer. In all cases, the topological spin of a fermion parity flux is $\theta_\sigma=e^{i\nu\pi/8}$ or $\theta_v=\theta_{vf}=e^{i\nu\pi/8}$, and the mutual statistics between $f$ and fluxes is $M_{f,\sigma}=-1$ or $M_{f,v}=-1$.  Hence, there is a 16-fold periodicity in $\nu$. 

With $\Z_2^f$ gauged, there remains a global symmetry group $G$. Hence, we obtain an SET state of topological order $\mathcal{C}_\nu$ and symmetry group $G$. We will see that it is not an arbitrary SET, but one with certain symmetry properties fixed by the 2-cocycle $\lambda\in\calH^2(G,\Z_2)$. Studying such a conditional SET leads to new consistency relations that do not exist in unconditional SETs. These conditional consistency relations, which we call \emph{conditional anomalies}, give rise to the $\mathcal{O}_2$ and $\mathcal{O}_3$ obstructions which otherwise do not exist. (There is also an $\tilde{\mathcal{O}}_3$ obstruction in general bosonic SETs, but $\mathcal{O}_3$ is different from, although related to, $\tilde{\mathcal{O}}_3$.) Together with an $\mathcal{O}_4$ obstruction of bosonic SETs, they can be used to derive the ESB criteria associated with the original fermionic iTOs.

\subsection{Basics of SETs}
\label{sec:set}
Before studying our specific SETs,  we review the basics of the general theory of bosonic SET states. For more details, one can refer to Ref.~\onlinecite{BarkeshliPRB2019_SET}. The description of SETs involves several layers of data which we briefly explain below one by one.

First, a bosonic topological order $\mathcal{C}$ contains anyons $1, a, b,\dots$, where $1$ is the trivial anyon. Mathematically, $\mathcal{C}$ is described by a unitary modular tensor category (UMTC).\cite{Kitaev2006} Physically, anyons are characterized by their fusion and braiding properties. They follow a set of fusion rules $a\times b=\sum_c N_{ab}^c c$, where the fusion coefficient $N_{ab}^c$ is a non-negative integer. There exists a unique anyon $\bar{a}$, namely the anti-particle of $a$, such that $N_{a\bar{a}}^1=1$. Two important quantities of each anyon $a$ are the quantum dimension $d_a$ and topological spin $\theta_a$. For Abelian anyons, $d_a=1$ and $\theta_a$ is its self-statistics. Associated with each $N_{ab}^c\neq 0$ there is a vector space ${V}_{c}^{ab}$, called the fusion or splitting space, whose dimension is $N_{ab}^c$. A key quantity that relates different fusion spaces is the so-called $F$ symbol, which is an isomorphism between fusion spaces of three anyons $\left(F^{abc}_{d}\right)_{ef}: \bigoplus_{e} V^{ab}_e \otimes V^{ec}_d\rightarrow \bigoplus_{f} V^{af}_d \otimes V^{bc}_f$. If one exchanges two anyons, it corresponds to another isomorphism, the $R$ symbol, $R^{ab}_c: V^{ab}_c \rightarrow V_{c}^{ba}$. 

Every $\mathcal{C}$ processes a set of topological symmetries, which form a group denoted as $\Aut(\mathcal{C})$. We only consider those that are unitary and orientation-preserving. A topological symmetry is an invertible map from $\mathcal{C}$ to itself. It contains two parts: (i) a permutation of anyons 
\begin{align}
a \rightarrow a'\equiv \varphi(a),
\label{eq:autolabel}
\end{align}
and (ii) an action on the fusion space $V_c^{ab}$ 
\begin{align}
\varphi(|a,b;c\rangle)= u_{c'}^{a'b'}|a',b';c'\rangle,
\label{eq:phi-u}
\end{align}
where $u_{c'}^{a'b'}$ is a phase factor and $|a,b; c\rangle\in V_c^{ab}$. For simplicity, we have assumed $N_{ab}^c=0$ or $1$, as all our cases satisfy this assumption. Topological symmetries keep all the data of $\mathcal{C}$ invariant, e.g.,
\begin{align}
N_{ab}^c & = N_{a'b'}^{c'}, \nonumber\\
\theta_{a} & = \theta_{a'}, \nonumber\\
\left[F^{abc}_d\right]_{ef} & =u^{a'b'}_{e'} u^{e'c'}_{d'} \left[F_{d'}^{a'b'c'}\right]_{e'f'}\left[u^{a'f'}_{d'}\right]^* \left[u^{b'c'}_{f'}\right]^*, \nonumber\\
R^{ab}_c & = u^{b'a'}_{c'}R_{c'}^{a'b'} \left[u^{a'b'}_{c'}\right]^*.
\label{eq:autoequi}
\end{align}
One can see that if $u_{c}^{ab}$ is multiplied by  $ {\gamma_a\gamma_b}/{\gamma_c}$, equations \eqref{eq:autoequi} remain unchanged.  Such a transformation is called a natural isomorphism. It is regarded as an equivalence relation between different sets of $\{u_{c}^{ab}\}$. Group multiplication of $\Aut(\mathcal{C})$ in the aspect of $u_{c}^{ab}$ is upto natural isomorphisms. It is believed that $\{u_{c}^{ab}\}$ is fixed up to natural isomorphisms once anyon permutation in \eqref{eq:autolabel} is given, which is indeed the case in our examples. 

Second, consider a system with the microscopic symmetries forming a group $G$. If it is unbroken at low energy, $G$ must be mapped into $\Aut(\mathcal{C})$. This is characterized by a group homomorphism $\rho:G\rightarrow \text{Aut}(\mathcal{C})$. 
That is, every $\gbf\in G$ is associated with a topological symmetry, denoted as $\rho_\gbf$.  We denote the anyon permutation $\rho_\gbf(a)={}^{\gbf}a$ for short. For anyon permutations, $\rho$ is an exact group homomorphism. On the other hand, the action on states in fusion space is upto natural isomorphisms:
\begin{align}
\rho_\mathbf{gh}=\kappa_\mathbf{g,h}\circ \rho_\mathbf{g}\circ\rho_\mathbf{h},
\end{align}
where $\kappa_\mathbf{g,h}$ is a natural isomorphism. Let $u_{\gbf}(a,b,c)$ be the phase factor associated with $\rho_{\gbf}$'s action on fusion spaces. Then, the explicit expression of $\kappa_{\gbf,\hbf}$ is 
\begin{align}
\kappa_\mathbf{g,h}(a,b,c)=\frac{u_{\mathbf{gh}}(a,b,c)}{u_{\mathbf{g}}(a,b,c)u_{\mathbf{h}}({}^{\bar{\mathbf{g}}}a,{}^{\bar{\mathbf{g}}}b,{}^{\bar{\mathbf{g}}}c)},
\label{eq:kappa1}
\end{align}
where $\bar{\gbf}$ is a short-hand notation for $\gbf^{-1}$. Since $\kappa_\mathbf{g,h}$ is a natural isomorphism, we can decompose it as
\begin{align}
\kappa_\mathbf{g,h}(a,b,c)=\frac{\beta_a(\mathbf{g,h})\beta_b(\mathbf{g,h})}{\beta_c(\mathbf{g,h})}.
\label{eqn_decomp}
\end{align}
The quantity $\beta_a(\gbf,\hbf)$ is subject to two kinds of ambiguities:  (i) the decomposition \eqref{eqn_decomp} is not unique and  a shift $\beta_a(\mathbf{g,h}) \rightarrow  \nu_a(\mathbf{g,h})\beta_a(\mathbf{g,h})$ is also valid as long as $\nu_a(\mathbf{g,h})\nu_b(\mathbf{g,h})=\nu_c(\mathbf{g,h})$ if $N_{ab}^c\neq 0$; (ii) a natural isomorphism in $u_\gbf(a,b,c)$ induces a shift $\beta_a(\gbf,\hbf)\rightarrow \beta_a(\gbf,\hbf) \gamma_a(\gbf\hbf)/\gamma_a(\gbf)\gamma_{{}^{\bar{\gbf}}a}(\hbf)$.

With $\beta_a(\gbf,\hbf)$, we define an important quantity
\begin{align}
\Omega_a(\mathbf{g,h,k})=\frac{\beta_{{}^{\bar{\mathbf{g}}}a}(\mathbf{h,k})\beta_{a}(\mathbf{g,hk})}{\beta_{a}(\mathbf{g,h})\beta_{a}(\mathbf{gh,k})}.
\label{eqn_obstution_H3_def}
\end{align}
By definition, $\Omega_a$ is a coboundary in $\calH^3_{\rho}(G,U(1))$ for every $a$. Also, associativity of $\rho_\gbf$ can be used to show that $\Omega_a(\mathbf{g,h,k})\Omega_b(\mathbf{g,h,k})=\Omega_c(\mathbf{g,h,k})$ if $N_{ab}^c\neq 0$. The latter implies that
\begin{align}
\Omega_a(\mathbf{g,h,k})=M_{a, \tilde{\mathcal{O}}_3(\mathbf{g,h,k})}^*
\label{eqn_obstution_H3_def2}
\end{align}
where $\tilde{\mathcal{O}}_3(\gbf,\hbf,\kbf)$ is an Abelian anyon in $\mathcal{C}$ and $M_{a,b}$ is the mutual statistical phase between $a$ and an Abelian anyon $b$. Moreover, $\tilde{\mathcal{O}}_3(\mathbf{g,h,k})$ is a 3-cocycle in $\mathcal{H}_\rho^3(G,\mathcal{A})$, where $\mathcal{A}$ denotes the group of Abelian anyons in $\mathcal{C}$. However, it is important to note that $\tilde{\mathcal{O}}_3(\gbf,\hbf,\kbf)$ might be a nontrivial cocycle. The ambiguities in $\beta_a(\gbf,\hbf)$ induce coboundary transformations in $\tilde{\mathcal{O}}_3(\gbf,\hbf, \kbf)$. Accordingly, only the cohomology class $[\tilde{\mathcal{O}}_3]$ in $\mathcal{H}_\rho^3(G,\mathcal{A})$ matters.

Third, other than topological actions, symmetries in $G$ also act on local degrees of freedom around each anyon. Consider a state that contains a set of anyons $\{a_i\}$, which are spatially well-separated. The overall action of $\gbf\in G$ is a combination of local and topological actions:
\begin{align}
\mathcal{R}_\mathbf{g}=\prod_{i} U_\mathbf{g}^{(i)} \rho_\mathbf{g},
\label{eq:symaction}
\end{align}
where $U_\mathbf{g}^{(i)}$ is a local unitary operator, supported in the neighborhood of anyon $a_i$, and $\rho_{\gbf}$ is the topological action in the fusion space of $\{a_i\}$ which can be decomposed into those in \eqref{eq:phi-u}. The local operators form a projective representation
\begin{align}
U_\mathbf{g}^{(i)}{}^{\mathbf{g}}U_\mathbf{h}^{(i)} & =\eta_{a_i}(\mathbf{g,h})U_\mathbf{gh}^{(i)},
\label{eq:U-proj}
\end{align}
where ${}^{\gbf}U^{(i)}_\hbf = \rho_\gbf U_\hbf^{(i)} \rho_\gbf^{-1} $, and $\eta_a(\gbf,\hbf)$ is a 2-cocycle in $\calH^2_\rho(G,U(1))$ for every $a$. The local unitary operator $U_\gbf^{(i)}$ has a phase ambiguity. At the same time, $\rho_\gbf$ also has a phase ambiguity due to natural isomorphisms. The two ambiguities shall be correlated such that $\mathcal{R}_{\gbf}\mathcal{R}_{\hbf} = \mathcal{R}_{\gbf\hbf}$. It was shown in Ref.~\cite{BarkeshliPRB2019_SET} that a nice quantity to look at is the ratio
 \begin{align}
\omega_a(\gbf,\hbf) = \frac{\beta_{a}(\mathbf{g,h})}{\eta_a(\gbf,\hbf)},
\label{eqn_sf_relation}
 \end{align}
which satisfies $\omega_a(\gbf,\hbf)\omega_b(\gbf,\hbf) = \omega_c(\gbf,\hbf)$ if $N_{ab}^c\neq 0$. That means, we have
\begin{align}
\omega_a(\gbf,\hbf) = M_{a,w(\gbf,\hbf)}^*, 
\label{eqn_sf_relation2}
\end{align}
where $w(\mathbf{g,h})$ is an Abelian anyon in $\mathcal{A}\subset\mathcal{C}$. It is equivalent to say that
\begin{align}
\frac{\eta_a(\gbf,\hbf)\eta_b(\gbf,\hbf)}{\eta_c(\gbf,\hbf)}=\frac{\beta_a(\gbf,\hbf)\beta_b(\gbf,\hbf)}{\beta_c(\gbf,\hbf)},
\label{eq:eta_beta}
\end{align}
whenever $N_{ab}^c\neq 0$. Accordingly, gauge transformations of $\eta_a(\gbf, \hbf)$ and $\beta_a(\gbf,\hbf)$ shall be correlated such that \eqref{eq:eta_beta} always holds. Combining (\ref{eqn_obstution_H3_def}), (\ref{eqn_obstution_H3_def2}), (\ref{eqn_sf_relation}), (\ref{eqn_sf_relation2}) and that $\eta_a(\gbf,\hbf)$ is 2-cocycle, one can show that
\begin{align}
\tilde{\mathcal{O}}_3 & (\mathbf{g,h,k}) =\text{d}w(\mathbf{g,h,k})  \nonumber\\& =\rho_\mathbf{g}[w(\mathbf{h,k})] \overline{w(\mathbf{gh,k})} w(\mathbf{g,hk})\overline{w(\mathbf{g,h})},
\label{eqn_obstruction_sf}
\end{align}
where $\bar{x}$ stands for the anti-particle of anyon $x$. This implies that $\tilde{\mathcal{O}}_3(\mathbf{g,h,k})$ should be a 3-coboundary in $\mathcal{H}^3_\rho(G,\mathcal{A})$ to give rise to a valid $w(\gbf,\hbf)$. If $\tilde{\mathcal{O}}_3(\mathbf{g,h,k})$, obtained from its definitions \eqref{eqn_obstution_H3_def} and \eqref{eqn_obstution_H3_def2}, is a nontrivial 3-cocycle,  Eq.~\eqref{eqn_obstruction_sf} can never have a solution for $w(\gbf,\hbf)$, implying that the form of symmetry action in \eqref{eq:symaction} does not hold. Then, it is said that the topological action $\rho$ has a \emph{symmetry localization} obstruction.

When the obstruction $\tilde{\mathcal{O}}_3(\gbf,\hbf,\kbf)$ is trivial, one then look for $w(\gbf,\hbf)$ that satisfies (\ref{eqn_obstruction_sf}). It is not hard to see that given $w(\gbf,\hbf)$ a solution, $w(\gbf,\hbf)t(\gbf,\hbf)$ is also a valid solution if $t(\gbf,\hbf)\in\mathcal{H}_\rho^2(G,\mathcal{A})$. At the same time, $w(\gbf,\hbf)$ is subject to an ambiguity $w(\gbf,\hbf)\rightarrow w(\gbf,\hbf)\rho_\gbf[\zeta(\hbf)]\zeta(\gbf)\overline{\zeta(\gbf\hbf)}$, where $\zeta(\gbf)$ is an arbitrary Abelian anyon in $\mathcal{A}$. This ambiguity is due to an independent phase shift that we can perform on $U_\gbf^{(i)}$ or equivalently on $\eta_a(\gbf,\hbf)$ for a fixed $\beta_a(\gbf,\hbf)$.  Then, different solution classes $[w]$ to \eqref{eqn_obstruction_sf} are said to describe different \emph{symmetry fractionalization} classes. They are related to each other by a cohomology class $[t]\in \calH^2_\rho(G,\mathcal{A})$. Once $w(\gbf,\hbf)$ and $\beta_a(\gbf,\hbf)$ are given, the projective phase factor $\eta_a(\gbf,\hbf)$ is determined by \eqref{eqn_sf_relation} and \eqref{eqn_sf_relation2}.

Fourth, for a given symmetry action $\rho\in \Aut(\mathcal{C})$ and a symmetry fractionalization class described by $w(\gbf,\hbf)$, there are additional consistency conditions to satisfy. For an SET to be valid, one shall be able to insert symmetry defects (i.e., couple to a background gauge field) and these defects shall form a fusion category.  It was shown in Ref.~\cite{chen14} that given $\rho$ and $\omega$, one can construct a quantity $\tilde{\mathcal{O}}_4$ out of them. It is a 4-cocycle in $\mathcal{H}^4(G,U(1))$ and subject to coboundary transformations. For the defects to form a consistent fusion theory, it is required the cohomology class $[\tilde{\mathcal{O}}_4]$ is trivial. If it is nontrivial, we say there is an $\tilde{\mathcal{O}}_4$ obstruction.  The general expression $\tilde{\mathcal{O}}_4$ is complicated. A simple one  with a trivial $\rho$ is given by
\begin{align}
\tilde{\mathcal{O}}_4(\mathbf{g}, \mathbf{h}, & \mathbf{k}, \mathbf{l}) 
  = R_{w\left(\mathbf{k},\mathbf{l}\right) ,w\left(
\mathbf{g},\mathbf{h}\right)}\frac{F_{w\left(\mathbf{h},\mathbf{k}\right),w\left(\mathbf{g},\mathbf{h}\mathbf{k}\right)
,w\left(\mathbf{g}\mathbf{h}\mathbf{k},\mathbf{l}\right) }}{ F_{w\left(
\mathbf{h},\mathbf{k}\right)  ,w\left( \mathbf{h}\mathbf{k},\mathbf{l}\right)  ,w\left(\mathbf{g},\mathbf{h}\mathbf{k}\mathbf{l}\right)
}} \nonumber \\
& \times\frac{F_{w\left(\mathbf{g},\mathbf{h}\right), w\left(
\mathbf{k},\mathbf{l}\right)  ,w\left( \mathbf{g}\mathbf{h},\mathbf{k}\mathbf{l}\right)  }}{F_{w\left(\mathbf{g},\mathbf{h}\right)  ,w\left(\mathbf{g}\mathbf{h},\mathbf{k}\right)  ,w\left(
\mathbf{g}\mathbf{h}\mathbf{k},\mathbf{l}\right)  }} \frac{ F_{w\left( \mathbf{k},\mathbf{l}\right)
,w\left( \mathbf{h},\mathbf{k}\mathbf{l}\right)  ,w\left( \mathbf{g}, \mathbf{h}\mathbf{k}\mathbf{l}\right)  }}{F_{w\left(\mathbf{k},\mathbf{l}\right)  ,w\left(\mathbf{g},\mathbf{h}\right)
,w\left(\mathbf{g}\mathbf{h},\mathbf{k}\mathbf{l}\right)}},\label{eq:o4}
\end{align}
where the $R$ and $F$ symbols are associated with Abelian anyons in $\mathcal{C}$. If $[\tilde{\mathcal{O}}_4]$ is trivial, one can study the fusion theory of defects, which involves additional $F$ symbols among the defects. It was shown that different defect theories can be obtained by twisting the defect $F$ symbols with a 3-cocycle $\alpha_3\in \calH^3(G,U(1))$. Physically, it corresponds to stack an SET with a 2D SPT state which is indeed classified by $\calH^3(G,U(1))$. In this work, we will not explore the data $\alpha_3$ as it is irrelevant to the phenomenon of enforced symmetry breaking.

To summarize, there are two obstruction classes: (i) existence of a valid symmetry fractionalization class $[w]$ requires $[\tilde{\mathcal{O}}_3]$ to be trivial; (ii) a consistent defect theory further requires $[\tilde{\mathcal{O}}_4]$ to be trivial.  A valid bosonic SET state requires both obstruction classes to be trivial.

\subsection{$\mathcal{O}_2$ and symmetry fractionalization condition}
\label{sec:O2andSF}

We now apply the above general theory to our case, the topological order $\mathcal{C}_\nu$ enriched by  symmetry group $G$. In this subsection, we consider the case of odd $\nu$. In this case, there is no nontrivial topological symmetry, i.e., $\Aut(\mathcal{C}_\nu)= \Z_1$. Then, no symmetries in $G$ permute anyons. It is known that the $\tilde{\mathcal{O}}_3$ obstruction is always trivial without anyon permutations. We can take the gauge $u_{c}^{ab}=1$, $\beta_a(\bfg,\bfh)=1$, such that $\tilde{\mathcal{O}}_3(\gbf,\hbf,\kbf)=1$. 

We make some important comments here before proceeding. The fermion $f$ is a local excitation in the original iTO, so the fusion space ${V}^{ff}_1$ is not topological but instead local. Then, one should impose the condition that $u^{ff}_1=1$ after $\Z_2^f$ is gauged. This condition was previously discussed in Ref.~\cite{Tata2021arXiv}. It is a necessary condition for establishing an inverse mapping from the bosonic SET back to the fermionic iTO. With this condition imposed, natural isomorphisms on $f$ can only be $\gamma_f = \pm1$. Then, the quantity $\beta_f(\gbf,\hbf)$ in the decomposition \eqref{eqn_decomp} shall also take a value $+1$ or $-1$. Under an appropriate gauge choice, we find that we can set $\beta_f(\gbf,\hbf)=1$ 
for all $\mathcal{C}_\nu$'s (the cases of even $\nu$'s are discussed in Appendix \ref{app:localization_anomaly}). In fact, we postulate that the most general case is $\beta_f(\gbf,\hbf)=\gamma_f(\bfg\bfh)/[\gamma_f(\bfg)\gamma_f(\bfh)]$, and a shift $\beta_f(\gbf,\hbf)\rightarrow \beta_f(\gbf,\hbf)\nu_f(\gbf,\hbf)$ is not allowed in general (see a discussion below).  The $\calO_2$ and $\calO_3$ obstructions in Sec.~\ref{sec:def-n1n2} are correctly recovered only under this assumption. We believe that it is due to the locality of $f$ in the original fermionic systems, such that \eqref{eq:f-lambda} below describes the absolute symmetry fractionalization on $f$ and disallows a shift $\beta_f(\gbf,\hbf)\rightarrow \beta_f(\gbf,\hbf)\nu_f(\gbf,\hbf)$. We will take $\beta_f(\gbf,\hbf)=\gamma_f(\bfg\bfh)/[\gamma_f(\bfg)\gamma_f(\bfh)]$ as an assumption in this work, and we refer the readers to recent works \cite{fset2021a,fset2021b} on general fermionic SET states for more systematic discussions.



Symmetry fractionalization is characterized by an Abelian anyon $w(\gbf,\hbf)\in \mathcal{A}\subset\mathcal{C}_\nu$.  Again, it is special in our SET as it originates from a fermionic theory with symmetry group $G_f$ determined by $G$ and a 2-cocycle $\lambda\in\calH^2(G, \Z_2)$. From our study of 0D iTOs (Sec.~\ref{sec:deri-0D}), we know that the fermion $f$ obeys a projective representation of $G$, with the projective factor being $(-1)^{\lambda(\gbf,\hbf)}$. Accordingly, when $\Z_2^f$ is gauged, the local action of $G$ on $f$ shall be associated with a projective phase factor
\begin{align}
{\eta}_f(\gbf,\hbf)=(-1)^{\lambda(\gbf,\hbf)}.
\label{eq:f-lambda}
\end{align}
It is an important condition, constraining the possible symmetry fractionalization classes $[w]$. In fact, for odd $\nu$, we show below that the condition \eqref{eq:f-lambda} is so strong that it is actually not compatible with any symmetry fractionalization pattern when $\lambda$ is a nontrivial cocycle in $\calH^2(G,\Z_2)$. We remark that due to  \eqref{eq:eta_beta} and the condition that natural isomorphism $\gamma_f = \pm1$, gauge transformations on $\eta_f(\gbf,\hbf)$ are also constrained with values $\pm 1$. This makes sense, as the right-hand side of Eq.~\eqref{eq:f-lambda} only has a $\pm1$ ambiguity from coboundary transformations of $\lambda(\gbf,\hbf)$. We also remark that the condition \eqref{eq:f-lambda} implies that we interpret $\bfg\in G$ as $\bfg_0\in G_f$. Generally speaking, $\bfg\in G$ corresponds to the coset $\{\gbf_0, \gbf_1\}$ in the quotient group $G_f/\Z_2^f$. 

We now show that $\mathcal{C}_\nu$ with an odd $\nu$ is incompatible with \eqref{eq:f-lambda} when $\lambda$ is a non-trivial cocycle. Combining Eqs.~\eqref{eqn_sf_relation}, \eqref{eqn_sf_relation2} and \eqref{eq:f-lambda}, we obtain
\begin{align}
M_{f,w(\gbf,\hbf)}^* = \beta_f(\gbf,\hbf)(-1)^{\lambda(\gbf,\hbf)}.
\label{eq:mlambda-odd}
\end{align}
As $w(\gbf,\hbf)=1$ or $f$, the left-hand side must be equal to 1. Under the assumption that $\beta_f(\gbf,\hbf)=\gamma_f(\gbf\hbf)/[\gamma_f(\gbf)\gamma_f(\hbf)]$, the right-hand side can be equal to $1$ if and only if $\lambda(\gbf,\hbf)$ is a trivial cocycle in $\calH^2(G, \Z_2)$. That is, Eq.~\eqref{eq:mlambda-odd} has no solution $w(\gbf,\hbf)$, if $\lambda$ is a nontrivial 2-cocycle in $\calH^2(G,\Z_2)$. Instead, for even $\nu$, the fermion-parity fluxes are Abelian so that $w(\gbf,\hbf)$ has choices other than $1$ and $f$, making solutions of Eq.~\eqref{eq:mlambda-odd} to exist even if $\lambda$ is a non-trivial cocycle (see Sec.~\ref{sec_derivation_h3_nu2}). Accordingly, we recover the obstruction $\mathcal{O}_2(\gbf,\hbf) = \nu \lambda(\gbf,\hbf)$, first discussed in Sec.~\ref{sec:def-n1n2}. We remark that if we allow $\beta_f(\bfg,\bfh)= \nu_f(\gbf,\hbf)\gamma_f(\gbf\hbf)/[\gamma_f(\gbf)\gamma_f(\hbf)]$, with $\{\nu_a(\bfg,\bfh)\}$ respecting anyon fusion, the $\calO_2$ obstruction becomes $\nu[\lambda(\gbf,\hbf)+\nu_f(\gbf,\hbf)]$. Since $\nu_f(\gbf,\hbf)$ may be a non-trivial 2-cocycle in general, it is inconsistent with the result from Sec.~\ref{sec:def-n1n2}. This justifies the assumption on $\beta_f(\bfg,\bfh)$.

We note that $\calO_2$ does not exist in usual SETs. It exists in our SETs due to the condition \eqref{eq:f-lambda}. We will call it a \emph{conditional} obstruction or anomaly. The obstruction $\mathcal{O}_2$ alone is enough to establish Criterion \ref{criterion3}, which is complete for the ESB physics in all odd-$\nu$ iTOs. So, we will move on to the even-$\nu$ cases below. However, it does not mean that there is no other obstructions for the odd-$\nu$ cases. The $\tilde{\calO}_4$ obstruction of bosonic SETs will generally be there even if $\lambda$ is trivial.

\subsection{$\mathcal{O}_3$ and $\mathcal{O}_4$ for $\nu=4k+2$}
\label{sec_derivation_h3_nu2}

We now consider even $\nu$, by starting with $\nu=4k+2$.  Different from the odd-$\nu$ case, the group of topological symmetries $\Aut(\mathcal{C}_\nu)= \Z_2$ for even $\nu$.\footnote{For $\nu=8$, there exist other topological symmetries that permute $f$ with the fermion-parity vortices. However, our SETs originate from fermionic iTOs so that $f$ should be unpermuted. We only consider those topological symmetries that keep $f$ unpermutated.} The nontrivial one is associated with the permutation $v\leftrightarrow vf$.  The symmetry action on $\mathcal{C}_{\nu}$ is given by a group homomorphism
\begin{align}
n_1: G\rightarrow \Z_2.
\label{eq:n1-SET}
\end{align}
Equivalently, $n_1\in \calH^1(G,\Z_2)$. If $n_1(\gbf)=1$, the symmetry $\gbf$ permutes $v$ and $vf$; if $n_1(\gbf)=0$, it does not. In other words, the anyon permutation is given by 
\begin{align}
    \rho_\gbf(v) = v f^{n_1(\gbf)}, \quad \rho_\gbf(f) = f.
\label{eq:4k+2_permute}
\end{align}
For $n_1$ to be valid, it requires the $\tilde{\mathcal{O}}_3$ obstruction to be trivial. We verify this explicitly in Appendix \ref{app:localization_anomaly} that $\tilde{\mathcal{O}}_3$ is indeed trivial for even $\nu$. In fact, we show in Appendix \ref{app:localization_anomaly} that $\tilde{\mathcal{O}}_3(\gbf,\hbf,\kbf)=1$ under certain gauge choice. 

The quantity $n_1$ in \eqref{eq:n1-SET} is the same $n_1$ defined in Sec.~\ref{sec:def-n1n2} that characterizes the Majorana zero mode on the $\gbf_0$ and $\gbf_1$ defects (for even $\nu$, if $\gbf_0$ carries MZM, so is $\gbf_1$). This correspondence has already been known before in the context of fermionic SPT phases\cite{Cheng2018PRB}. To see that, imagine inserting a $\gbf$ defect in the bosonic SET and winding a $v$ anyon around defect. If $n_1(\gbf)=1$, $v$ becomes $vf$ after winding around the $\gbf$ defect, implying that an anyon $vf\times \bar{v} = f$ is absorbed into the defect. That is, a $\gbf$ defect has the ability to absorb $f$ and it must carry a Majorana zero mode in the original fermionic iTO.

After the topological symmetry action is specified by $n_1$, we now discuss the local symmetry action characterized by the symmetry fractionalization  $w(\mathbf{g,h})\in \mathcal{A}=\mathcal{C}_\nu$. Again, $w(\mathbf{g,h})$ is subject to the condition \eqref{eq:mlambda-odd}.  However, different from the odd-$\nu$ case, $w(\gbf,\hbf)$ can now be any anyon in $\mathcal{C}_\nu=\{1,f,v,vf\}$. For convenience, we take $\beta_f(\bfg,\bfh)=1$ and make a comment on general $\beta_f(\bfg,\bfh)$ later. Then, for \eqref{eq:mlambda-odd} to be satisfied, $w(\gbf,\hbf)$ must take the following form
\begin{align}
w(\mathbf{g,h})=v^{\lambda(\mathbf{g,h})} f^{n_2(\mathbf{g,h})},
\label{eqn_sf_data}
\end{align}
where $n_2(\mathbf{g,h})=0, 1$ is any 2-cochain in $\mathcal{C}^2(G,\Z_2)$. This makes $\eta_f(\gbf,\hbf)=(-1)^{\lambda(\gbf,\hbf)}$. Note that \eqref{eqn_sf_data} is already a restricted form of $w(\gbf,\hbf)$, since $\lambda(\gbf,\hbf)$ is required to be a cocycle due to the original fermionic iTO. We claim that the quantity $n_2(\gbf,\hbf)$ is the same $n_2$ that we define in the fermionic language. Intuitively, $n_2(\gbf,\hbf)=0$ or $1$ represents whether $f$ appear in the symmetry fractionalization anyon $w(\gbf,\hbf)$.  More importantly, we show below that it satisfies the same condition $\dd n_2 =\calO_3$ as the one in Sec.~\ref{sec:def-n1n2}.


Recall that $v\times v = f$ and $f\times f =1$ for $\nu=4k+2$. More generally, the fusion rules are given by
\begin{align}
v^{a_1}f^{b_1}\times v^{a_2}f^{b_2}=v^{[a_1+a_2]} f^{[b_1+b_2]+a_1a_2},
\label{eq:fusion_4k+2}
\end{align}
where $a_i,b_i =0, 1$ and $[x] = x \modulo{2}$. With the form of $w(\gbf,\hbf)$ in  \eqref{eqn_sf_data}, the anyon permutation \eqref{eq:4k+2_permute} and the gauge choice $\tilde{\mathcal{O}}_3(\gbf,\hbf,\kbf)=1$, the obstruction condition \eqref{eqn_obstruction_sf} becomes: 
\begin{align}
1  
=&v^{\lambda(\mathbf{h,k})} f^{n_1(\mathbf{g})\lambda(\mathbf{h,k})}f^{n_2(\mathbf{h, k})}\cdot \bar{v}^{\lambda(\mathbf{gh,k})} f^{n_2(\mathbf{gh,k})}\nonumber \\
&\cdot v^{\lambda(\mathbf{g,hk})} f^{n_2(\mathbf{g,hk})}\cdot {\bar v}^{\lambda(\mathbf{g,h})} f^{n_2(\mathbf{g,h})} 
\nonumber\\
 =& v^{\left[\lambda(\mathbf{h,k})+ \lambda(\mathbf{g,hk})\right]} f^{\lambda(\mathbf{h,k}) \lambda(\mathbf{g,hk})} \cdot {\bar v}^{\left[\lambda(\mathbf{g,h})+ \lambda(\mathbf{gh,k})\right]}\nonumber \\
 &\cdot f^{\lambda(\mathbf{g,h}) \lambda(\mathbf{gh, k})} f^{n_1(\mathbf{g})\lambda(\mathbf{h},\mathbf{k})} f^{dn_2(\mathbf{g,h,k})} \nonumber\\
 =& f^{\text{d}n_2(\mathbf{g,h,k})}f^{\{n_1\cup\lambda+\lambda\cup_1\lambda\}(\mathbf{g,h,k})}.
\label{eqn_H3_obstr_nu_2}
\end{align}
From the second to the third line, we have used the relation $\left[\lambda(\mathbf{h,k})+ \lambda(\mathbf{g,hk})\right]=\left[\lambda(\mathbf{g,h})+ \lambda(\mathbf{gh,k})\right]$ as $\lambda(\gbf,\hbf)$ is a 2-cocycle, making the fusion product of the terms associated with $v$ to be 1. Let us define 
\begin{align}
\mathcal{O}_3(\mathbf{g,h,k})&=\{n_1\cup\lambda+\lambda\cup_1\lambda\}(\mathbf{g,h,k}).\label{eqn_o3obs}
\end{align}
which is always 3-cocycle in $\calH^3(G, \Z_2)$ according to general properties of the cup products $\cup$ and $\cup_1$  (see a review in Appendix \ref{app:cohomology}). Then,  Eq.~\eqref{eqn_H3_obstr_nu_2} imposes 
\begin{align}
    \mathcal{O}_3(\gbf,\hbf,\kbf) = \dd n_2(\gbf,\hbf,\kbf).
\label{eq:o3obs2}
\end{align}
This indicates a new level of obstruction: given $n_1$ and $\lambda$, if $\mathcal{O}_3(\mathbf{g,h,k})$ is a nontrivial 3-cocycle in $\mathcal{H}^3(G,\Z_2)$,  then there is no solution $n_2(\mathbf{g,h})$ to Eq.~\eqref{eq:o3obs2}. That is, there is no valid symmetry fractionalization in the form \eqref{eqn_sf_data}.  Furthermore, if Eq.~\eqref{eq:o3obs2} has no solution for all possible $n_1$, ESB occurs in fermionic iTOs with $\nu=4k+2$. This is the first half of Criterion \ref{criterion4}. Hence, we recover our result on $\calO_3$ in the fermionic theory in Sec.~\ref{sec:def-n1n2}. We note that $n_1\cup \lambda$ and $\lambda \cup n_1$ are equivalent cocycles in $\mathcal{H}^3(G, \Z_2)$ when $\lambda$ is a 2-cocycle.

A few remarks on gauge transformations are in order. First, $w(\gbf,\hbf)$ is defined up to an ambiguous anyon $\dd \zeta(\gbf,\hbf) = \rho_{\gbf}[\zeta(\hbf)]\zeta(\gbf)\overline{\zeta(\gbf\hbf)}$. In general,  $\zeta(\gbf)=v^{a(\gbf)}f^{b(\gbf)}$, with arbitrary $a(\gbf), b(\gbf)=0,1$. However, to make a one-to-one correspondence between bosonic SETs and fermionic iTOs,  we need to put some restrictions on $\zeta(\gbf)$. Recall that $\zeta(\gbf)$ originates from an extra phase ambiguity of the local operator $U^{(i)}_\gbf$  near anyon $a_i$  after \eqref{eq:eta_beta} is satisfied. When $a(\gbf)=1$, we will have $U_\gbf^{(i)} \rightarrow -U_\gbf^{(i)}$ if the anyon $a_i = f$. This means, it switches $\gbf_0$ and $\gbf_1$ in the fermionic language. We shall not allow such a transformation as $\gbf_0$ and $\gbf_1$ are distinct symmetries in $G_f$, as discussed in Sec.~\ref{sec:sym}. In other words, we do not allow coboundary transformation on $\lambda(\gbf,\hbf)$ and $\bfg\in G$ has been taken to be $\bfg_0\in G_f$. The remaining ambiguity, $\zeta(\gbf) = f^{b(\gbf)}$, brings a coboundary transformation $n_2(\gbf,\hbf)\rightarrow n_2(\gbf,\hbf)+\dd b(\gbf,\hbf)$. Accordingly, distinct solutions $n_2(\gbf, \hbf)$ to Eq.~\eqref{eq:o3obs2} are defined up to a coboundary $\dd b(\gbf, \hbf)$ in $\mathcal{B}^2(G, \Z_2)$. Then, given $n_2(\gbf,\hbf)$ being a solution, $n_2(\gbf,\hbf)+t(\gbf,\hbf)$ is a distinct solution if $t(\gbf,\hbf)$ is a non-trivial cocycle in $\calH^2(G, \Z_2)$. Second,  we have taken the gauge $\beta_f(\gbf,\hbf)=1$ and $\tilde{\mathcal{O}}_3=1$. Under other gauges, one may expect $\tilde{\mathcal{O}}_3$ to be any possible anyons in $\mathcal{A}=\{1,f,v,vf\}$. However, this contradicts with the fact that the right-hand side of \eqref{eqn_H3_obstr_nu_2} can only be 1 or $f$. Here, we show that $\tilde{\mathcal{O}}_3$ can only be $1$ or $f$. In general, we have $\beta_f(\gbf,\hbf) = \gamma_f(\gbf\hbf)/[\gamma_f(\gbf)\gamma_f(\hbf)]$ with $\gamma_f(\gbf) = \pm 1$. Let $\gamma_f(\gbf) = (-1)^{\sigma(\gbf)}$. Then, inserting this into the definition \eqref{eqn_obstution_H3_def} of $\Omega_a(\gbf,\hbf,\kbf)$ and taking $a=f$, we find
\begin{align}
    \Omega_f(\gbf,\hbf,\kbf) = (-1)^{\dd ^2\sigma(\gbf,\hbf,\kbf)} = 1.
\end{align}
Then, $\tilde{\mathcal{O}}_3(\gbf,\hbf,\kbf)$ must be $1$ or $f$ according to the definition \eqref{eqn_obstution_H3_def2}. It is not hard to check that gauge transformations on $\beta_v(\gbf,\hbf)$ result in a shift in $n_2(\gbf,\hbf)$, which is general any way. So, $\mathcal{O}_3$ in \eqref{eqn_o3obs} is the general form. Third, it is straightforward to show that a coboundary transformation on $\lambda$ does not change the cohomology class of $\mathcal{O}_3=n_1\cup\lambda+\lambda \cup_1 \lambda$. It is expected because coboundary transformations on $\lambda$ correspond to isomorphisms of $G_f$ and $\mathcal{O}_3$ obstructions should be the same for isomorphic $G_f$'s.

For certain choices of $n_1$, the obstruction $\calO_3$ may be trivial, allowing a valid $n_2$ and thereby a valid symmetry fractionalization. In these cases, we need to further consider the $\tilde{\mathcal{O}}_4$ obstruction. As discussed in Sec.~\ref{sec:set}, the general expression of $\tilde{\calO}_4$ obstruction in bosonic SETs with non-trivial anyon permutations is complicated.
In principle, for the current specific topological order $\mathcal{C}_\nu$ that is simple enough, one may try to solve the consistency equations of defects to obtain the general $\tilde{\calO}_4$.
Here, we avoid this calculation and instead derive $\tilde\calO_4$ for the case where $\lambda\cup_1\lambda$ is a trivial coboundary, using the stacking trick to be discussed in Sec.~\ref{sec:stack}. From now on, we will drop the ``$\tilde{\ } $'' on $\calO_4$ to imply that it is specific for the topological order $\mathcal{C}_\nu$.  To use the stacking trick, we will need two parts, $\calO_4^A$ and $\calO_4^B$. Here we give $\calO_4^A$.

Consider the case that there is no anyon permutation under $G$, i.e., $n_1(\gbf)=0$ for every $\gbf$. Assuming that $\mathcal{O}_3=\lambda \cup_1 \lambda$ is trivial, there exists $n_2(\gbf,\hbf)$ that satisfies $\calO_3=\lambda\cup_1\lambda=\dd n_2$. The symmetry fractionalization is then given by the anyon $w(\gbf,\hbf)$ in \eqref{eqn_sf_data}. Since there is no anyon permutation,   the formula  (\ref{eq:o4}) of $\tilde{\calO}_4$ is applicable. We still need the $F$ and $R$ symbols of $\mathcal{C}_\nu$. It is known that $F$ and $R$ symbols of general Abelian topological orders can be parametrized as follows\cite{WangPRX2016}:
\begin{align}
F_{x,y,z} &  =e^{ i\sum_{i}\Phi_{i}x^{i}\left(y^{i}+z^{i
}-\left[  y^{i}+z^{i}\right]  \right) }, \nonumber\\
R_{x,y} &  =e^{ i\sum_{i}\Phi_{i}x^{i}y^{i}+i\sum_{i<j}K
_{ij}x^{i}y^{j}}, \label{eq:fr_symbol}
\end{align}
where an integer vector $x=(x^1,x^2,...)$ with  $x^i=0,1,...,N_i-1$ denotes an Abelian anyon,  $\Phi_i$ is the self-statistical phase of the $i$th generating anyon, $K_{ij}$ is the mutual statistical phase between the $i$th and $j$th generating anyons, and $[x^i+y^i]=x^i+y^i \modulo{N_i}$.  The anyons form a group $\mathcal{A}=\prod_i \Z_{N_i}$ under fusion.  For $\nu=4k+2$,  we have $\mathcal{A}=\Z_4$ and $\Phi_1=\nu\pi/8$ and $K_{11}=\nu\pi/4$. 
Under this notation, the symmetry fractionalization anyon $w(\gbf,\hbf)$ in (\ref{eqn_sf_data}) becomes  $\omega(\gbf,\hbf)=\lambda(\gbf,\hbf)+2n_2(\gbf,\hbf)$. Inserting it into the formula \eqref{eq:o4} of $ \tilde{\mathcal{O}}_4 $, we obtain
\begin{align}
\mathcal{O}_{4}^A(\gbf,\hbf,\kbf,\lbf) =& e^{i\nu\frac{\pi}{8} [\lambda\cup \lambda +\hat{\text{d}} \lambda \cup_1\lambda](\gbf,\hbf,\kbf,\lbf)}(-1)^{n_2\cup n_2 (\gbf,\hbf,\kbf,\lbf)}\nonumber \\
&\times  e^{i\frac{\pi \nu }{4}[\lambda \cup n_2+n_2\cup \lambda+\hat{\text{d}}  n_2\cup_1 \lambda +\hat{\text{d}}  \lambda\cup_1 n_2] (\gbf,\hbf,\kbf,\lbf)}\nonumber\\
=&e^{i\nu\frac{\pi}{8} [\lambda\cup \lambda +\hat{\text{d}} \lambda \cup_1\lambda+4 n_2\cup n_2 +4 \lambda \cup n_2](\gbf,\hbf,\kbf,\lbf)}
\label{eqn_H4_nu2}
\end{align}
where we have  added a superscript ``$A$'' to imply that it holds only for $n_1=0$ \emph{and} trivial $\lambda\cup_1\lambda$.

A more general case with $n_1\neq0$ (but still with a trivial $\lambda\cup_1\lambda$) can be obtained using the stacking trick in Sec.~\ref{sec:stack}.
As explained there,  $\calO_4$ is given by adding the $\calO_4^A$ in Eq.~\eqref{eqn_H4_nu2}, and the $\calO_4^B[n_1,n_2]$ in Eq.~\eqref{eq:O4B}. For the most general case that even $\lambda\cup_1\lambda$ is nontrivial (but with a trivial $\calO_3=n_1\cup\lambda + \lambda\cup_1\lambda$), it requires a formula of $\calO_4^A$ that works for nonzero $n_1$, which unfortunately we do not have yet.

\subsection{Stacking trick}
\label{sec:stack}

\begin{table*}[t]
\begin{tabular}{|c|c|c|c|c|}
\hline
Physical quantities &  $n_1\in \mathcal{H}^1(G,\Z_2)$       &  $n_2\in \mathcal{C}^2(G,\Z_2)$                                              & $\mathcal{O}_3\in\mathcal{H}^3(G,\Z_2)$                                                                & $\mathcal{O}_4\in {H}^4(G,U(1))$                                     \\ \hline
$A: \nu$ (chiral)                   & $n_1^A=0$   & $n_2^A: dn_2^A=\mathcal{O}_3^A$                    & $\mathcal{O}_3^A=\frac{\nu}{2} \lambda \cup_1\lambda$                          & $\mathcal{O}_4^A[\nu]$                               \\ \hline
$B:\nu=0$ (nonchiral)                   & $n_1^B$     & $n_2^B: dn_2^B=\mathcal{O}_3^B$                    & $ \mathcal{O}_3^B= \lambda \cup n_1^B $ & $\mathcal{O}_4^B[n_1^B,n_2^B]$                               \\ \hline
$A \boxtimes B: \nu$ (chiral)     & $n_1=n_1^B$ & $n_2=F[n_2^A, n_2^B, n_1,\nu]: dn_2=\mathcal{O}_3$ & $\mathcal{O}_3 =\mathcal{O}_3^A+\mathcal{O}_3^B$                              & $\mathcal{O}_4=\mathcal{O}_4^A\times \mathcal{O}_4^B$ \\ \hline
\end{tabular}
\caption{Relations of quantities in the stacking trick. $F[n_2^A, n_2^B, n_1,\nu]$ is some functional that depends on $n_2^A,n_2^B,n_1$ and $\nu$  To use the stacking trick, we necessarily require $\mathcal{O}_3^A$ to be a trivial cocycle, so that we have a special solution of $n_2^A$, which together with $\lambda$ determines $\mathcal{O}_4^A$.  }
\label{table:stacking}
\end{table*}

We now describe the stacking trick. Given $\nu$, the obstruction $\calO_4$ for cases with arbitrary choices of $n_1$ and $n_2$ can be computed using the following stacking trick. First, we need the formula for $\calO_4$ for a special choice of $n_1=n_1^A$ and $n_2=n_2^A$.
This is denoted by $\calO_4^A$, or $\calO_4^A[\nu]$ to indicate its dependence on $\nu$. In this work, such special choice is always constructed with $n_1^A=0$ to avoid the complication of anyon permutation, and such $\calO_4^A$ is given in Eqs.~\eqref{eqn_H4_nu2}, \eqref{eqn_H4_nu4} and \eqref{eqn_H4_nu8}, for cases of $\nu=4k+2$, $8k+4$ and $16k+8$, respectively. In the case of $\nu=4k+2$, $n_2^A$ is one of the solutions to $\dd n_2=\calO_3 = \lambda\cup_1\lambda$. (That means, if $\lambda\cup_1\lambda$ is nontrivial, we will not have $n_2^A$ so that we cannot use the stacking trick.) The latter two cases will be discussed in Sec.~\ref{sec:derive_4k}.
Ungauging the fermion parity, this SET state corresponds to a particular anomalous 2D fermionic iTO state $A$, with chiral central charge $\nu/2$ and an anomaly indicated by $\calO_4^A$. It lives on the boundary of a 3D SPT bulk.\cite{ASPT}

Next, to obtain all possible fermionic iTO states with index $\nu$ and generic choices of $n_1$ and $n_2$, we stack the above chiral iTO state with a generic nonchiral fermionic SPT state $B$, which may also be anomalous with another $\calO_4^B$ obstruction.
The state $B$ is characterized by a 1-cocycle $n_1\in \mathcal{H}^1(G, \mathbb Z_2)$ and a 2-cochain $n_2\in C^2(G, \mathbb Z_2)$, satisfying $\dd n_2=\lambda\cup n_1$. 
Its anomaly is given by\cite{WangGu2020PRX}
\begin{align}
  \label{eq:O4B}
   & \calO_4^B[n_1, n_2]\nonumber \\
  =&(-1)^{\{\lambda\cup n_2 + n_2\cup n_2 + n_2\cup_1 \hat{\text{d}}n_2\} (\gbf,\hbf,\kbf,\lbf) + \lambda(\gbf,\hbf \kbf)\hat{\text{d}}n_2(\hbf,\kbf,\lbf)}\nonumber \\
  &(-1)^{ \hat{\text{d}}n_2(\gbf,\hbf,\kbf\lbf)\hat{\text{d}}n_2(\gbf\hbf,\kbf,\lbf)}(-i)^{\hat{\text{d}}n_2(\gbf,\hbf,\kbf)[1-\hat{\text{d}}n_2(\gbf,\hbf,\kbf\lbf)](\mathrm{mod}\ 2)}.
\end{align}
We will also denote $n_1=n_1^B$ and $n_2=n_2^B$, but both can vary. 

Stacking the two anomalous states together, their $\calO_4$ obstructions also add up. This is because the two states $A$ and $B$ can be realized on the surface of 3D bulk states that are bosonic SPT states characterized by cocycles $\calO_4^{A}$ and $\calO_4^B$, respectively\cite{chen14}.
When we stack $A$ and $B$, the stacked SPT state can then be realized on the surface of the bulk state obtained by stacking the two bulk states, which carries the bosonic SPT state $\calO_4^A\times \calO_4^B$.
Therefore, the resulting SPT state has obstruction $\calO_4^A\times \calO_4^B$.
The 2D fermionic iTOs with $\nu$ has ESB if and only if $\calO_4=\calO_4^A[\nu]\times \calO_4^B[n_1,n_2]$ is a nontrivial cocycle for all possible combinations of $n_1$ and $n_2$.

We notice that the 2-cochain $n_2$ describing the stacked anomalous cocycle may not be $n_2^A + n_2^B$, as stacking two SPT states with nontrivial $n_1^B$ produces a twist on the total $n_2$. However, such twists are irrelevant to our goal of finding all possible iTOs for a given $\nu$. For fixed $n_1^A$ and $n_2^A$, by varying $n_1^B$ and $n_2^B$, we shall be able to exhaust all possible valid symmetry-enriched iTOs. 

\subsection{{$\mathcal{O}_3$ and $\mathcal{O}_4$ for $\nu=4k$}}
\label{sec:derive_4k}

We now consider the $\mathcal{O}_3$ and $\mathcal{O}_4$ obstructions for $\nu=4k$ and derive ESB Criteria \ref{criterion5} and \ref{criterion6}. The discussion is in parallel to the above subsection, so we will be brief in places that are not very different. 

In this case, the topological order is again denoted as $\mathcal{C}_\nu =\{1, f, v,vf\}$ but with a different fusion rule $v\times v = 1$. To begin, we define the quantity $n_1:G\rightarrow \Z_2$, which determines the same permutation as in \eqref{eq:4k+2_permute}. Again, it has the physical meaning of whether $\gbf_0$ and $\gbf_1$ defects carry Majorana zero modes in the original fermionic iTOs. Similarly to the $\nu=4k+2$ case,  we show in Appendix \ref{app:localization_anomaly} that the symmetry localization anomaly  $\Tilde{\mathcal{O}}_3$ is trivial, regardless of the choice of $n_1$. Moreover, there exists a gauge that $\beta_f(\gbf,\hbf)=1$ and $\tilde{\mathcal{O}}_3=1$.  

Next, symmetry fractionalization $w(\gbf,\hbf)$ is again given in the form \eqref{eqn_sf_data} in the gauge $\beta_f(\bfg,\bfh)=1$, so that the conditions \eqref{eq:f-lambda} and \eqref{eq:mlambda-odd} are satisfied.  The general fusion rules for $\nu=4k$ are
\begin{align}
v^{a_1}f^{b_1}\times v^{a_2}f^{b_2}=v^{[a_1+a_2]} f^{[b_1+b_2]},
\label{eqn_4_fusion}
\end{align}
which is different from (\ref{eq:fusion_4k+2}).
With the form of $w(\gbf,\hbf)$ in  \eqref{eqn_sf_data}, the anyon permutation \eqref{eq:4k+2_permute} and the gauge $\tilde{\mathcal{O}}_3(\gbf,\hbf,\kbf)=1$, the obstruction condition \eqref{eqn_obstruction_sf} becomes: 
\begin{align}
1  =& v^{\lambda(\mathbf{h,k})} f^{n_1(\mathbf{g})\lambda(\mathbf{h,k})}f^{n_2(\mathbf{h, k})}\cdot {v}^{\lambda(\mathbf{gh,k})} f^{n_2(\mathbf{gh,k})}\nonumber \\
&\cdot v^{\lambda(\mathbf{g,hk})} f^{n_2(\mathbf{g,hk})}\cdot v^{\lambda(\mathbf{g,h})} f^{n_2(\mathbf{g,h})} 
\nonumber\\
 =& f^{\text{d}n_2(\mathbf{g,h,k})}f^{n_1\cup\lambda(\mathbf{g,h,k})}.
\label{eqn_H3_obstr_nu_4}
\end{align}
Accordingly, we define $\mathcal{O}_3$ for $\nu=4k$ as
\begin{align}
\mathcal{O}_3(\mathbf{g,h,k})&=\{n_1\cup\lambda\}(\mathbf{g,h,k}). \label{eqn_H3_4_obst}
\end{align}
General properties of the cup product guarantee $\mathcal{O}_3$ to be a cocycle in $\calH^3(G,\Z_2)$. Then,  Eq.~\eqref{eqn_H3_obstr_nu_4} becomes
\begin{align}
\mathcal{O}_3(\gbf,\hbf,\kbf)=\dd n_2(\gbf,\hbf, \kbf).
\label{eqn_SF_4_2}
\end{align}
Then, if $\mathcal{O}_3$ is a nontrivial cocycle, there is no solution $n_2(\gbf,\hbf)$ to Eq.~\eqref{eqn_SF_4_2}. Nevertheless, the obstruction $\calO_3$ alone cannot guarantee ESB to occur:  the case that $n_1=0$ makes $\calO_3=1$, which always allows a solution $n_2(\gbf,\hbf)$. So, we will need to move on to $\calO_4$ to look for ESB physics. Gauge transformations on $w(\gbf,\hbf)$ and $\tilde\calO_3$ are similar to the $\nu=4k+2$ case. That is, gauge transformations on $w(\gbf,\hbf)$ make the solutions $n_2(\gbf,\hbf)$  to Eq.~\eqref{eqn_SF_4_2} distinct up to a coboundary $\dd b(\gbf,\hbf)$ in $\mathcal{B}^2(G, \Z_2)$, and $\tilde\calO_3$ can only be $1$ or $f$ in general. Hence, the $\calO_3$ obstruction in Sec.~\ref{sec:def-n1n2} is recovered.

We now use the stacking trick in Sec.~\ref{sec:stack} to derive $\calO_4$. The obstruction $\calO_4^B$ is already given in \eqref{eq:O4B}, so we only need $\calO_4^A$ here. First, we notice that since $n_1=0$ always makes $\calO_3$ vanish, we can choose $n_1^A=0$ and $n_2^A=0$ for the special state $A$. This greatly simplifies our problem. Below we describe $\calO_4^A$ for different $\nu$'s.

For $\nu=8k+4$, explicit parametrizations of the $F$ and $R$ symbols are  given by (\ref{eq:fr_symbol}) where now $x=(x^1, x^2)$ labels $\Z_2\times \Z_2$ anyons, and similarly for $y, z$. The self-statistics $\Phi_i$ and the mutual statistics $K_{ij}$ of the generating anyons are
\begin{equation}
\Phi = \left( \frac{\nu\pi}{8}, \pi\right), \quad K = \left(
\begin{matrix}
\pi & \pi \\
\pi & 0 
\end{matrix}
\right),
\end{equation}
where we take $v$ and $f$ as the first and second generators of $\Z_2\times\Z_2$.  More explicitly, by inserting the values of $\Phi$ and $K$ and Eq.~\eqref{eq:fr_symbol} into the expression \eqref{eq:o4}, we have
\begin{align}
\mathcal{O}_4(\mathbf{g},\mathbf{h},\mathbf{k},\mathbf{l}) = & e^{i\frac{\pi\nu}{8}\lambda(\mathbf{g},\mathbf{h})\lambda(\mathbf{k},\mathbf{l})}\nonumber\\
& \times   e^{i\pi [n_2(\mathbf{g},\mathbf{h})n_2(\mathbf{k},\mathbf{l}) + n_2(\mathbf{g},\mathbf{h})\lambda(\mathbf{k},\mathbf{l})]} \nonumber\\
&\times e^{i\frac{\pi\nu}{8}\lambda(\mathbf{h},\mathbf{k})[\lambda(\mathbf{g},\mathbf{h}\mathbf{k})+\lambda(\mathbf{g}\mathbf{h}\mathbf{k},\mathbf{l})-\lambda(\mathbf{h}\mathbf{k},\mathbf{l})-\lambda(\mathbf{g},\mathbf{h}\mathbf{k}\mathbf{l})]} \nonumber\\
& \times e^{ i\frac{\pi\nu}{8}\lambda(\mathbf{g},\mathbf{h})[\lambda(\mathbf{k},\mathbf{l}) + \lambda(\mathbf{g}\mathbf{h},\mathbf{k}\mathbf{l})-\lambda(\mathbf{g}\mathbf{h},\mathbf{k})-\lambda(\mathbf{g}\mathbf{h}\mathbf{k},\mathbf{l})]} \nonumber\\
& \times e^{ i\frac{\pi\nu}{8}\lambda(\mathbf{k},\mathbf{l})[\lambda(\mathbf{h},\mathbf{k}\mathbf{l})+\lambda(\mathbf{g},\mathbf{h}\mathbf{k}\mathbf{l})-\lambda(\mathbf{g},\mathbf{h})-\lambda(\mathbf{g}\mathbf{h},\mathbf{k}\mathbf{l})]}\nonumber \\
=& e^{i\frac{\pi\nu}{8}[\lambda\cup\lambda+\hat{\dd}\lambda\cup_1\lambda+2n_2\cup( n_2+\lambda)](\mathbf{g},\mathbf{h},\mathbf{k},\mathbf{l})}.
\label{eqn_H4_nu4-0}
\end{align}
For our special choice of $n_2^A=0$, we further have 
\begin{align}
\mathcal{O}_4^A(\mathbf{g},\mathbf{h},\mathbf{k},\mathbf{l}) 
=& e^{i\frac{\pi \nu }{8}[\lambda\cup\lambda+\hat{\text{d}}\lambda\cup_1\lambda](\mathbf{g},\mathbf{h},\mathbf{k},\mathbf{l})}  
\label{eqn_H4_nu4}
\end{align}
where $a\cup b$ and $a \cup_1 b$ are the (higher) cup product of two cochains. 

For $\nu=16k+8$, the difference lies in the $F$ and $R$ symbols. They are given by the parameters:
\begin{equation}
\Phi = \left(\pi, \pi\right), \quad K = \left(
\begin{matrix}
0 & \pi \\
\pi & 0
\end{matrix}
\right)
\end{equation}
Plugging $\Phi$ and $K$ into (\ref{eq:o4}), we have
\begin{align}
\mathcal{O}_4(\mathbf{g},\mathbf{h},\mathbf{k},\mathbf{l}) =  (-1)^{[\lambda\cup\lambda+n_2\cup n_2+n_2\cup\lambda](\mathbf{g},\mathbf{h},\mathbf{k},\mathbf{l})}.
\label{eqn_H4_nu8-0}
\end{align}
Then, for our special choice of $n_1^A=n_2^A=0$, we have
\begin{align}
\mathcal{O}_4^A(\mathbf{g},\mathbf{h},\mathbf{k},\mathbf{l}) =  (-1)^{[\lambda\cup\lambda](\mathbf{g},\mathbf{h},\mathbf{k},\mathbf{l})} 
\label{eqn_H4_nu8}.
\end{align}

To summarize, we can write the expressions of $\calO_4$ with $n_1=0$ compactly as
\begin{align}
    \calO_4 = e^{i\frac{\nu\pi}{8}(\lambda\cup\lambda+\hat{\dd}\lambda\cup_1\lambda)+ i\pi n_2\cup (n_2+\lambda)}.
\end{align}
This expression applies to all $\nu$'s. One can check that it is the same as \eqref{eqn_H4_nu2}, \eqref{eqn_H4_nu4-0}, and \eqref{eqn_H4_nu8-0} for the cases discussed above. Finally, the result of $\calO_4^A[\nu]$ from Eqs.~\eqref{eqn_H4_nu4} and \eqref{eqn_H4_nu8} are combined with $\calO_4^B[n_1,n_2]$ in Eq.~\eqref{eq:O4B} to obtain the most general $\calO_4[n_1,n_2]=\calO_4^A[\nu]+\calO_4^B[n_1,n_2]$ that appears in Criterion \ref{criterion5} and Criterion \ref{criterion6}.  

\section{Discussions}
\label{sec: discussion}
To summarize, we have studied the phenomenon of enforced symmetry breaking (ESB) by 0D, 1D and 2D fermionic invertible topological orders (iTOs). We have obtained a set of criteria for asserting the existence or non-existence of ESB for finite groups, and illustrated the ESB physics with examples in all cases. We give both fermionic and bosonic descriptions (via gauging fermion parity) of symmetry-enriched fermionic iTOs.

It is very interesting to generalize the current study to ESB by general 2D fermionic topological orders, i.e., those with anyon excitations.  For example, our dimensional reduction argument in Sec.~\ref{sec:criterion_odd_nu} can be easily used to show that  $G_f$ from a nontrivial cocycle $\lambda(\gbf,\hbf)\in \calH^2(G,\Z_2)$ is incompatible with those fermionic topological orders that support both Majorana and non-Majorana vortices after gauging the fermion parity (e.g., the $SO(3)_3$ fermionic topological order). Also, it is interesting to study enforced breaking of anti-unitary symmetries, such as time reversal. In this work, we only study 2D fermionic iTOs which are chiral, so that enforced breaking of time-reversal is obvious. However, for non-chiral fermionic topological orders, it is a challenging problem. Moreover, ESB might also exist for 3D fermionic topological order, and it would be very interesting to explore such examples.

For ESB phenomenon to occur, it is required that, by definition, the symmetries should have some nontrivial action in the Hilbert space,  such as non-trivial extension by fermion parity or time reversal symmetry. In other words, there are conditions imposed at the very beginning. It has been clearly seen in our exploration of bosonic SETs that the obstruction functions $\calO_2$ and $\calO_3$ are due to the imposed conditions. We have named them \emph{conditional anomalies}. The conditions on the bosonic SETs in our study follow from the conditions in the original fermionic iTOs. However, generally speaking, conditions on properties  (such as symmetry fractionalization) of topological orders may be imposed by other reasons. For example, in the study of spin liquids, the spinon should always carry a half-integer spin in systems with an odd number of spin-$\frac12$s per unit cell~\cite{Meng2016LSM} , and this indeed causes ESB phenomenon for certain topological orders~\cite{Zaletel2015DS}.  Therefore, a general study on conditional anomalies or conditional obstructions are very important and helpful. Finally, we conjecture that the novel concept of ESB can also be defined for gapless systems and might have important implications in fundamental physics, such as CP violation problem in Standard Model. We will leave these potential directions for future study.

\emph{Note added.} Some preliminary results, regarding 0D and 1D ESB physics and 2D ESB physics from $\calO_2$ and $\calO_3$ obstructions, were previously reported in the Workshop on Strongly Correlated Systems held in January 2020 in Shenzhen, China, by one of the authors (C.W.). While we are preparing for the manuscript, we become aware of the work Ref.~\cite{Barkeshli2021arxiv} which studies 2D symmetry-enriched fermionic iTOs and derives the general classification, including the most general form of the $\calO_4$ obstruction function. However, Ref.~\cite{Barkeshli2021arxiv} does not study the phenomenon of enforced symmetry breaking. We also become aware of the works Refs.~\cite{fset2021a,fset2021b,fset2021c} which study the general theory of fermionic SET phases.

\begin{acknowledgements}
We are grateful to Qing-Rui Wang for enlightening discussions. SQN and CW were supported by Research Grant Council of Hong Kong (GRF 17300220). YQ was supported by National Natural Science Foundation of China (NSFC 11874115). ZCG was supported by Research Grant Council of Hong Kong(GRF 14306420,
ANR/RGC Joint Research Scheme no. A-CUHK402/18).
\end{acknowledgements}

\bibliographystyle{apsrev4-2}
\bibliography{esb}

\appendix

\section{Group cohomology}
\label{app:cohomology}

In this appendix, we review some basic knowledge of group cohomology, including cup products and Pontryagin square. 

\subsection{Definition}

Consider a finite group $G$ and  a $G$-module $M$. A $G$-module $M$ is an Abelian group equipped with a $G$ action $a \rightarrow g\cdot a$, where $a, g\cdot a \in M$ and $g\in G$. The action is compatible with the multiplication in $M$,
\begin{align}
g\cdot (a+b)=g\cdot a+g\cdot b, 
\end{align}
where we have used additive notation for the multiplication of Abelian group. Note that $g\cdot 0 = 0$, where $0$ is the group identity of $M$. Simple examples of $M$ include $U(1)=\mathbb{R}/\Z=\{ \alpha|\alpha\in[0,2\pi)\}$ and $\Z_n=\{ 0,1,2,\dots ,n-1\}$ with a trivial action $g\cdot a = a$. In the presence of anti-unitary symmetries in $G$ and for $M=U(1)$, the group action is given by $g\cdot \alpha ={s(g)}\alpha \modulo{2\pi}$, where $s(g)=1$ for unitary $g$ and $s(g)=-1$ for anti-unitary $g$.

A function $w_n:G^n\rightarrow M$, i.e., $w_n(g_1,g_2,\dots, g_n)$ with $g_i\in G$, is called an $n$-cochain. All $n$-cochains form an Abelian group $\mathcal{C}^n(G,M)$ under the function multiplication
$w_n(g_1,g_2,...,g_n)=w_n'(g_1,g_2,...,g_n)+w_n''(g_1,g_2,...,g_n)$, which inherits from the multiplication of $M$. The differential map $\dd: \mathcal{C}^n(G,M)\rightarrow \mathcal{C}^{n+1}(G,M)$  is defined as follows, 
\begin{align}
&\text{d}w_n(g_1,g_2,...,g_{n+1})\nonumber \\
=&g_1\cdot w_n(g_2,g_3,..., g_{n+1})+(-1)^{n+1}
 w_n(g_1,g_2,...,g_n) \nonumber \\
 +&\sum_{i=1}^n(-1)^i w_n(g_1,...,g_{i-1},g_ig_{i+1},g_{i+2},...,g_{n+1}).
\end{align}
One can check that the differential operator satisfies the nice property 
$\text{d}^2=1$. 

With the differential map, one can define two special types of cochains: an $n$-coboundary is defined as $w_n$ that satisfies $w_n = \dd w_{n-1}$ for some $w_{n-1}$, and an $n$-cocycle is defined as $w_n$ that satisfies $\dd w_n=0$. The $n$-coboundaries form the group  $\mathcal{B}^n(G,M)$ and the $n$-cocycles form the group $\mathcal{Z}^n(G,M)$. The property $\text{d}^2=1$ implies that any coboundary is also a cocycle. Therefore, $\mathcal{B}^n(G,M)\subset \mathcal{Z}^n(G,M)\subset \mathcal{C}^n(G,M)$. Two cocycles $w_n$ and $w_{n}'$ are said to be equivalent, or belonging to the same cohomology class, if $w_n = w_n' + \dd w_{n-1}$. Then, inequivalent cocycles are classified by the quotient group
\begin{align}
\mathcal{H}^n(G,M)=\frac{\mathcal{Z}^n(G,M)}{\mathcal{B}^n(G,M)}
\end{align}
which is called the $n$-th cohomology group of $G$ over $M$.

\subsection{Cup product}

It is useful to define maps between different cochain groups $\mathcal{C}^n(G,M)$, beyond the differential map $\dd$. In the main text, we have used the cup products and Pontryagin square to simplify some expressions. Here, we give a brief review on the two maps and refer to Ref.~\cite{Steenrod1947} for more general discussions.

The cup product is a map
\begin{align}
\cup: \mathcal{C}^n(G,M_1)\times \mathcal{C}^m(G,M_2)\rightarrow \mathcal{C}^{n+m}(G,M_3).
\end{align}
To define it, we need a bilinear map $B: M_1\times M_2 \rightarrow M_3$ such that $B(a+a', b)=B(a,b)+B(a',b)$ and $B(a,b+b')=B(a,b)+B(a,b')$. We will give a specific example of $B$ below. Given $B$ and two cochains $w_n\in\mathcal{C}^n(G,M_1)$ and $w_m\in \mathcal{C}^m(G,M_2)$, the cup product is defined as 
\begin{align}
&w_n\cup w_m(g_1,...,g_{n+m})\nonumber \\
=&B[w_n(g_1,...,g_n), w_{m}(g_{n+1},...,g_{n+m})].
\end{align}
Under the differential map, it satisfies 
\begin{align}
\text{d}(w_n\cup w_m)=\text{d}w_n\cup w_m+(-1)^n w_n\cup\text{d}w_m.
\label{eq:differential_cup}
\end{align}
This is a very nice property, which implies that if $w_n$ and $w_m$ are both cocycles, so is $w_n\cup w_m$. Moreover, one can show that if either $w_n$ or $w_m$ is a coboundary and the other is a cocycle, then $w_n\cup w_m$ is a coboundary.  Therefore, the cup product can actually be understood as a cohomological map $\cup: \mathcal{H}^n(G,M_1)\times \mathcal{H}^m(G,M_2)\rightarrow \mathcal{H}^{n+m}(G,M_3)$.

One can generalize it to higher cup products. For this work, only the cup-1 product will be used.  It is a map of degree $-1$, namely
\begin{align}
\cup_1:\mathcal{C}^n(G,M_1)\times \mathcal{C}^m(G,M_2)\rightarrow \mathcal{C}^{n+m-1}(G,M_3)
\end{align}
Given a bilinear map $B$, the product $\cup_1$ is defined by
\begin{align}
 w_n\cup_1w_m & (g_1,...,g_{n+m-1})
=\sum_{i=0}^{n-1}(-1)^{(n-i)(m+1)} \nonumber \\
&B[w_n(g_1,...,g_i,\prod_{j=1}^m g_{i+j},
g_{m+i+1}...,g_{n+m-1}), \nonumber\\ 
& \quad w_m(g_{i+1},...,g_{i+m})].
\end{align}
Under the differential map, it satisfies
\begin{align}
&w_n\cup w_m-(-1)^{nm} w_m\cup w_n
=(-1)^{n+m}\text{d}w_n\cup_1 w_m\nonumber \\&\quad+(-1)^mw_n\cup_1\text{d}w_m-(-1)^{n+m}\text{d}(w_n\cup_1w_m)
\label{eq:cup_product_commute}
\end{align}
Different from the cup product, we see that even if $w_n$ and $w_m$ are cocycles,  $w_n\cup_1 w_m$ might not be a cocycle. Nevertheless, for equivalence classes of cocycles, this relation implies the super-commutative relation  $[w_n]\cup [w_m]=(-1)^{nm} [w_m]\cup [w_n]$.

In this work, we mostly consider the case $M_1=M_2=M_3=\Z_{N}$ with a trivial $G$ action. Let $\Z_N=\{0,1,\dots, N-1\}$ and the group multiplication is the usual addition modulo $N$. We take the bilinear map $B$ to be
\begin{align}
B(a,b) = a b \modulo{N},
\end{align}
where $ab$ is the usual  multiplication. Then, $w_n\cup w_m = w_n w_m$ and $w_n\cup_1 w_m$ becomes
\begin{align}
 &w_n\cup_1  w_m  (g_1,...,g_{n+m-1})
\nonumber \\
&=\sum_{i=0}^{n-1}(-1)^{(n-i)(m+1)}w_m(g_{i+1},...,g_{i+m}) \nonumber \\
&\quad \,\, w_n(g_1,...,g_i,\prod_{j=1}^m g_{i+j},
g_{m+i+1},...,g_{n+m-1}).
\end{align}
For 2-cochains $w_2$ and $v_2$, we have
\begin{align}
    w_2\cup_1 v_2(g_1,g_2,g_3) = &  v_2(g_1,g_2) w_2(g_1g_2,g_3) \nonumber\\
    & - v_2(g_2,g_3)w_2(g_1,g_2g_3).
\end{align}
where ``modulo $N$'' is assumed on the right-hand side.

\subsection{Pontryagin square}

Another cohomological operation used in the main text is the Pontryagin square, denoted as $\mathcal{P}$. It is a map
\begin{align}
\mathcal{P}:\mathcal{H}^{2m}(G,\Z_{2N})\rightarrow \mathcal{H}^{4m}(G,\Z_{4N}).
\end{align}
To define $\mathcal{P}$, consider a cocycle $w_{2m}\in \mathcal{Z}^{2m}(G,\Z_{2N})$. It can be viewed as a cochain in $\mathcal{C}^{2m}(G,\Z_{4N})$ by embedding $\Z_{2N}=\{0,1,\dots,2N-1\}$ into $\Z_{4N}=\{0,1,\dots,4N-1\}$. For clarity, let us denote $w_{2m}$ as $\hat w_{2m}$ when it is lifted to a cochain in  $\mathcal{C}^{2m}(G,\Z_{4N})$. We have $\hat w_{2m}=w_{2m}$ (mod $2N$) and $\text{d}\hat w_{2m}= 2N c_{2m+1}$, where $c_{2m+1}\in\mathcal{C}^{2m+1}(G,\Z_{4N})$. The Pontryagin square of $w_{2m}$ is defined as
\begin{align}
\mathcal{P}(w_{2m})=\hat w_{2m}\cup \hat w_{2m}+ \hat w_{2m}\cup_1 \text{d} \hat w_{2m}.
\end{align}
According to  Eqs.\eqref{eq:differential_cup} and \eqref{eq:cup_product_commute}, we have 
\begin{align}
\text{d}\mathcal{P}(w_{2m})& =2 \hat w_{2m} \cup \dd \hat w_{2m}+\dd  \hat w_{2m}\cup_1 \dd \hat w_{2m} \nonumber\\
& =4N \hat w_{2m} c_{2m+1} + 4N^2 c_{2m+1}\cup_1c_{2m+1}\nonumber\\
& = 0 \modulo{4N}.
\end{align}
Accordingly, $\mathcal{P}(w_{2m})$ is a cocycle in $\mathcal{Z}^{4m}(G, \Z_{4N})$. In addition, it can be checked that the cohomology class of $\mathcal{P}(w_{2m})$ does not change under the coboundary transformation $w_{2m}\rightarrow w_{2m}+\text{d}c_{2m-1}$, or equivalently $\hat w_{2m}\rightarrow \hat w_{2m}+\widehat{\dd  c}_{2m-1} + 2N c_{2m}$, where $\widehat{\dd  c}_{2m-1}\in \mathcal{B}^{2m}(G,\Z_{4N})$ is the lift of ${\dd  c}_{2m-1} \in \mathcal{B}^{2m}(G,\Z_{2N})$. Therefore, $\mathcal{P}$ is a well-defined cohomological map from $\mathcal{H}^{2m}(G,\Z_{2N})$ to $\mathcal{H}^{4m}(G,\Z_{4N})$.

\section{Absence of symmetry localization anomaly}
\label{app:localization_anomaly}
In this appendix, we show that the $\tilde{\calO}_3$ obstruction (symmetry localization anomaly) is always trivial for those bosonic SETs obtained by gauging fermion parity of the symmetry-enriched fermionic iTOs with even $\nu$ (i.e., integer chiral central charge $c_-=\nu/2$). As discussed in the main text, the topological order $\mathcal{C}_\nu$ from gauging fermion parity is Abelian when $\nu$ is even. We will discuss the cases $\nu = 4k+2$ and $\nu=4k$ separately. In both cases, $\tilde{\calO}_3$ obstruction is trivial and actually we obtain $\tilde{\calO}_3=1$ under our gauge choice.


\subsection{$\nu=4k+2$}


For $\nu=4k+2$, the topological order $\mathcal{C}_\nu$ contains four Abelian anyons: $1, v, f$ and  $vf$. They form a $\Z_4$ group under fusion, with fusion rules $v\times v = f$ and $v\times f = vf$. In this appendix, we will use alternative labels $0,1,2,3$ to denote $1, v, f, vf$ respectively. Then,  the fusion rules are given by $x\times y =[x+y]$, where $x,y=0,1,2,3$ and $[...]$ takes modulo 4.  From the expressions \eqref{eq:fr_symbol}, we have the $F$ and $R$ symbols given by
\begin{align}
F_{x,y,z}&=e^{i\frac{\pi \nu}{8} x (y+z-[y+z])}\label{eqn_F_nu_2}, \\
R_{x,y}&=e^{i\frac{\pi \nu}{8} xy}.
\end{align}

Now we consider the topological symmetries of $\mathcal{C}_\nu$, which form the group $\Aut(\mathcal{C}_\nu)$. Recall from Sec.~\ref{sec:set} that a topological symmetry contains two pieces of data: (1) a permutation of anyons $x\rightarrow x'=\varphi(x)$ and (2) an action in the fusion space $\varphi(|x,y;z
\rangle) = u^{x'y'}_{z'}|x',y';z'\rangle$. For the current $\mathcal{C}_\nu$, there is one and only one nontrivial permutation: $\phi(1)=1$, $\phi(v)=vf$, $\phi(f)=f$ and $\phi(vf)=v$. Equivalently,
\begin{align}
    \phi(x) = 4\Theta(x)-x =\bar{x},
    \label{eq:phi}
\end{align}
where $\Theta(x)=0$ if $x=0$, $\Theta(x)=1$ if $x=1,2,3$, and $\bar{x}$ is the antiparticle of anyon $x$. Together with the trivial topological symmetry, we have $\Aut(\mathcal{C}_\nu) = \Z_2 = \{1, \phi\} $.


We still need to specify the action of the nontrivial permutation $\phi$ in fusion spaces, i.e., to specify the phase factor $u^{xy}_z$. Since $z$ is uniquely determined by $x$ and $y$ in Abelian topological orders, let us denote the action as 
\begin{align}
\phi(|x,y;z\rangle) = u(x',y')|x',y';z'\rangle
\label{eqn_baem_non}
\end{align}
where $x'=\phi(x)$. The condition that $F$ symbol is invariant under $\phi$ action (see Eq.~\eqref{eq:autoequi} in the main text) leads to
\begin{align}
\delta u(x,y,z)\equiv \frac{u( x,  y)u([x+y], z)}{u(y, z)u( x, [y+z])}=\frac{F_{\bar x,\bar y,\bar z}}{F_{ x, y,  z}},
\label{eqn_U_commute_F_2}
\end{align}
where we have used $\phi(x)= \bar{x}$. Plugging the $F$ symbol above into the last piece, we have
\begin{align}
\frac{F_{ \bar x, \bar y, \bar z}}{F_{ x,  y,  z}}=e^{i\frac{\pi \nu}{8} [ \bar x(\bar y+\bar z-[\bar y+\bar z])-x(y+z-[y+z])]}.
\label{eqn_exp_U}
\end{align}
By inserting \eqref{eq:phi}, we can simplify this expressions to
\begin{align}
\frac{F_{ \bar x, \bar y, \bar z}}{F_{ x,  y,  z}}=e^{i\frac{\pi\nu}{2} x\{ \Theta(y)+\Theta(z)-\Theta([y+z])\}}.
\label{eqn_exp_U1}
\end{align}
Introducing the function $f(x,y)=e^{i\frac{\pi\nu}{2} x \Theta(y)}$, we have
\begin{align}
\delta f(x,y,z)&\equiv \frac{f(x,y)f([x+y],z)}{f(y,z)f(x,[y+z])}\nonumber \\
&=e^{i\frac{\pi\nu}{2} \{x\Theta(y)+[x+y]\Theta(z)-y\Theta(z)-x\Theta([y+z])\}}\nonumber\\
&=e^{i\frac{\pi\nu}{2} x \{\Theta(y)+\Theta(z)-\Theta([y+z])\}}
\label{eqn_u_coboundary}
\end{align}
where in the second equality, we have used the fact that $[x+y]-y=x\modulo{2}$. Comparing (\ref{eqn_U_commute_F_2}) and (\ref{eqn_u_coboundary}), we find that one choice of $u(x,y)$ is
\begin{align}
u(x,y)=f(x,y)=e^{i\frac{\pi\nu}{2} x \Theta(y)}.
\label{eq:o3-u}
\end{align}
A general solution $u(x,y)$ that satisfies \eqref{eqn_U_commute_F_2} can be expressed as $u(x,y)=f(x,y)w(x,y)$, where $w(x,y)$ satisfies $\delta w=1$, i.e., it is a 2-cocycle in $\mathcal{Z}^2(\Z_4, U(1))$. However, $\calH^2(\Z_4, U(1))$ is trivial, so $w(x,y)$ is always a coboundary. That is, $w(x,y)=\delta v(x,y)=\frac{v(x)v(y)}{v([x+y])}$. Therefore, a phase factor $u(x,y)$ always differs from $f(x,y)$ by a natural isomorphism. Accordingly, we will set $w(x,y)=1$ and take $u(x,y)=f(x,y)$ below.

Next, we consider a group homomorphism from $G$ to $\Aut(\mathcal{C}_\nu)$, namely
\begin{align}
\rho:G\rightarrow \text{Aut}(\mathcal{C_\nu})=\Z_2.
\end{align}
If $\rho_\gbf$ is nontrivial in $\text{Aut}(\mathcal{C}_\nu)$, the symmetry $\gbf\in G$ acts as $\phi$ in (\ref{eqn_baem_non}), with $u(x,y)$ fixed in \eqref{eq:o3-u}; otherwise it acts as a natural isomorphism, which we fix to be $u^{xy}_z=1$ for any $x,y,z$. More specifically, for a state $|x,y;z\rangle$, we have
\begin{align}
\rho_\bfg(|x,y;z\rangle)=U_\bfg({}^\bfg x, {}^\bfg y) |{}^\bfg x, {}^\bfg y; {}^\bfg z\rangle,
\label{eqn_state_transf}
\end{align}
where ${}^\bfg x:=\rho_\bfg(x)$ denotes the anyon permutation and $U_\bfg( x,  y)$ can be unified as
\begin{align}
U_\bfg( x,  y)=e^{i\frac{\pi \nu}{2} n_1(\bfg) x \Theta(y)},
\end{align}
where $n_1(\bfg)=0$ or 1, for $\rho_\bfg$ being trivial or nontrivial in $\Aut(\mathcal{C}_\nu)$ respectively. Since $\rho$ is a group homomorphism, we have
\begin{align}
\dd n_1(\bfg,\bfh)=n_1(\bfg)+n_1(\bfh)-n_1(\bfg\bfh)=0, \modulo{2}.
\label{eqn_rho_cocy}
\end{align}
Recall from Sec.~\ref{sec:set} that $\rho_{\bfg\bfh} = \kappa_{\bfg,\bfh}\circ\rho_{\bfg}\circ\rho_{\bfh}$. Applying this relation to the current case, we obtain
\begin{align}
\kappa_{\bfg,\bfh}(x,y) & =\frac{U_{\bfg\bfh}(x,y)}{U_{\bfg}(x,y)U_{\bfh}(^{\bbfg}x,{}^{\bbfg}y)}\nonumber\\
& =e^{i\frac{\pi \nu}{2} \dd n_1(\bfg,\bfh) x\Theta(y)} \nonumber\\
& =1, 
\end{align}
where we used the property that $x\Theta(y) = {}^{\bbfg} x \Theta({}^{\bbfg} y) \modulo{2}$ holds for any $\bfg$. Accordingly, the phase factor $\beta_x(\bfg,\bfh)$ defined in \eqref{eqn_decomp} can be set to 1, making the $\tilde{\calO}_3$ obstruction defined in \eqref{eqn_obstution_H3_def} and \eqref{eqn_obstution_H3_def2} to be 1. This concludes the $\nu=4k+2$ case. 

\subsection{$\nu=4k$}
For $\nu=4k$, the four anyons in $\mathcal{C}_\nu$ form a $\Z_2\times \Z_2$ fusion group. The fusion rules are $v\times v =1$, $f\times f=1$, and $ v\times f =vf$. 
Alternatively, we label the anyons by a two-component vector $x=(x_1,x_2)$ with $x_1,x_2=0, 1$. For convenience and different to the main text,  we take $1=(0,0)$, $v=(1,0)$, $vf=(0,1)$ and $f=(1,1)$. As such, the fusion rules are given by $x\times y=([x_1+y_1], [x_2+y_2])$ where $[...]$ takes modulo 2 in this subsection.  From the expressions \eqref{eq:fr_symbol}, we have the $F$ and $R$ symbols given by
\begin{align}
F_{x,y,z}&=e^{i\frac{\pi \nu }{8} \sum_{\mu=1,2} x_\mu(y_\mu+z_\mu-[y_\mu+z_\mu])}\label{eqn_F_nu_2}\\
R_{x,y}&=e^{i\frac{\pi \nu}{8} \sum_{\mu=1,2} x_\mu y_\mu+i\pi K x_1 y_2},
\end{align}
where $K=1+\nu/4 \modulo{2}$. 

Let us consider topological symmetries in $\mathcal{C}_\nu$.  There is a nontrivial topological symmetry, again denoted as $\phi$, that gives the permutation $\phi(1)=1$, $\phi(v)=vf$, $\phi(vf)=v$, and $\phi(f)=f$. Equivalently, the permutation is given by
\begin{align}
    \phi((x_1,x_2)) = (x_2,x_1).
\end{align}
In the case that $\nu=8$, there exist other topological symmetries that permute $f$ with the fermion-parity vortices. However, we are only concerning those SETs of $\mathcal{C}_\nu$ that keep $f$ unpermuted. Accordingly, for our purpose, we only need to consider the autoequivalences $\{1,\phi\}$, denoted as ${\Aut}(\mathcal{C}_\nu) = \Z_2$. 

Similar to the above subsection, the action of $\phi$ in the fusion space is given by (\ref{eqn_baem_non}), and the phase factor $u(x,y)$ satisfies
\begin{align}
\delta u (x,y,z)=\frac{F_{x', y', z'}}{F_{ x,  y,  z}}, 
\label{eqn_u_equation2}
\end{align}
where $x'=\phi(x)=(x_2,x_1)$.  Plugging the above expression of the $F$ symbol, we have
\begin{align}
\frac{F_{ x',  y', z'}}{F_{ x,  y,  z}}=e^{i \frac{\pi \nu}{4} \sum_{\mu=1,2}  x_\mu(y_\mu+z_\mu-[y_\mu+z_\mu])}=1,
\end{align}
when $\nu=4k$. Therefore, $\delta u(x,y,z)=1$, making $u(x,y)$ a 2-cocycle in $\mathcal{Z}^2(\Z_2\times \Z_2,U(1))$. At the same time, the invariance of $R$ symbol imposes the following condition (see Eq.~\eqref{eq:autoequi} in the main text):
\begin{align}
    \frac{u(y,x)}{u(x,y)} = \frac{R_{x',y'}}{R_{x,y}} = e^{i\pi K(x_1y_2-x_2y_1)}. 
\end{align}
Any 2-cocycle $u(x,y)\in \mathcal{Z}^2(\Z_2\times \Z_2,U(1))$ that satisfies this condition can be written as 
\begin{align}
u(x,y)=e^{i\pi K x_1y_2} \delta v(x,y),
\end{align}
where $\delta v(x,y)=\frac{v(x)v(y)}{v(xy)}$ is a coboundary. Multiplying $\delta v(x,y)$ only gives a natural isomorphism on $u(x,y)$, so we will set $\delta v(x,y)=1$ below.

Next, we introduce a group homomorphism 
\begin{align}
\rho:G\rightarrow {\Aut}(\mathcal{C}_\nu)=\Z_2.
\end{align}
Similar to (\ref{eqn_state_transf}), let $U_\bfg({}^{\bfg}x,{}^\bfg y)$ be the phase factor for the action of $\rho_\bfg$ on the sate $|x,y;z\rangle$. Using $u(x,y)$ obtained above, we have the following expression
$U_\bfg( x,  y)$ can be unified as
\begin{align}
U_\bfg( x,  y)=e^{i\pi K n_1(\bfg) x_1y_2},
\end{align}
where $n_1(\bfg)=0$ or 1, for $\rho_\bfg$ being trivial or nontrivial in ${\Aut}(\mathcal{C}_\nu)$ respectively. Using the condition $\dd n_1(\bfg,\bfh)=0$, we have
\begin{align}
\kappa_{\bfg,\bfh}(x,y) & =\frac{U_{\bfg\bfh}(x,y)}{U_{\bfg}(x,y)U_{\bfh}(^{\bbfg}x,{}^{\bbfg}y)} \nonumber\\
& =e^{i\pi K \dd n_1(\bfg,\bfh) x_1y_2 + i\pi K n_1(\bfh) n_1(\bfg)(x_1y_2+x_2y_1)} \nonumber\\
& =e^{i\pi K n_1(\bfh) n_1(\bfg)(x_1y_2+x_2y_1)},
\end{align}
where we have used ${}^{\bbfg}x_1 {}^{\bbfg} y_2 = x_1y_2 + n_1(\bfg)(x_1y_2+x_2y_1) \modulo{2}$. Then, $\beta_x(\bfg,\bfh)$ can be chosen as
\begin{align}
    \beta_x(\bfg,\bfh)= e^{i\pi K n_1(\bfg)n_1(\bfh)(x_1+1)x_2},
    \label{eq:betax}
\end{align}
such that $\kappa_{\bfg,\bfh}(x,y) = \beta_x(\bfg,\bfh)\beta_y(\bfg,\bfh)/\beta_{xy}(\bfg,\bfh)$. Inserting $\beta_{x}(\bfg,\bfh)$ into \eqref{eqn_obstution_H3_def}, we have
\begin{align}
    \Omega_x(\bfg,\bfh,\bfk) &  = \frac{\beta_{{}^{\bar{\mathbf{g}}}x}(\mathbf{h,k})\beta_{x}(\mathbf{g,hk})}{\beta_{x}(\mathbf{g,h})\beta_{x}(\mathbf{gh,k})}\nonumber\\
    & = \frac{e^{i\pi K (x_1+1)x_2 [n_1(\bfh)n_1(\bfk)+n_1(\bfg)n_1(\bfh\bfk)]}}{e^{i\pi K (x_1+1)x_2 [n_1(\bfg)n_1(\bfh)+n_1(\bfg\bfh)n_1(\bfk)]}}\nonumber\\
    & =1,
\end{align}
where we have used $\dd n_1(\bfg,\bfh)=0 \modulo{2}$ to obtain the last line. Therefore, according to \eqref{eqn_obstution_H3_def2}, we can set $\tilde{\calO}_3(\gbf,\hbf,\kbf)=1$. We comment that Eq.~\eqref{eq:betax} is set such that $\beta_f(\bfg,\bfh)=1$ for any $\bfg$ and $\bfh$.


\section{Evaluating obstructions for Abelian $G$}
\label{sec_calculation_of_H3_nu2_abelian}

Here we present some calculations related to the $\calO_3$ and $\calO_4$ obstructions for Abelian group $G=\prod_i \Z_{N_i}$. Without loss of generality, we take $N_i = 2^{k_i}$, $i=1,2,\dots, K$.  We also collect the topological invariants of $\calH^2(G,\Z_2), \calH^3(G,\Z_2)$ for Abelian group $G$, which are very useful for the calculations in different parts of the paper.

\subsection{Topological invariants}
\label{app:invariant}

Let us discuss some general aspects of cohomology groups $\calH^n(G,\Z_2)$ ($n=1,2,3$) and $\calH^4(G,U(1))$. In particular, we discuss the so-called topological invariants, which are certain combinations of cocycles that are invariant under coboundary transformations. Many of them are discussed in the main text and we summarize them here.  For a cochain $\omega \in \mathcal{C}^n(G,\Z_2)$, we find it useful to define the  differential operator $\hat{\dd}$, which has the same expression as $\dd$ but without taking modulo $2$. Accordingly,
\begin{align}
\dd\omega= \hat{\dd}\omega \modulo{2}.
\end{align}
We will use $a=(a_1,a_2,\dots,a_K)$ to denote the group elements of $G$, with $a_i=0,1,\dots, N_i-1$. We will also use $[a_i]$ to denote ``$a_i \modulo{N_i}$'' for short.

First, the cohomology group $\calH^1(G,\Z_2) = \prod_{i} \Z_2$. There are $K$ root cocycles 
\begin{align}
v_i(a) = [a_i] \modulo{2},
\label{eqn:root_1_cocycle}
\end{align}
where $i=1,2,\dots, K$. A general cocycle can be constructed from these root cocycles 
\begin{equation}
v(a) = \sum_i q_i v_i(a),
\label{eqn:1_cocycle_paramet}
\end{equation}
where an overall ``modulo 2'' is assumed, and $q_i=0,1$ are the parameters for the general 1-cocycle. Equation \eqref{eqn:1_cocycle_paramet} gives a complete parametrization of 1-cocycles in $\calH^2(G,\Z_2)$.

Next, the second cohomology group $\calH^2(G,\Z_2) = \prod_i \Z_2 \prod_{i<j} \Z_2$. A general 2-cocycle can be generated by root cocycles, which are
\begin{align}
w_i(a,b) & = \frac{1}{N_i} ([a_i] + [b_i] - [a_i + b_i]) = \frac{\hat{\dd} v_i(a,b)}{N_i},\nonumber \\
w_{ij}(a,b) & = [a_i][b_j] = v_i \cup v_j(a,b),
\label{eqn:root_2_cocycle}
\end{align}
where again an overall ``modulo 2'' is assumed. Note that $w_{ij}$ and $w_{ji}$ are equivalent cocycles. A general cocycle is given by
\begin{equation}
w(a,b) = \sum_i p_i w_i(a,b) + \sum_{ij} p_{ij} w_{ij}(a,b)
\label{2cocycle}
\end{equation}
where $p_i, p_{ij} = 0,1$. One can define the following complete set of topological invariants:
\begin{align}
\Omega_i  & = \sum_{n=0}^{N_i-1} w(e_i, n e_i), \nonumber\\
\Omega_{ij}  & = w(e_i, e_j) -w(e_j, e_i),
\end{align}
where $e_i = (0,\dots, 0,1,0,\dots,0)$ with only the $i$th entry being $1$ and others being $0$. One can straightforwardly check that $\Omega_i$ and $\Omega_{ij}$ are invariant under coboundary transformations. We remark that $\Omega_{ij} = -\Omega_{ji}$, so only one of them is independent. For the cocycle in  \eqref{2cocycle}, we have
\begin{equation}
\Omega_i = p_i + \frac{N_i(N_i-1)p_{ii}}{2}, \quad \Omega_{ij} = p_{ij}-p_{ji}. \label{eqn:invariant_exp_2_cocycle}
\end{equation}
We observe that the possible values that $\{\Omega_i,\Omega_{ij}\}$ can take saturate $|\calH^2(G,\Z_2)|$, so this is a complete set of topological invariants. By computing the topological invariants of a general cocycle (not necessarily parametrized as in \eqref{2cocycle}), one can easily assert its cohomology class. 

The third cohomology group $\calH^3(G, \Z_2) = \prod_{ij} \Z_2 \prod_{i<j<k} \Z_2 $. A general cocycle can be generated by the following root cocycles
\begin{align}
u_{ij} (a,b,c) & = \frac{[a_i]}{N_j}([b_j]+[c_j]-[b_j+c_j])  \nonumber\\
& = v_i\cup w_j(a,b,c),\nonumber\\
\tilde u_{ij} (a,b,c) & = ([a_i]+[b_i]-[a_i+b_i])\frac{[c_j]}{N_i}  \nonumber \\
& = w_i\cup v_j(a,b,c),\nonumber \\
u_{ijk}(a,b,c) & = [a_i][b_j][c_k] \nonumber\\
& = v_i\cup v_j\cup v_k(a,b,c).
\label{eqn:root_3_cocycle}
\end{align}
This set of root cocycles is over-complete. In particular, $u_{ij}$ and $\Tilde{u}_{ji}$ are equivalent cocycles. We list them here for later convenience. With these generators, a general cocycle is given by
\begin{align}
u = \sum_{ij} ( t_{ij} u_{ij} +\tilde{t}_{ij}\tilde{u}_{ij}) + \sum_{ijk} t_{ijk} u_{ijk}
\label{3cocycle}
\end{align}
where $t_{ij}, \tilde{t}_{ij}, t_{ijk} = 0, 1$  are the parameters. One can define the following topological invariants: let $\chi_a(b,c) = u(a,b,c)-u(b,a,c)+u(b,c,a)$, and define
\begin{align}
\Xi_{ij} & = \sum_{i=0}^{N_j-1} \chi_{e_i}(e_j, ne_j) \nonumber\\
\Xi_{ijk} & = \chi_{e_i}(e_j,e_k) - \chi_{e_i}(e_k,e_j)
\end{align}
where again ``modulo 2'' is assumed. One can check that they are invariant under coboundary transformations. For the cocycle in \eqref{3cocycle}, we obtain
\begin{align}
\Xi_{ij} & = t_{ij} + \tilde{t}_{ji} + \frac{N_j(N_j-1)}{2}(t_{ijj}-t_{jij}+t_{jji}) \nonumber\\
\Xi_{ijk} &  = t_{ijk}-t_{ikj}+t_{jki}-t_{jik}+t_{kij}-t_{kji}.
\label{eqn:invariant_3_cocycle}
\end{align}
By varying the parameters, one can see that the possible values that $\{\Xi_{ij}, \Xi_{ijk}\}$ can take saturate $\calH^3(G, \Z_2)$, so this is a complete set of topological invariants. Note that $\Xi_{ij}$ and $\Xi_{ji}$ are independent and  $\Xi_{ijk}$ is a fully anti-symmetric tensor.  Using topological invariants, it is sufficient to assert the cohomology class that a general 3-cocycle belongs to. In particular, if all these topological invariants are zero, it is a trivial cocycle.

\subsection{Evaluation of $\mathcal{O}_3$}
\label{app:O3}

\begin{table*}[t]
\caption{Examples of enforced symmetry breaking due to $\calO_3$ obstruction, for $\nu=4k+2$, obtained by solving conditions $\Xi_{ij}=\Xi_{ijk}=0$. We have used the topological invariants $\{\Omega_i, \Omega_{ij}\}$ to label $\lambda\in \calH^2(G,\Z_2)$. If a topological invariant is not shown, its value is arbitrary.}
\begin{tabular}{c|lllllll}
\hline
\multirow{2}{*}{\qquad \qquad \qquad \qquad  $G$\qquad \qquad  \qquad \qquad}   & \multicolumn{7}{c}{\multirow{2}{*}{\qquad \qquad \qquad    ESB occurs if the topological invariants of $\lambda$ satisfy: \qquad \qquad   \qquad }}  \\ &                                                                                                       \\ \hline
\multirow{2}{*}{$\Z_2\times \Z_2$}    &  \multicolumn{7}{c}{\multirow{2}{*}{$\Omega_{1}=\Omega_2=\Omega_{12}=1$}  }  
\\ 
& \\
\hline
\multirow{2}{*}{$\Z_2\times \Z_4$}   & \multicolumn{7}{c}{\multirow{2}{*}{$\Omega_2=\Omega_{12}=1$} }    
\\ 
&\\
\hline
\multirow{9}{*}{$\Z_2\times \Z_2\times \Z_2$}   & \multicolumn{7}{c}{\multirow{2}{*}{(1) $\Omega_1=\Omega_2=\Omega_3=1$ and any $\Omega_{ij}=1$; or }}    \\ 
&\\
& \multicolumn{7}{c}{\multirow{2}{*}{(2) $\Omega_1=\Omega_2=1, \Omega_3=0$, and $\Omega_{12}=1$; or }} \\ &   \\
 & \multicolumn{7}{c}{\multirow{2}{*}{(3) $\Omega_1=\Omega_2=1, \Omega_3=0$, $\Omega_{12}=0$ and $\Omega_{13}\neq \Omega_{23}$; or }}\\& \\
 & \multicolumn{7}{c}{\multirow{2}{*}{(4) $\Omega_1=1, \Omega_2=\Omega_3=0$, $\Omega_{23}=1$ and $\Omega_{12}=0$; or }} \\ & \\
  & \multicolumn{7}{c}{\multirow{2}{*}{(5) $\Omega_1=\Omega_2=\Omega_3=0$, and $\Omega_{12}=\Omega_{23}=\Omega_{13}=1$; or } } \\&   \\
   & \multicolumn{7}{c}{\multirow{2}{*}{(6) permutations of indices 1, 2 and 3 of the above cases} } \\&   \\ \cline{1-8}
\multicolumn{1}{c|}{\multirow{5}{*}{$\Z_2\times \Z_2\times \Z_4$}} & \multicolumn{7}{c}{\multirow{2}{*}{(1) $\Omega_3=1$ and $(\Omega_{12}-1)(\Omega_{13}-1)=0$; or }} \\&      \\
\multicolumn{1}{c|}{}   & \multicolumn{7}{c}{\multirow{2}{*}{(2) $\Omega_3=0$ and $\Omega_1=\Omega_2=\Omega_{12}=1$; or}} \\& \\
\multicolumn{1}{c|}{} & \multicolumn{7}{c}{\multirow{2}{*}{ (3) $\Omega_3 =0$, $\Omega_1=1$, $\Omega_2\Omega_{12}=0$ and $\Omega_{13}\Omega_2 + (\Omega_{12}-1)\Omega_{23}=1$; or }}\\ & \\
\multicolumn{1}{c|}{} & \multicolumn{7}{c}{\multirow{2}{*}{ (4) permutations of indices 1 and 2 of the above cases} }\\ & \\ \cline{1-8} 
\multicolumn{1}{c|}{\multirow{2}{*}{$\Z_2\times \Z_4\times \Z_4$}} & \multicolumn{7}{c}{\multirow{2}{*}{ $(\Omega_2-1)(\Omega_3-1)=0$ and $(\Omega_{12}-1)(\Omega_{13}-1)=0$}} \\ & \\
\hline
\end{tabular}
\label{table:app_O3_abelian}
\end{table*}

Here we apply the above topological invariants to  study the $\calO_3$ obstruction for $\nu=4k+2$, where $\mathcal{O}_3 = n_1\cup\lambda + \lambda\cup_1\lambda$. We will also discuss $\calO_3$ for $\nu=4k$, which is $\mathcal{O}_3 = n_1\cup\lambda$. We will give the sufficient and necessary conditions on when $\calO_3$ obstruction vanishes for Abelian $G$ and show the condition on $\lambda\in\calH^2(G,\Z_2)$ for ESB to occur for some examples. 

First, we give an explicit parameterization of the 1-cocycle $n_1$ and 2-cocycle $\lambda$. According to Eq.~(\ref{eqn:1_cocycle_paramet}) and Eq.(\ref{2cocycle}), we  take the following choices:
\begin{align}
n_1 = \sum_{i}q_i v_i, \quad \lambda = \sum_{i}p_i w_i +\sum_{ij} p_{ij} w_{ij}
\end{align}
where $v_i$, $w_{i}$ and $w_{ij}$ are the root cocycles in (\ref{eqn:root_1_cocycle}) and (\ref{eqn:root_2_cocycle}). Note that the parametrization of $\lambda$ is over-complete. The same parametrization is used in the main text. 

Secondly, we make a simplification on the expression of $\mathcal{O}_3$ such that it can be expressed in terms of the root cocycles given in (\ref{eqn:root_3_cocycle}). For a cocycle $\lambda\in \calH^2(G,\Z_2)$, one can show this very helpful relation:
\begin{align}
\lambda \cup_1 \lambda = \frac{1}{2}\hat{\dd}\lambda.
\end{align}
(Recall that $\hat{\dd}$ is defined without taking modulo 2.) At the same time, one can check that $\hat{\dd}w_i = 0$, and
\begin{align}
\frac{1}{2}\hat{\dd}w_{ij} & = \frac{1}{2}(\hat{\dd}v_i \cup v_j + v_i\cup\hat{\dd}v_j) \nonumber\\
&= \frac{N_j}{2} v_i \cup w_j + \frac{N_i}{2}w_i\cup v_j.
\end{align}
Accordingly, we have
\begin{align}
\mathcal{O}_3 & = n_1\cup\lambda + \lambda \cup_1\lambda \nonumber\\
& = \sum_{ij}\left(q_ip_j v_i\cup w_j + p_{ij} \frac{N_j}{2} v_i \cup w_j + p_{ij} \frac{N_i}{2}w_i\cup v_j\right) \nonumber \\
&\qquad + \sum_{ijk}q_ip_{jk}v_i\cup w_{jk} \nonumber \\
&= \sum_{ij}\bigg[ \left(q_ip_j+p_{ij} \frac{N_j}{2}\right)u_{ij}+ p_{ij} \frac{N_i}{2}\tilde u_{ij}\bigg]\nonumber \\
&\qquad + \sum_{ijk}q_ip_{jk}u_{ijk}. 
\end{align}
Comparing to \eqref{3cocycle} and using \eqref{eqn:invariant_3_cocycle}, we immediately have
\begin{align}
\Xi_{ij} & = q_ip_j + (p_{ij}+p_{ji})\frac{N_j}{2} \nonumber \\
&+ \frac{N_j(N_j-1)}{2}(q_ip_{jj}-q_jp_{ij}+q_jp_{ji}) \nonumber\\
\Xi_{ijk} & = q_i(p_{jk} -p_{jk}) + q_j(p_{ki} -  p_{ik}) + q_k (p_{ij} -p_{ji} ).
\end{align}
Making use of the expressions of $\Omega_i$ and $\Omega_{ij}$ in \eqref{eqn:invariant_exp_2_cocycle}, we have
\begin{align}
\Xi_{ij} & = q_i \Omega_j + \frac{N_j}{2}(q_j-1)\Omega_{ij}, \nonumber\\
\Xi_{ijk} & = q_i \Omega_{jk} + q_j\Omega_{ki} +q_k \Omega_{ij} 
\label{eq:appXI-1}
\end{align}
where ``$\modulo{2}$'' is assumed and $N_i=2^{k_i}$ is used. For $\calO_3=\lambda\cup n_1$ at $\nu=4k$, the result is very similar: $\Xi_{ijk}$ is the same, while $\Xi_{ij}$ becomes
\begin{align}
    \Xi_{ij} & = q_i \Omega_j + \frac{N_j}{2}q_j\Omega_{ij}.
    \label{eq:appXI-2}
\end{align}
The equations in \eqref{eq:appXI-1} and \eqref{eq:appXI-2} are the main results of the calculations in this subsection. 

We can now assert that $\calO_3$ is trivial if and only if $\Xi_{ij} = \Xi_{ijk}=0$ for all $i,j,k$. Recall that $\Omega_i,\Omega_{ij}$ (i.e., $\lambda$) are given in our problem to generate the fermionic symmetry group $G_f$. For given $\{\Omega_i, \Omega_{ij}\}$, the condition $\Xi_{ij}=\Xi_{ijk}=0$ is a set of equations with $\{q_i\}$ being the unknowns. Then, $G_f$ is enforced to break if these equations have no solutions. That is,  ESB of $G_f$ occurs if $\mathcal{O}_3$ is nontrivial regardless of $n_1$. 

For $\nu=4k$, there is always a solution to $\Xi_{ij}=\Xi_{ijk}=0$:  $q_i=0$ for all $i$, i.e., a trivial $n_1$. So, $\calO_3$ obstruction alone cannot lead to ESB. For $\nu=4k+2$, we do have situations that ESB can occur. With straightforward analysis on equations in \eqref{eq:appXI-1}, we find a few examples of ESB, which are summarized in Table \ref{table:app_O3_abelian}. Two general properties of \eqref{eq:appXI-1} are: (1) If $\Omega_{ij}=0$ for all $i$ and $j$ and $\Omega_i$ is nonzero for at least one $i$, say $\Omega_I \neq 0$ of a fixed $I$, the requirement $\Xi_{iI}=0$ gives $q_i = 0$ for all $i$. That is, there is one and only one solution $n_1=0$. (2) If $N_j =0 \modulo{4}$ for all $j$,  then $n_1=0$ is always a solution.

\subsection{A result for $\nu=4k+2$}

We have only considered $\calO_3$ obstruction above. In this subsection, we consider $\calO_4$ in a special case. We show the following corollary for finite Abelian groups.
\begin{col}
\textit{For 2D fermionic iTOs with $\nu=4k+2$ and a finite Abelian group $G$, ESB can not occur if $\lambda$ is a type-I 2-cocycle in $\calH^2(G,\Z_2)$.}
\end{col}

Recall that we have defined type-I 2-cocycles in $\calH^2(G,\Z_2)$ as those with $\Omega_{ij}=0$ for all $i,j$. If $\Omega_i=0$ for all $i$, i.e., $\lambda=0$, then ESB can never occur. If at least one $\Omega_i$ is nonzero, then, according to the result at the end of Appendix \ref{app:O3},  we must have $q_i=0$ for all $i$ to have a vanishing $\calO_3$. To show that ESB does not occur, it remains to check if $\calO_4$ vanishes when $n_1=0$ and $\lambda$ is type-I.

In that case, the $\calO_4$ obstruction formula in  (\ref{eqn_H4_nu2}) is applicable. For type-I $\lambda$, we can write it as $\lambda =\sum_i p_i \hat{\dd}v_i/N_i$. It leads to two consequences: $\hat{\dd} \lambda=0$, and 
\begin{align}
    \lambda \cup \lambda=\sum_{ij} p_ip_j \hat{\dd }\left( \frac{v_i\cup\hat{\dd}v_j }{N_iN_j} \right),
\end{align}
which means the piece $e^{i\nu\pi\lambda\cup\lambda/8}$ is a $U(1)$-valued 4-coboundary. Then, one can easily check that  $n_2=0$ makes the $\mathcal{O}_4$ in (\ref{eqn_H4_nu2}) a $U(1)$-valued 4-coboundary. Accordingly, we find a solution that $n_1=n_2=0$ such that both $\calO_3$ and $\calO_4$ vanish. Hence, there is no ESB and the corollary is proven.

\end{document}